\documentclass[a4paper,fleqn]{cas-dc}

\usepackage[square,numbers,sort&compress]{natbib}

\usepackage{ifpdf}
\usepackage{graphicx}
\DeclareGraphicsExtensions{.pdf}
\usepackage[squaren]{SIunits}
\usepackage{sistyle}
\usepackage{amsmath,amsfonts,amssymb,amsxtra}
\usepackage{upgreek}
\usepackage{hyperref}
\hypersetup{colorlinks=false,bookmarksopen=true,linkbordercolor={1 1 1}}
\usepackage{subfig}
\usepackage{xcolor}
\usepackage{array}
\usepackage{multirow}
\usepackage{verbatim}

\def\tsc#1{\csdef{#1}{\textsc{\lowercase{#1}}\xspace}}
\tsc{WGM}
\tsc{QE}
\tsc{EP}
\tsc{PMS}
\tsc{BEC}
\tsc{DE}

\begin{document}
\let\WriteBookmarks\relax
\def\floatpagepagefraction{1}
\def\textpagefraction{.001}

\shorttitle{HV-Splitter: Design and Performance}
	
\shortauthors{P Walker et~al.}
	
\title [mode = title]{The High Voltage Splitter board for the JUNO SPMT system}              

\renewcommand{\printorcid}{}

\author[1,2]{Pablo Walker}[type=editor, auid=000, bioid=1]
\cormark[1]
\ead{pwalker1@uc.cl}

\author[1,2,3]{Juan Pedro Ochoa-Ricoux}
\author[1,2]{Angel Abusleme}
\author[1,2]{Agustin Campeny}

\author[4]{Mathieu Bongrand}
\author[5,6]{Cl\'{e}ment Bordereau}
\author[7]{Jos\'{e} Busto}
\author[8]{Anatael Cabrera}
\author[5]{St\'{e}phane Callier}
\author[4]{Steven Calvez}
\author[5]{C\'{e}dric Cerna}
\author[5]{Thomas Chabot}
\author[6]{Po-An Chen}
\author[9]{Guoming Chen}
\author[10,11]{Ziliang Chu}
\author[5]{G\'{e}rard Claverie}
\author[8]{Christophe De La Taille}
\author[5]{Charles-Edouard Demonchy}
\author[8]{Selma Conforti Di Lorenzo}
\author[5]{Fr\'{e}d\'{e}ric Druillole}
\author[10]{Lei Fan}
\author[5]{Am\'{e}lie Fournier}
\author[8]{Yang Han}
\author[10]{Miao He}
\author[5]{Patrick Hellmuth}
\author[1]{Rafael Herrera}
\author[6]{Yee Hsiung}
\author[6]{Bei-Zhen Hu}
\author[10]{Jun Hu}
\author[12]{Guihong Huang}
\author[9]{Yongbo Huang}
\author[5]{C\'{e}dric Huss}
\author[4]{L\'{e}onard Imbert}
\author[3]{Adrienne Jacobi}
\author[1]{Ignacio Jeria}
\author[10]{Xiaoshan Jiang}
\author[10]{Xiaoping Jing}
\author[5]{C\'{e}cile Jollet}
\author[5]{Jean Jouve}
\author[3]{Sindhujha Kumaran}
\author[5,14]{Loïc Labit}
\author[5]{S\'{e}bastien Leblanc}
\author[4]{Victor Lebrin}
\author[5]{Matthieu Lecocq}
\author[9]{Hongbang Liu}
\author[12]{Kainan Liu}
\author[9]{Xiwen Liu}
\author[3]{Roberto Carlos Mandujano}
\author[5]{Anselmo Meregaglia}
\author[8]{Diana Navas-Nicolas}
\author[10]{Zhe Ning}
\author[10,11]{Yujie Niu}
\author[5]{Fr\'{e}d\'{e}ric Perrot}
\author[4]{Rebin Karaparambil Rajan}
\author[5]{Reem Rasheed}
\author[5]{Abdel Rebii}
\author[5]{Mathieu Roche}
\author[8]{Cayetano Santos}
\author[4]{Mariangela Settimo}
\author[10]{Yunhua Sun}
\author[9]{Jingzhe Tang}
\author[1]{Giancarlo Troni}
\author[4]{Benoit Viaud}
\author[13]{Chung-Hsiang Wang}
\author[10]{Jian Wang}
\author[10]{Yangfu Wang}
\author[10]{Zhimin Wang}
\author[10]{Diru Wu}
\author[14]{Jacques Wurtz}
\author[10]{Wan Xie}
\author[10]{Jilei Xu}
\author[9]{Jinghuan Xu}
\author[10]{Meihang Xu}
\author[9]{Chengfeng Yang}
\author[4]{Fr\'{e}d\'{e}ric Yermia}
\author[9]{Junwei Zhang}
\author[9]{Siyuan Zhang}

\affiliation[1]{organization={Pontificia Universidad Cat\'{o}lica de Chile}, city={Santiago}, postcode={7820436}, country={Chile}}
\affiliation[2]{organization={Millennium Institute for Subatomic Physics at the High-energy Frontier (SAPHIR)}, city={Santiago}, postcode={7591538}, country={Chile}}
\affiliation[3]{organization={Department of Physics and Astronomy, University of California}, city={Irvine}, postcode={CA 92697}, country={USA}}
\affiliation[4]{organization={SUBATECH, Universit\'{e} de Nantes, IMT Atlantique, CNRS-IN2P3}, city={Nantes}, country={France}}
\affiliation[5]{organization={University of Bordeaux, CNRS, LP2i, UMR 5797}, postcode={F-33170 Gradignan}, country={France}}
\affiliation[6]{organization={Department of Physics, National Taiwan University}, city={Taipei}, country={Taiwan}}
\affiliation[7]{organization={Aix Marseille Univ, CNRS/IN2P3, CPPM}, city={Marseille}, country={France}}
\affiliation[8]{organization={Laboratoire des Physique des 2 Infinis Irène Joliot-Curie}, city={Paris}, country={France}}
\affiliation[9]{organization={Guangxi University}, city={Nanning}, country={China}}
\affiliation[10]{organization={Institute of High Energy Physics}, city={Beijing}, country={China}}
\affiliation[11]{organization={University of Chinese Academy of Sciences}, city={Beijing}, country={China}}
\affiliation[12]{organization={Wuyi University}, city={Jiangmen}, country={China}}
\affiliation[13]{organization={National United University}, city={Miao-Li}, country={Taiwan}}
\affiliation[14]{organization={Institut Pluridisciplinaire Hubert Curien, Université de Strasbourg}, city={Strasbourg}, postcode={F-67037}, country={France}}

\cortext[cor1]{Corresponding author}

\begin{abstract}
The Jiangmen Underground Neutrino Observatory (JUNO) in southern China is designed to study neutrinos from nuclear reactors and natural sources to address fundamental questions in neutrino physics. Achieving its goals requires continuous operation over a 20-year period. The small photomultiplier tube (small PMT or SPMT) system is a subsystem within the experiment composed of 25,600 3-inch PMTs and their associated readout electronics. The High Voltage Splitter (HVS) is the first board on the readout chain of the SPMT system and services the PMTs by providing high voltage for biasing and by decoupling the generated physics signal from the high-voltage bias for readout, which is then fed to the front-end board. The necessity to handle high voltage, manage a large channel count, and operate stably for 20 years imposes significant constraints on the physical design of the HVS. This paper serves as a comprehensive documentation of the HVS board: its role in the SPMT readout system, the challenges in its design, performance and reliability metrics, and the methods employed for production and quality control.
\end{abstract}


\maketitle

\clearpage
\section{Introduction}
\label{section:intro}

\subsection{The JUNO experiment}

The Jiangmen Underground Neutrino Observatory, or simply JUNO, is a multi-purpose neutrino experiment located in southern China, installed within a cavern with a $650\mathrm{\:m}$ overburden (1800 m.w.e.). The main objective of the experiment is to determine the neutrino mass ordering with a $3\sigma$ significance within roughly six years of data taking and to measure three oscillation parameters to sub-percent precision \cite{An2016, Li2013, Abusleme2022, Abusleme2025_2}. Both of these goals will be achieved through the detection of reactor antineutrinos from two nuclear power plants located at $\sim$$53\mathrm{\:km}$ from the experiment. JUNO will also study neutrinos from several natural sources and will search for new physics \cite{Juno2021, Abusleme2023_2, Abusleme2022_2, Abusleme2023_3, Abusleme2021_4, Zhao2024, Abusleme2021_3, Molla2023, Abusleme2023_4, Juno2025, Abusleme2023_5}.

\begin{figure}[t]
	\centering
	\includegraphics[width=0.9\linewidth]{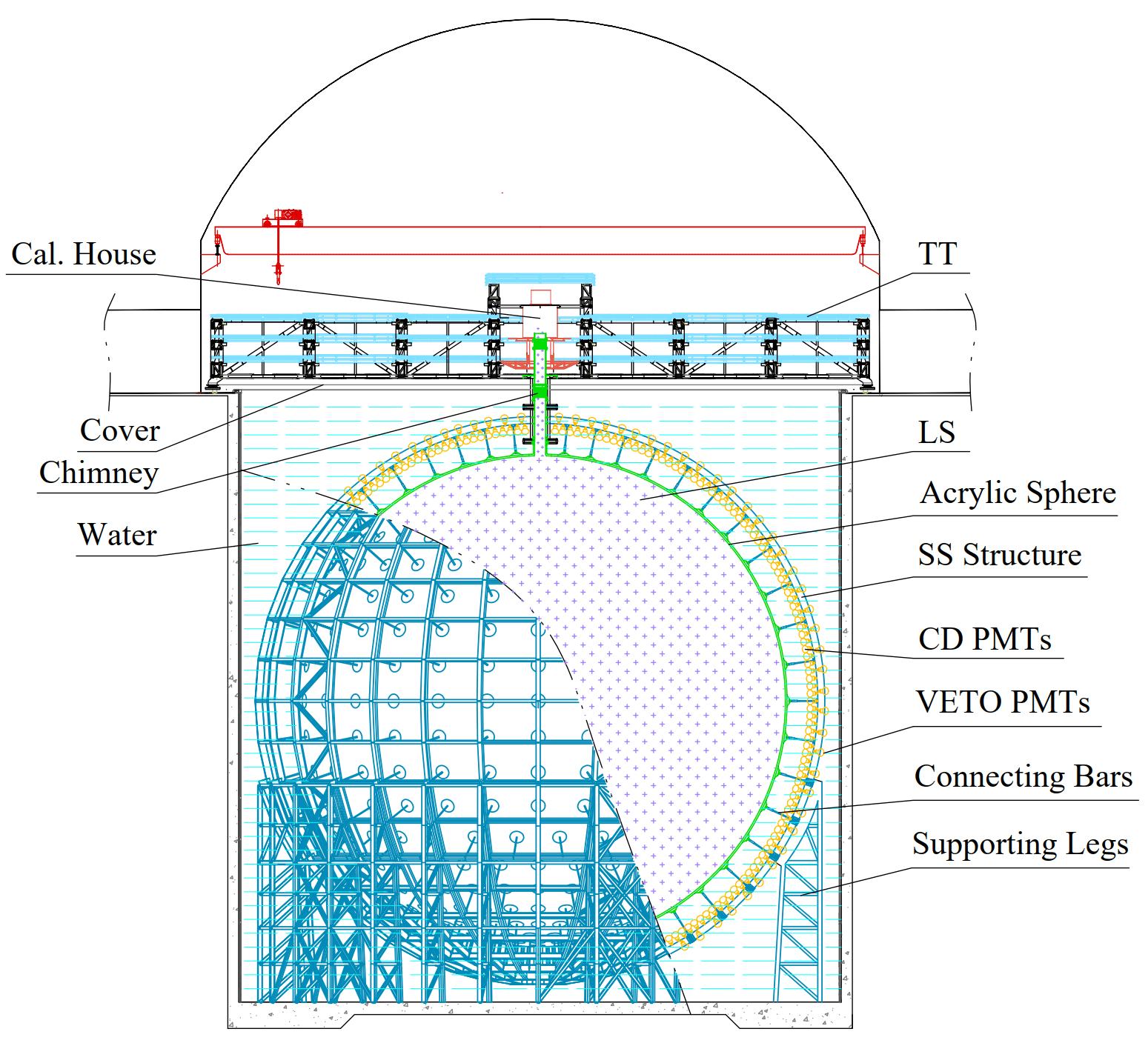}
	\caption{Schematic view of the JUNO detector \cite{Juno2021}.}
	\label{juno_schematic}
\end{figure}

Figure \ref{juno_schematic} shows a schematic view of the JUNO experiment. At the heart of it is the Central Detector (CD), a spherical 35.4-meter acrylic vessel holding $20\mathrm{\:kton}$ of liquid scintillator (LS), held and supported by a stainless steel frame \cite{Abusleme2023}. Photodetection of particle interactions within the CD is achieved through two independent and complementary systems: 17,612 20-inch photomultiplier tubes (PMTs) and 25,600 3-inch PMTs, which, together with their associated readout electronics, comprise the large PMT (LPMT) and small PMT (SPMT) systems, respectively \cite{Abusleme2022_3, Cerna2025}. The CD is submerged in a pool filled with $35\mathrm{\:kton}$ of ultrapure water, which forms part of an outer water Cherenkov detector \cite{Lu2023}, and provides passive shielding from natural radioactivity from the rock and neutrons from cosmic rays. The water Cherenkov detector is instrumented by 2,400 20-inch PMTs and supplemented by a Top Tracker (TT) placed atop consisting of three layers of plastic scintillator---repurposed from the Opera experiment---that provide precision tracking of cosmic ray muons \cite{Sandanayake2024}.

The LPMT system, thanks to its large photocoverage (75\%), drives the energy resolution of the JUNO detector. The 3-inch PMTs, installed in the gaps between the 20-inch PMTs, only slightly increase the photocoverage of the detector to 78\%, but more importantly, provide a complementary and independent set of sensors looking at the same events as the 20-inch PMTs. Due to their size, the SPMTs operate mostly in photon-counting mode in the energy region of interest for reactor antineutrino physics [1--10 MeV], providing charge measurements with essentially no instrumental non-linearities and reducing the impact of systematic uncertainties in the energy response of the LPMT system. Additionally, the SPMT system will extend the energy dynamic range of the detector and, due to the excellent timing resolution of the 3-inch PMTs (e.g., transit spread time (TTS) of $1.6\textrm{\:ns}$) \cite{Cao2021}, will help improve event reconstruction, including for cosmic-ray muons and supernova neutrinos.

\subsection{Overview of this work}

\begin{figure}[t]
	\centering
	\subfloat[][Top side]
	{\resizebox{0.45\textwidth}{!}{\includegraphics{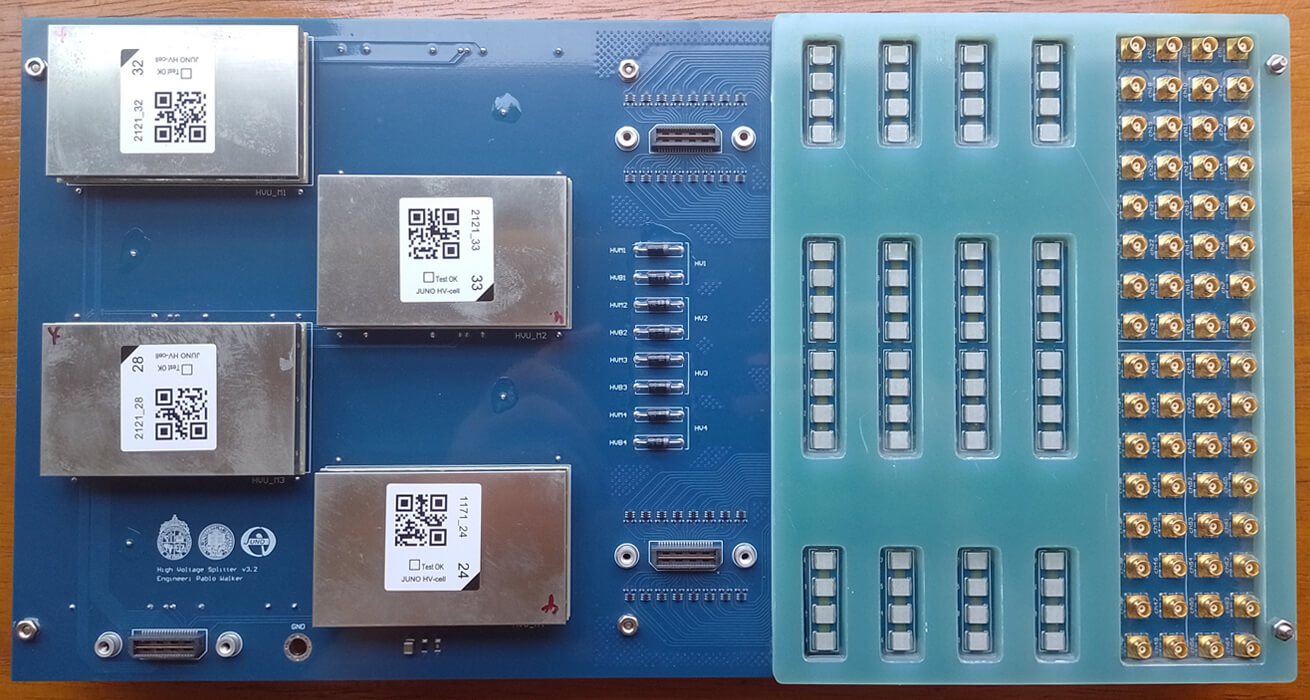}}}\\
	\subfloat[][Bottom side]
	{\resizebox{0.45\textwidth}{!}{\includegraphics{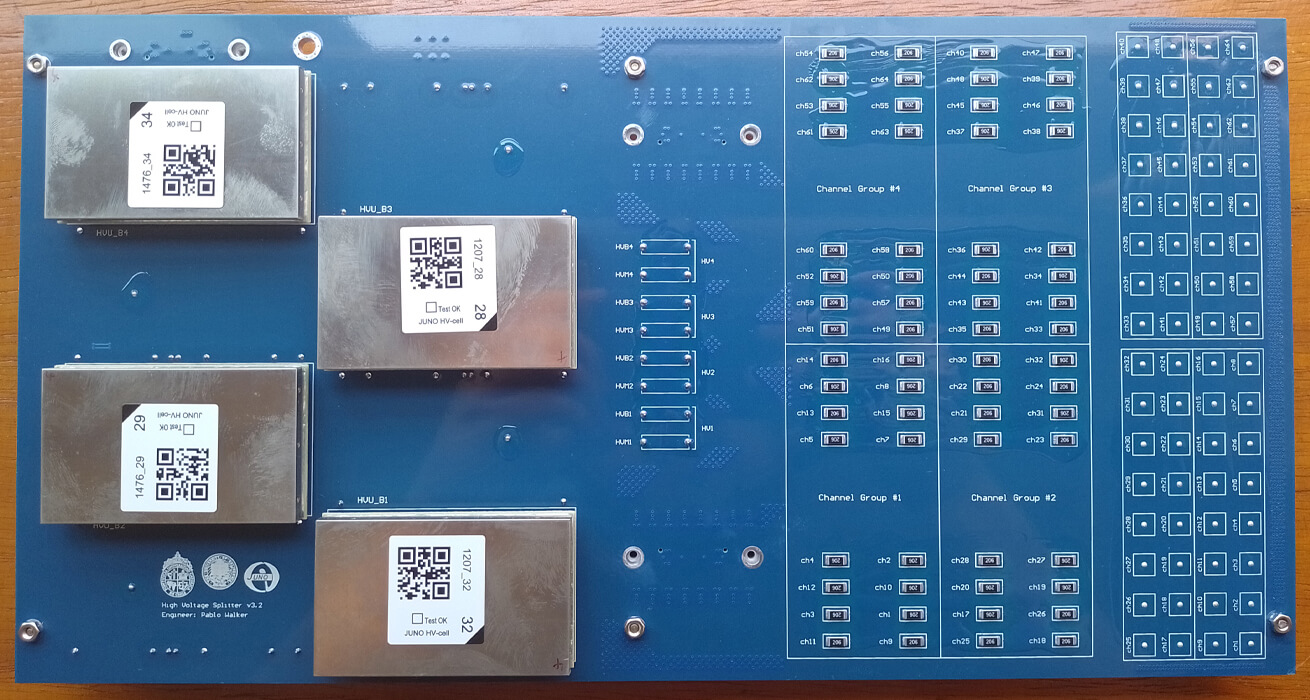}}}
	\caption{The HVS version 3.2. Glued to the right side of the board, as visible on the topside view, is an FR-4 piece with cutouts around the decoupling capacitors and the MCX connectors, which is used as a mold for pouring the insulating compound around these critical components.}
	\label{HVS_top_bottom}
\end{figure}

This paper documents the design and performance of one of the printed circuit boards (PCBs) that forms part of the readout electronics in the SPMT system: the High Voltage Splitter (HV-Splitter or HVS) board. The HV-Splitter is a simple board, both in terms of functionality and circuit design. The electrical behavior of the circuit is straightforward and easy to understand. Yet, the physical implementation of the circuit is very nuanced, as it must comply with stringent constraints related to its size, channel density, voltage, and most importantly, reliability. Figure \ref{HVS_top_bottom} shows an image of the last iteration of the HVS PCB (version 3.2), which was installed in JUNO. An overview of the SPMT system can be found in Ref. \cite{Juno2021}, while a more comprehensive review of the readout electronics is presented in Ref. \cite{Cerna2025}.

The document is broadly divided into three parts. The rest of Section \ref{section:intro} provides an overview of the functionality of the SPMT readout system and the various boards that form part of it. Sections \ref{section:system}, \ref{section:circuit}, and \ref{section:layout} cover the design of the HVS board following a top-down approach, starting with the system-level constraints and requirements, followed by the circuit-level design and physical-level design. Finally, Sections \ref{section:performances}, \ref{section:reliability}, and \ref{section:production} focus on the board's performance, along with the methods employed for production and quality control.

\subsection{The SPMT readout system}

The 25,600 3-inch PMTs that comprise the SPMT system are powered and read out in groups of 128 channels, resulting in a total of 200 channel groups, each serviced by an Underwater Box (UWB), the local front-end readout unit. The UWB is a cylindrical, precisely machined, and watertight stainless steel box that houses the front-end electronics. It is mounted to the stainless steel structure holding the acrylic sphere of the CD, in the ultrapure-water phase of the detector. The functions of the UWB electronics include, in broad terms, the supply of high voltage and decoupling of the physics signal; the conditioning and digitization of the signal; and data acquisition and communication with the surface electronics room.

\begin{figure}[t]
	\centering
	\includegraphics[width=1\linewidth, trim={2.4cm 0 2cm 1cm},clip]{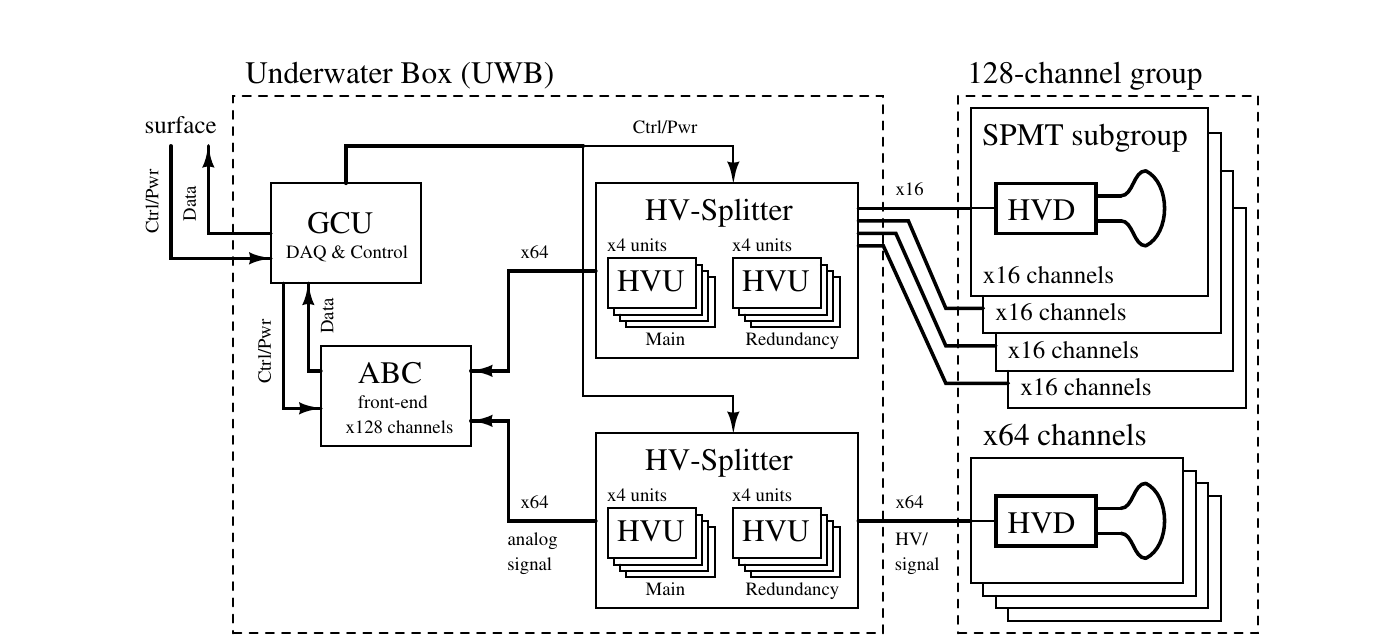}
	\caption{Block diagram of the SPMT readout system.}
	\label{block_diagram}
\end{figure}

Figure \ref{block_diagram} shows a block diagram of the SPMT readout system for a single channel group composed of 128 SPMTs. The front-end readout electronics consist of four PCBs: two copies of the High Voltage Splitter (HVS), each servicing 64 channels, one ASIC Battery Card (ABC) and one Global Control Unit (GCU) \cite{Cerna2025}.

The 3-inch PMTs, which were produced by Hainan Zhanchuang Photonics Technology Co. Ltd (HZC), were selected from the vendor to operate with a gain of $3\times10^6$, hereafter referred to as nominal gain, within a voltage range of $900$--$1300\mathrm{\:V}$ \cite{Cao2021}.

The HVS houses eight copies of the High Voltage Unit (HVU), a custom DC-DC converter board capable of generating the high voltage necessary to bias the SPMTs \cite{Bellato2021}. The HVUs are powered by $24\mathrm{\:V}$, and have a configurable high-voltage output which can be set within the range of $800$--$1500\mathrm{\:V}$. Under normal conditions, only four of the eight HVUs on the HVS output high voltage, while the remaining four serve as backups, implementing a 1:1 redundancy scheme.

In order to simplify integration, each 128-channel group is further divided into subgroups of 16 with a similar operating voltage at nominal gain, and all channels in that subgroup share the combined HV output of one HVU and its respective backup. The high voltage is delivered to the SPMTs through long, waterproof coaxial cables, custom-made in partnership with the Axon Cable Company, which can be of either $5\mathrm{\:m}$ or $10\mathrm{\:m}$ in length. The dynode chain of each SPMT is locally biased through a resistive divider network located on the High Voltage Divider (HVD), a PCB mounted on the base of the SPMT \cite{Wu2022, Xu2025}.

The signal path begins at the SPMT, which, when illuminated, generates a voltage signal with a relatively small amplitude---approximately $2\mathrm{\:mV}$ for a single photoelectron (SPE). The same coaxial cable used to carry the high voltage to the SPMT is also used to deliver the voltage signal to the HVS, where it is decoupled from the high-voltage DC component, and fed to the ABC board.

The ABC board is the heart of the readout system, as it houses the front-end readout ASIC, namely the CATIROC chip \cite{Conforti2021}. This custom chip is responsible for signal discrimination, filtering and sampling. A single ABC board services all 128 channels, which is accomplished by eight CATIROCs of 16 channels each. Local control is handled by a Kintex-7 field-programmable gate array (FPGA), which also handles the data transfer between the ABC board and the GCU.

The GCU has a multitude of functions, the main one being data readout from the ABC and communication with the data acquisition (DAQ) system at the surface facility for storage and analysis. It serves as the central control of the UWB, handling clock distribution and slow control of the ABC and the HVUs on the HVS boards. Additionally, it delivers power to the ABC and the HVS boards.

\subsection{Physical distribution of components}

\begin{figure}[t]
	\centering
	\includegraphics[width=1\linewidth]{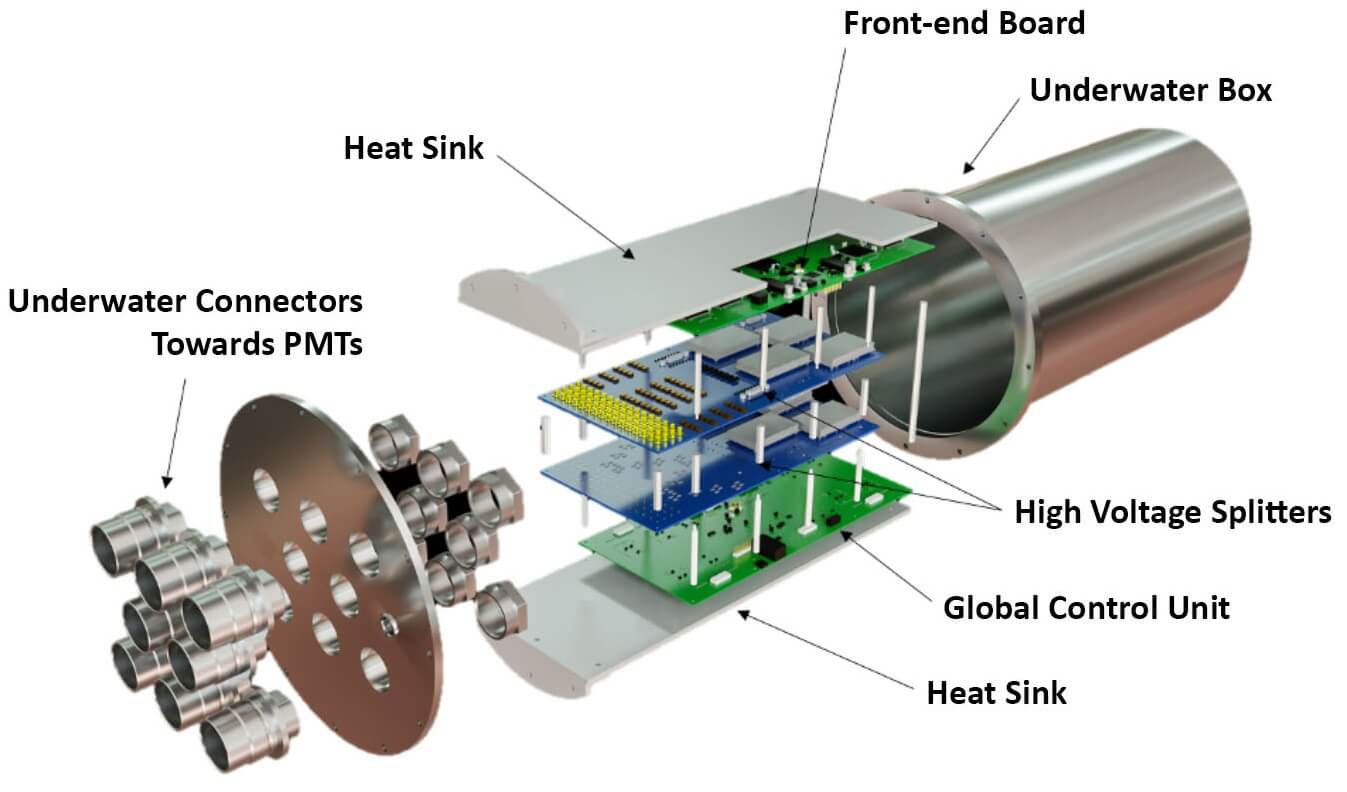}
	\caption{Exploded view of the main components of the UWB and associated electronics \cite{Juno2021}.}
	\label{UWB_diagram}
\end{figure}

\begin{figure}[t]
	\centering
	\includegraphics[width=0.6\linewidth]{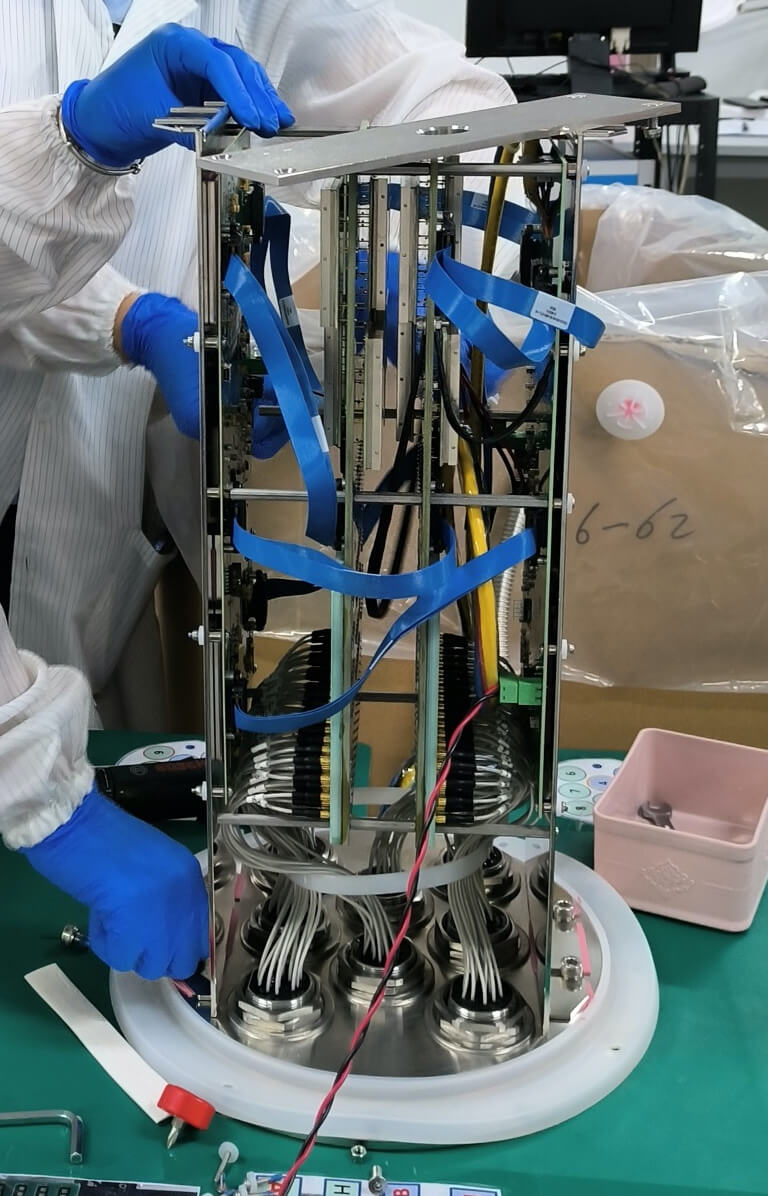}
	\caption{The readout electronics connected to the inside of the UWB lid during integration.}
	\label{UWB_integration}
\end{figure}

Figure \ref{UWB_diagram} shows an exploded view of the physical components of the UWB and readout electronics, which include the stainless steel box, the four PCBs, two heatsinks, and eight underwater connectors on the lid of the box. Figure \ref{UWB_integration} shows a photograph of the assembled board stack, connected and attached to the lid of the UWB.

Figures \ref{UWB_diagram} and \ref{UWB_integration} show that the PCBs are stacked through spacers and sandwiched between two heatsinks, which are thermally connected to the heat-generating components on the ABC and GCU boards on one side, and to the UWB lid on the other. This configuration allows the UWB to dissipate the heat generated by the electronics to the surrounding water, which is temperature-controlled at $21 \pm 1^\circ \mathrm{C}$ \cite{Lu2023}.

There are eight underwater connectors, each handling one of the 16-channel subgroups, which act as the interface between the submerged SPMTs and the dry electronics. On the outside of the UWB, they are connected to the SPMTs via long coaxial cables ($5\mathrm{\:m}$ or $10\mathrm{\:m}$), and on the inside, to the two HVS boards via short $15\mathrm{\:cm}$ coaxial cables. The underwater connectors, along with their matching receptacles, were custom-designed by Axon specifically for the JUNO experiment.

Not visible in the images is a flexible, waterproof stainless steel bellows, welded to the back of the UWB, on the opposite end to the underwater connectors. It houses the cables that connect the surface electronics room to the UWB, which are used for power delivery and communication.

\section{System-level design}
\label{section:system}

The purpose of the HVS board is twofold: to provide high voltage to the SPMTs for biasing, and to decouple the signal pulses generated by the SPMTs from their high-voltage DC component for readout. There are mainly four conflicting requirements that constrain the design of the HVS which are set at a system-level: reliability, high voltage, channel density, and size.

\subsection{Design constraints}

\subsubsection{Reliability}
\label{subsubsection:reliability}
Even though the JUNO experiment is planned to reach some of its research goals within roughly six years of data taking, it is designed for a 20-year expected lifetime, during which it will operate continuously. This means that each individual component, the HVS included, must have a robust design, with carefully selected components and materials, and a performance in terms of reliability that must be in compliance with the standards of the experiment. For the SPMT system, the system-level failure rate requirement has been set to a maximum allowable loss of 10\% after six years of operation.

\subsubsection{Operational voltage}
The operational voltage range for the HVS is determined by the voltage required to bias the 3-inch PMTs at nominal gain ($3\times 10^6$), which were specifically selected from the vendor within a cutoff window ranging from $900$--$1300\mathrm{\:V}$ \cite{Cao2021}. The HVUs, which supply high voltage to the SPMTs, were originally developed for the LPMT system, and were later adapted to the SPMT system by adjusting their voltage output range to $800$--$1500\mathrm{\:V}$. With a readily available design, and given the fact that they had already been tested to meet the JUNO reliability standards, it was decided to reuse and adapt the HVUs' design for the SPMT system.

\subsubsection{Channel density}
The waveforms generated by the SPMTs are not recorded, with only charge and timing information being sampled and stored for analysis. The electronic readout for the SPMT system is designed to be more compact than that of the LPMT system, using fewer electronic boards, leading to large 128-channel groups being serviced by a single set of electronics.

\subsubsection{Board size}
The size of the HVS board was influenced by the dimensions of the container and the arrangement of other boards in the stack. The UWB is cylindrical, with a diameter of $25.4\mathrm{\:cm}$ and a length of $50\mathrm{\:cm}$, which, along with cable positioning and clearance requirements, limited the maximum size of the HVS. Additionally, shared interfaces and mounting holes between the boards required proper clearances, indirectly influencing the size of the HVS board to ensure adequate high-voltage insulation. 

\subsubsection{Conflicting requirements}
Although the size of the UWB may seem considerable for a PCB, the high voltage, channel density, reliability, and available size were often in conflict during the design process of the HVS. The use of high voltage introduces a large clearance overhead between high and low voltage traces. The need for reliability during a long expected lifetime introduces the need to consider that any individual channel may fail in time, and therefore, a large clearance must be used between neighboring high-voltage channels as well. The clearance overhead and multiplicity of channels lead to a larger board, yet, the size of the board is limited by that of the container. In order to handle 128 channels, it was decided to use two identical HVS boards of 64 channels, as a single PCB with 128 channels was deemed impractical within the available space.

\subsubsection{Signal integrity}
\label{subsubsection:integrity}
Another important constraint on the design of the HVS is related to noise and distortion effects. The average voltage amplitude of an SPE is roughly $2\mathrm{\:mV}$, while the threshold of the trigger system on the ABC board is set to $1/3$ of that amplitude. Both electronic noise and signal distortions, including reflections and crosstalk, must be limited to reduce the risk of false positives. It was thus required that the RMS noise and distortion levels in the analog part of the electronics be kept below $1/10$ of an SPE. Although the design of the HVS does not inherently introduce noise, distortion effects are inevitable. To minimize reflections, the analog traces on the HVS were matched to the $50\:\Omega$ characteristic impedance of the analog path of the SPMT system.

\section{Circuit-level design}
\label{section:circuit}

\subsection{Single-channel circuit schematic}

\begin{figure}[t]
	\centering
	\includegraphics[width=1\linewidth]{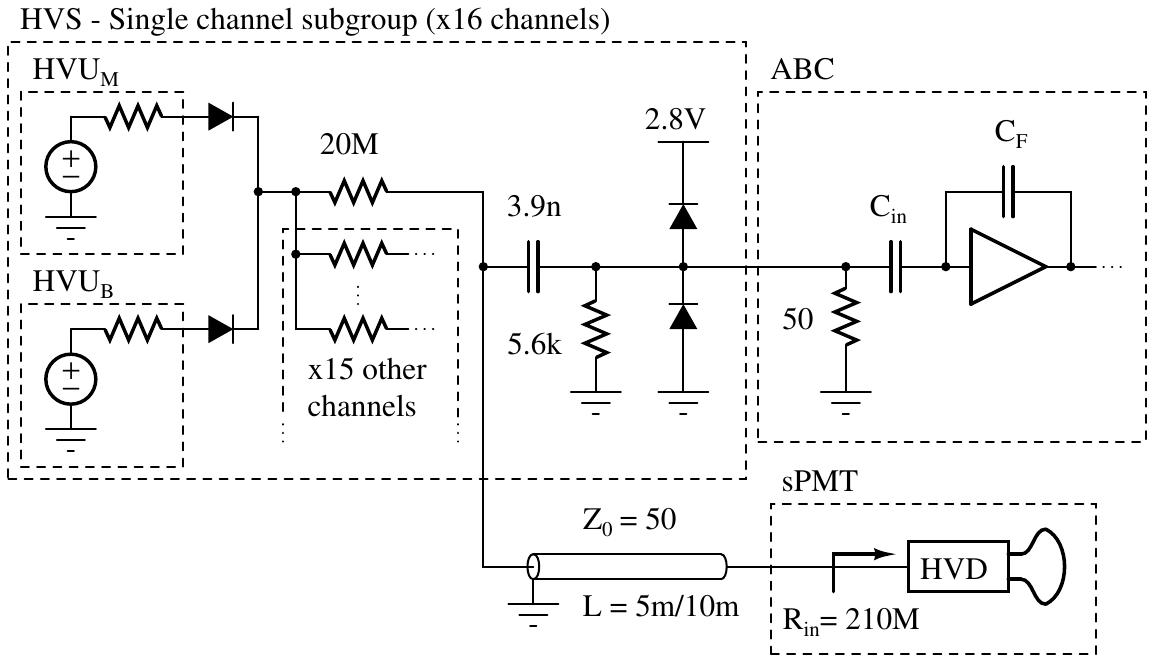}
	\caption{Simplified circuit schematic for a single channel of the HV-Splitter board.}
	\label{simplified_sch}
\end{figure}

Figure \ref{simplified_sch} shows a simplified circuit schematic for a single analog channel on the HVS board. The HVU, which is the DC-DC converter responsible for generating high voltage, is represented by a Th\'evenin equivalent circuit---a voltage source and an output resistance. The schematic includes two HVUs: the main unit ($\mathrm{HVU}_\mathrm{M}$) and the backup unit ($\mathrm{HVU}_\mathrm{B}$). Under normal operation, only $\mathrm{HVU}_\mathrm{M}$ outputs high voltage, while $\mathrm{HVU}_\mathrm{B}$ remains idle. Following the output of each HVU, there is a series diode capable of withstanding a high-voltage reverse bias. In the event of a failure where $\mathrm{HVU}_\mathrm{M}$ becomes unresponsive or unable to output voltage, $\mathrm{HVU}_\mathrm{B}$ takes over. This diode setup allows to combine the outputs of both HVUs, implementing a 1:1 redundancy scheme.

The combined output from the HVUs is then connected to 16 individual SPMT channels via series resistors. Following each series resistor, the HVS connects to the SPMT through a long coaxial cable, which can be either $5\mathrm{\:m}$ or $10\mathrm{\:m}$ in length. The HVD at the base of the SPMT is a resistive divider network that biases the dynode chain and has an input resistance of $210\mathrm{\:M\Omega}$.

The next component on the analog signal line is the high-voltage decoupling capacitor, used to decouple the high-voltage DC component from the AC signal pulse coming from the SPMT. On the low-voltage side of the capacitor, a $5.6\mathrm{\:k\Omega}$ shunt resistor is used locally on the HVS board to provide a grounding path for the capacitor, allowing it to charge. This resistor is connected in parallel with a much smaller $50\mathrm{\:\Omega}$ termination resistor on the ABC board, making it effectively redundant within the overall readout system, and relevant primarily for isolated functional testing of the HVS.

The final components on the analog signal line located on the HVS are the electrostatic discharge (ESD) protection diodes, placed near the interface with the ABC board. These diodes are designed to remain inactive during normal operation and only come into play during an ESD event transient.

\subsection{High Voltage Units}

\begin{figure}[t]
	\centering
	\includegraphics[width=0.9\linewidth]{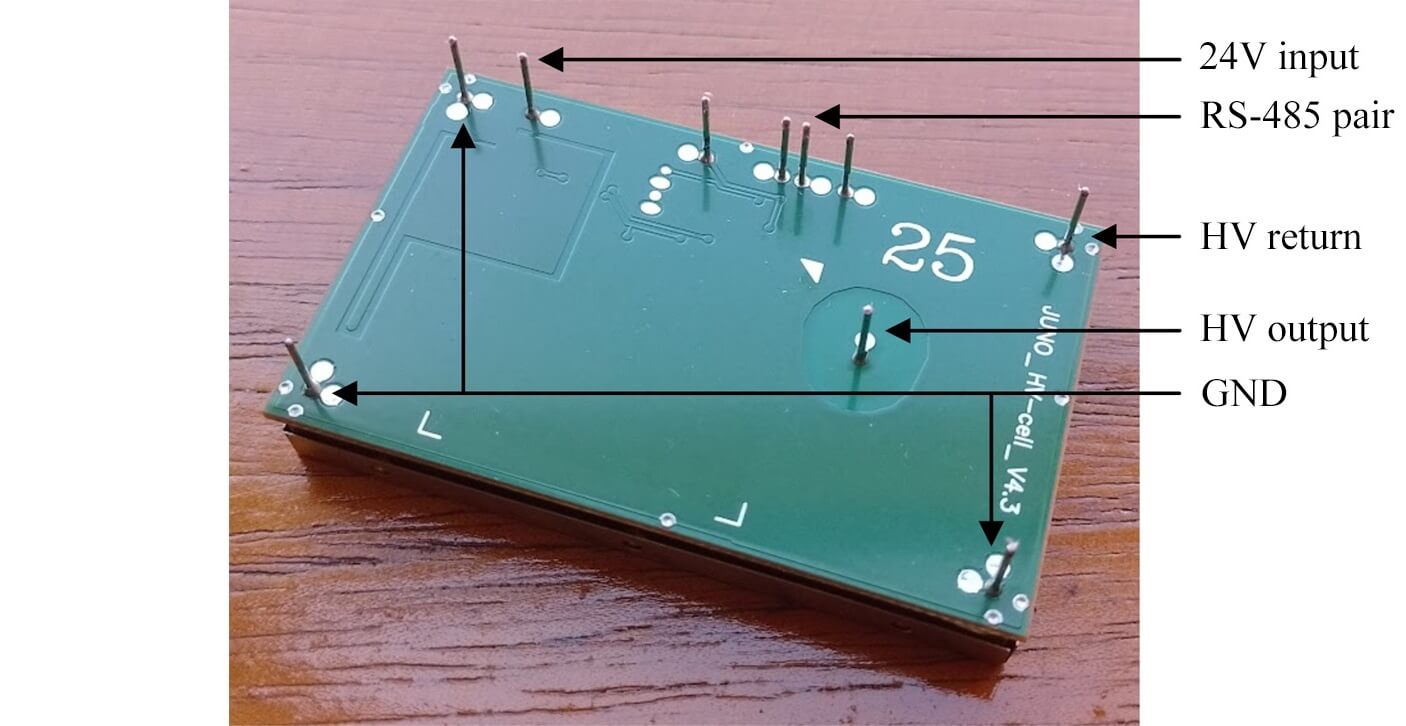}
	\caption{Underside and pinout of the HVU.}
	\label{hvu_bottom}
\end{figure}

The HVU is a PCB fully encased in Pentelast-712 \cite{Pentelast712}, a compound with excellent voltage insulation properties, and covered by a metal shield. Figure \ref{hvu_bottom} shows the underside and pinout of the HVU. It is designed to operate with a $24\mathrm{\:V}$ input, and the output voltage can range from $800\mathrm{\:V}$ to $3\mathrm{\:kV}$, having been originally designed to meet the LPMT system requirements. For the SPMT system, the maximum operational voltage is limited by firmware to $1.5\mathrm{\:kV}$—though this limit can be increased if necessary—with a maximum output current of $300\mathrm{\:\upmu A}$. The HVU has an RS-485 differential pair for communication, which is used to configure and monitor the output voltage.

\subsection{Diodes}

The high-voltage output of both the main and the backup HVUs are combined into a single node using diodes (PN\footnote{PN stands for Part Number, used to identify commercial electronics components}: R5000F) capable of withstanding a high-voltage reverse bias. The selected component has a peak reverse voltage of $5\mathrm{\:kV}$, and a roughly $2\mathrm{\:V}$ forward voltage drop at the nominal operational currents ($\sim$$60$-$100\mathrm{\:\upmu A}$).

\subsection{Current-limiting resistors}
\label{subsection:resistors}

The circuit behavior of an SPMT can be modeled as a current supply with a large output impedance, and effectively all generated current flows through the low-impedance branch containing the decoupling capacitor and the $50\mathrm{\:\Omega}$ shunt resistor, generating a voltage pulse. As long as the series resistor used to separate the channels has a large enough resistance, the value of the resistor no effect on the AC behavior of the circuit.

The value of the series resistor does have an impact on the DC behavior of the circuit, forming a resistive divider with the input resistance of the HVD. Let $R_\mathit{S}$ represent the series resistor, $R_\mathit{in}$ the input resistance of the HVD, $V_\mathit{HV}$ the high-voltage output of the HVU, and $V_\mathit{PMT}$ the bias voltage applied to the SPMT. The relationship between $V_\mathit{HV}$ and $V_\mathit{PMT}$ is expressed as 
\begin{displaymath}
V_\mathit{PMT} = V_\mathit{HV} \cdot R_\mathit{in}/(R_\mathit{S}+R_\mathit{in})
\end{displaymath}
meaning that the value of $R_\mathit{S}$ can have a significant effect on the bias voltage applied to the SPMT.

\begin{figure}[t]
	\centering
	\includegraphics[width=0.65\linewidth]{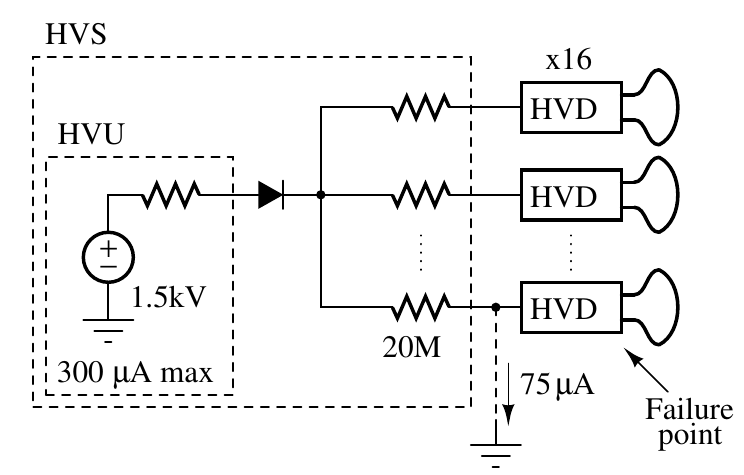}
	\caption{Worst case scenario for channel failure. The output of the HVS is shorted to the ground, while the corresponding HVU is outputting its maximum operational voltage.}
	\label{short_failure_scenario}
\end{figure}

The value of the series resistor also has an effect on system reliability. If an SPMT channel fails, the worst-case scenario is that is shorted to ground, as shown in Figure \ref{short_failure_scenario}. As the maximum output current of the HVU is $300\mathrm{\:\upmu A}$, current draw saturation due to the short causes the voltage to drop, thus affecting the carefully selected bias voltage of the other SPMTs connected to the same HVU. The series resistor can be used to limit the current consumption of a shorted channel, allowing for other channels in the same subgroup to operate normally despite the failure.

With the previously stated considerations, the value of the series resistor was selected to be $R_S = 20\mathrm{\:M\Omega}$, which allows some channel failures within each subgroup. For example, let us consider the scenario in which the HVU is outputting $V_\mathit{HV} = 1.5\mathrm{\:kV}$.

\begin{itemize}
	\item \textbf{Current consumption:} Given that the input resistance of the HVD is $R_\mathit{in} = 210\mathrm{\:M\Omega}$, the idle current consumption for a single channel under normal conditions is $6.5\mathrm{\:\upmu A}$, and $104\mathrm{\:\upmu A}$ for the whole 16-channel subgroup. The current consumption of a shorted channel is $75\mathrm{\:\upmu A}$, as shown in Figure \ref{short_failure_scenario}. With one, two and three channel failures, the current consumption from the HVU would be $172.5\mathrm{\:\upmu A}$, $241\mathrm{\:\upmu A}$ and $309.5\mathrm{\:\upmu A}$, respectively. Therefore, the $20\mathrm{\:M\Omega}$ resistor allows for two channels to fail while others continue operating, as three failures would cause HVU current saturation.
	\item \textbf{Applied voltage:} The voltage applied to the SPMT would be $V_\mathit{PMT} = 1.37\mathrm{\:kV}$. While the $130\mathrm{\:V}$ drop across the series resistor is considerable, the resulting voltage is enough to bias any SPMT.
\end{itemize}

Naturally, channel subgroups that use lower bias voltages can withstand a larger number of channel failures before HVU saturation. With the selected $20\mathrm{\:M\Omega}$ resistance value, the HVU needs to output $1.42\mathrm{\:kV}$ to provide an SPMT with the maximum operational voltage of $1.3\mathrm{\:kV}$.

The selected component (PN: HV732HTTE206J) has a $500\mathrm{\:mW}$ power rating, and a $2\mathrm{\:kV}$ voltage rating, both well above the expected maximums.

\subsection{Decoupling capacitors}

The HVS went through several layout iterations, allowing to assess what could be achieved in the allotted space and what form factors were preferable for some of the components. From this process, the requirements for the decoupling capacitors were defined as follows: a minimum voltage rating of $2\mathrm{\:kV}$ to ensure a safety margin; an SMD (surface-mount device) package to allow for unobstructed routing on the inner and bottom layers; and a maximum SMD package size of 2220.\footnote{All SMD package sizes referenced in this document are in imperial size coding} The only capacitor type that could fit those requirements and was commercially available at the time were Multilayer Ceramic Capacitors (MLCC).

\subsubsection{Dielectric considerations}

Multilayer ceramic capacitors can be broadly categorized by their performance into two categories: high stability capacitors (Class I) and high volumetric efficiency capacitors (Class II). The Class I and Class II classification originates from the EIA RS-198 standard, established by the now-defunct Electronics Industry Alliance (EIA); however, the terms defined in the standard are still used by commercial manufacturers today. As an example, manufacturers typically characterize the dielectric type of a MLCC by their temperature coefficient, a three-letter code established in the standard (e.g., C0G (Class I), X7R (Class II), etc.).

High volumetric efficiency capacitors (Class II), typically made from ferroelectric materials, can achieve higher capacitance values for the same form factor compared to Class I capacitors. However, they are susceptible to capacitance changes due to temperature, aging, and, most importantly, the applied DC voltage---a phenomenon known as the voltage coefficient of capacitance (VCC). High stability capacitors (Class I), typically made from paraelectric materials, exhibit minimal capacitance variations. However, they require a larger form factor for the same capacitance value compared to Class II capacitors.

\begin{figure}[t]
	\centering
	\includegraphics[width=0.9\linewidth]{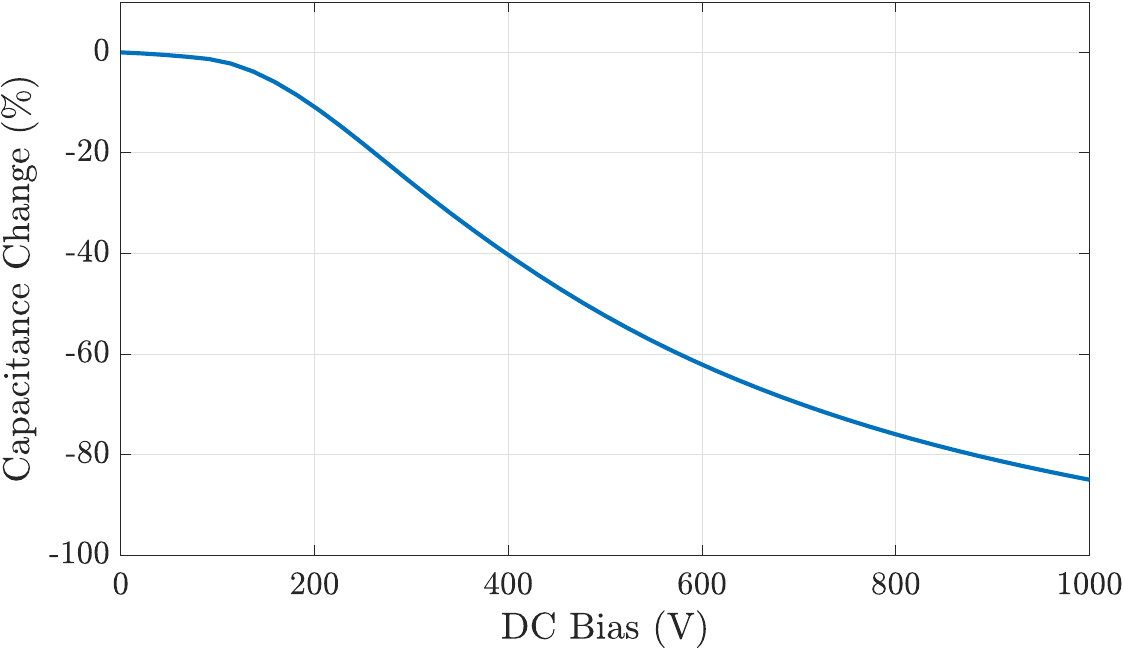}
	\caption{Illustration of the capacitance variation versus applied DC voltage of a generic 1~kV X7R (Class II) MLCC capacitor. The curve was derived from trends across various manufacturers.}
	\label{capacitance_variation}
\end{figure}

Capacitance changes due to VCC for a Class II X7R MLCC can be as severe as 90\% when operating at their rated voltage \cite{Kemet2024}. To illustrate the effects of VCC, Figure \ref{capacitance_variation} shows a generic example of a $1\mathrm{\:kV}$ X7R (Class II) MLCC capacitor. In a typical X7R MLCC, the capacitance decrease starts immediately as voltage increases from $0\mathrm{\:V}$, and the resulting effective-capacitance curve is highly non-linear. There are mainly two factors that influence VCC: dielectric type (e.g., X5R, X7R, etc.) and dielectric thickness, primarily determined by the SMD package size.

The effects of VCC are particularly relevant given the requirements of the SPMT system, as the voltage applied to the capacitor can be anywhere in the $900$--$1300\mathrm{\:V}$ range. Furthermore, capacitance changes due to aging could also become relevant given the 20-year lifetime of the JUNO experiment. 

\subsubsection{Signal waveform and overshoot}
\label{sec:cap_overshoot}

Depending on the value of the decoupling capacitor, there is noticeable overshoot on the SPMT waveform measured on the output of the HVS. This has been document in Refs. \cite{Luo2016, Wu2022}, and is attributed to the discharge of the decoupling capacitor and the capacitor connected to the anode of the SPMT on the HVD. Additionally, the duration of the overshoot is directly proportional to the propagation delay, i.e., cable length, of the coaxial cable used to connect the HVS to the HVD. The exact relation between the value of the capacitance and the amplitude of the overshoot are explored in Refs. \cite{Luo2016, Wu2022}, and is out of the scope of this document. The final selection for the decoupling capacitor was based in experimental results.

\begin{table*}[t]
	\centering	
	\caption{List of capacitor candidates that met the voltage and form factor requirements. These components were also selected based on brand, lead time and price. SMD package size is expressed in imperial size coding.}
	\resizebox{0.8\linewidth}{!}{
		\begin{tabular}{c|c|c|c|c|c|c}
			\textbf{Part Number} & \textbf{Brand} & \textbf{Capacitance} & \textbf{V-Rating} & \textbf{Tolerance} & \textbf{Packaging} & \textbf{Dielectric} \\ \hline\hline
			2220Y4K00472MXT & Knowles Syfer & 4.7~nF & 4~kV & 20\% & 2220 & X7R \\ \hline
			2220J6K00222MXT & Knowles Syfer & 2.2~nF & 6~kV & 20\% & 2220 & X7R \\ \hline
			1812N222M202NXT & Knowles Novacap & 2.2~nF & 2~kV & 20\% & 1812 & C0G \\ \hline
			C2220C392MGGACTU & KEMET & 3.9~nF & 2~kV & 20\% & 2220 & C0G \\ \hline
			C1825C302MGGACAUTO & KEMET & 3.0~nF & 2~kV & 20\% & 1825 & C0G \\
			\hline\hline
		\end{tabular}
	}
	\label{tested_caps}
\end{table*}

Table \ref{tested_caps} shows a list of the most promising capacitor candidates that were tested. The list is not exhaustive and is biased towards components that were readily available at the time of testing. Nevertheless, these components were carefully selected to meet the voltage and form factor requirements, with additional practical considerations based on brand, lead time and price. 

\begin{figure}[t]
	\centering
	\subfloat[][Average SPE signals for different capacitors.]
	{\resizebox{0.95\linewidth}{!}{\includegraphics[width=0.6\textwidth]{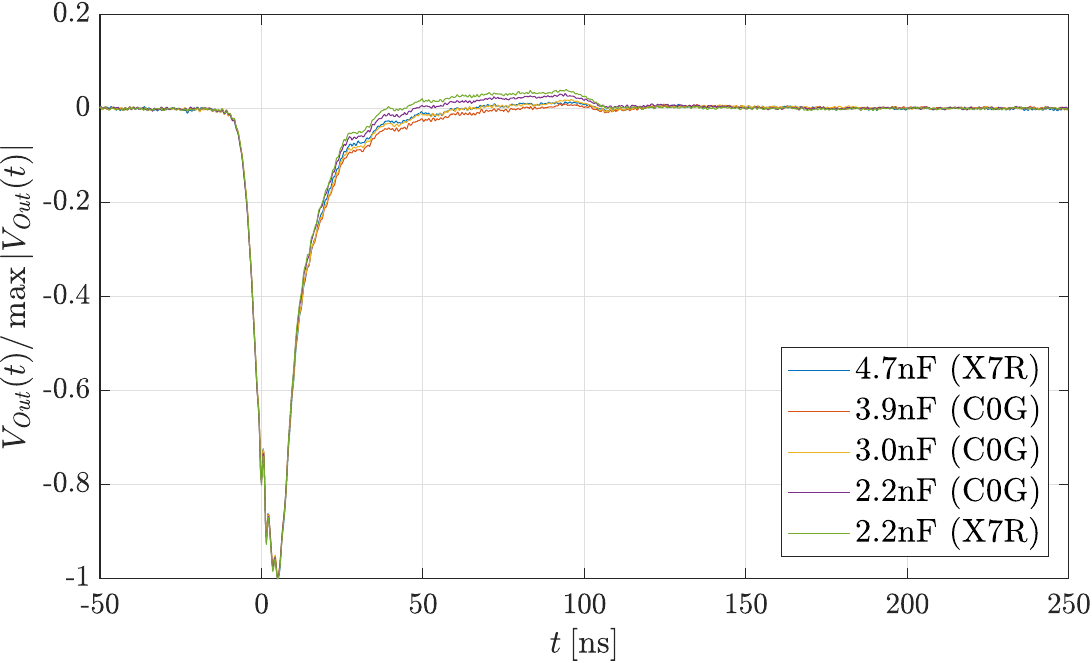}}}\\
	\subfloat[][Zoom-in of the signal overshoot.]
	{\resizebox{0.95\linewidth}{!}{\includegraphics[width=0.6\textwidth]{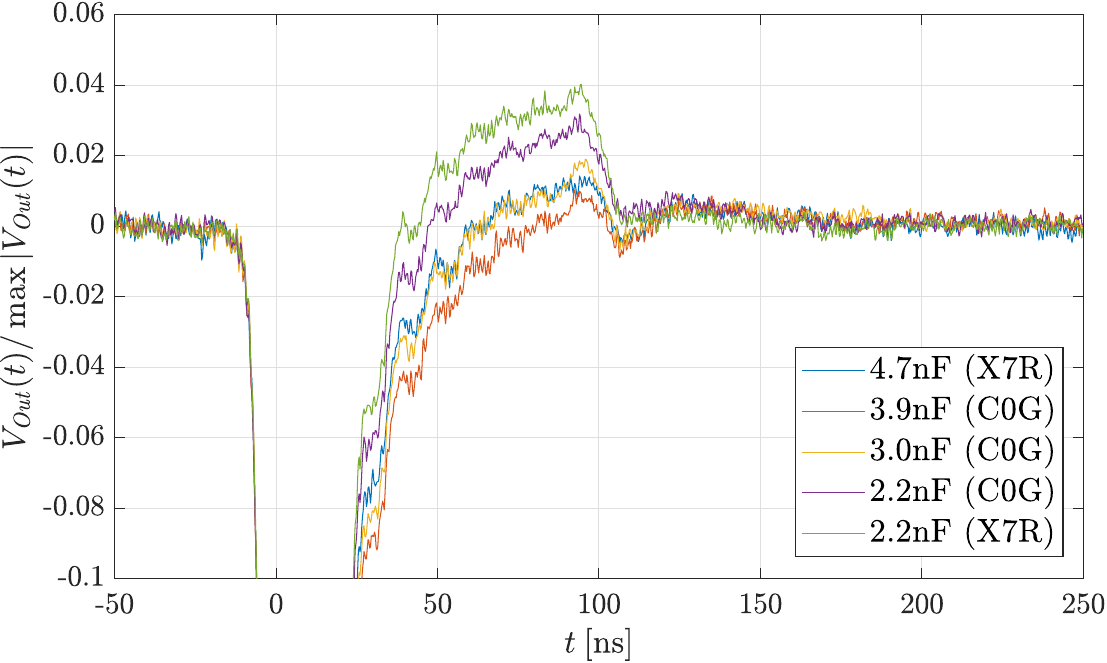}}}
	\caption{Average SPE signals, normalized to unit amplitude, for different capacitor candidates. The effective capacitance of X7R capacitors can be expected to be lower than their nominal value. Signal has been digitally filtered to remove some unwanted aggressor frequencies present during data-taking.}
	\label{waveform_comparison}
\end{figure}

The components were tested on an early prototype of the HVS, connected through a $10\mathrm{\:m}$ coaxial cable to a 3-inch PMT operating at nominal gain ($3 \times 10^6$) and biased at $900\mathrm{\:V}$. Dark noise single photoelectrons (SPEs) generated by the SPMT were used as signal for the measurements. The signal was measured on the output of the HVS by an oscilloscope using a $50\mathrm{\:\Omega}$ termination.

Figure \ref{waveform_comparison} shows the average SPE waveforms for different capacitor candidates, normalized to unit amplitude. Since individual SPE events can exhibit significant white noise, $5000$ independent events were averaged to improve the signal-to-noise ratio (SNR). Additionally, imperfections in the test setup introduced a systematic aggressor frequency with considerable amplitude, which was removed during post-processing through digital filtering. As a result, the presented waveforms are not exact representations of the measured SPEs.

From figure \ref{waveform_comparison} it can be observed that, for these values, the positive amplitude of the overshoot monotonically decreases for increasing values of effective capacitance. The green curve of the X7R $2.2\mathrm{\:nF}$ capacitor has the largest overshoot, and it can be safely assumed that the effective capacitance value of this component, given the Class II dielectric and $900\mathrm{\:V}$ bias voltage, is considerably smaller than the $2.2\mathrm{\:nF}$ nominal value. In a similar fashion, the blue curve of the X7R $4.7\mathrm{\:nF}$ capacitor overlaps with the yellow curve for the C0G $3.0\mathrm{\:nF}$ capacitor, suggesting that the effective capacitance of the X7R component is closer to this value, and at $900\mathrm{\:V}$ out of the $4\mathrm{\:kV}$ voltage rating the capacitor its operating with a significant effective capacitance loss.

\subsubsection{Component selection}

For the SPMT system, the positive amplitude of the overshoot was not deemed problematic in regular operation, as it cannot trigger false positives given that the discriminator on the CATIROC is tuned to the negative amplitude of a regular pulse. It does become more relevant in the event of consecutive pulses, as pile-up can occur, causing the measurement of the charge to be biased. Nevertheless, all the components tested were deemed sufficient for this application, and more practical considerations were prioritized in the selection of the component (e.g., price, lead time) over factors like capacitance value and dielectric type.

Ultimately, one component did stood out among the rest. The selected component (PN: C2220C392MGGACTU) is a C0G $3.9\mathrm{\:nF}$ capacitor with a $2\mathrm{\:kV}$ voltage rating. It meets the requirements of voltage rating and form factor, and has all the desirable characteristics: a stable capacitance and minimal signal distortion. From the red curve for the C0G $3.9\mathrm{\:nF}$ capacitor in Figure \ref{waveform_comparison}(b), it can be seen that both the positive and negative overshoot peaks are approximately $1\%$ of the peak signal amplitude, even in an older, unoptimized HVS design.

\subsection{Protections}

During a combined test of early versions of the HVS and ABC boards, one of the SPMT cable connectors was damaged mid-test, causing a short between the central conductor and the grounded sleeve of the cable. This incident damaged several CATIROC chips, which subsequently created a short between power and ground on the ABC board. Although the CATIROC chips have internal overvoltage protections, these were ineffective in preventing the damage, and their failure likely caused the short on the ABC board.

\subsubsection{Failure mode and circuit behavior}

\begin{figure}[t]
	\centering
	\subfloat[][Schematic representation of a sudden short and elecstrostatic discharge.]
	{\resizebox{0.95\linewidth}{!}{\includegraphics{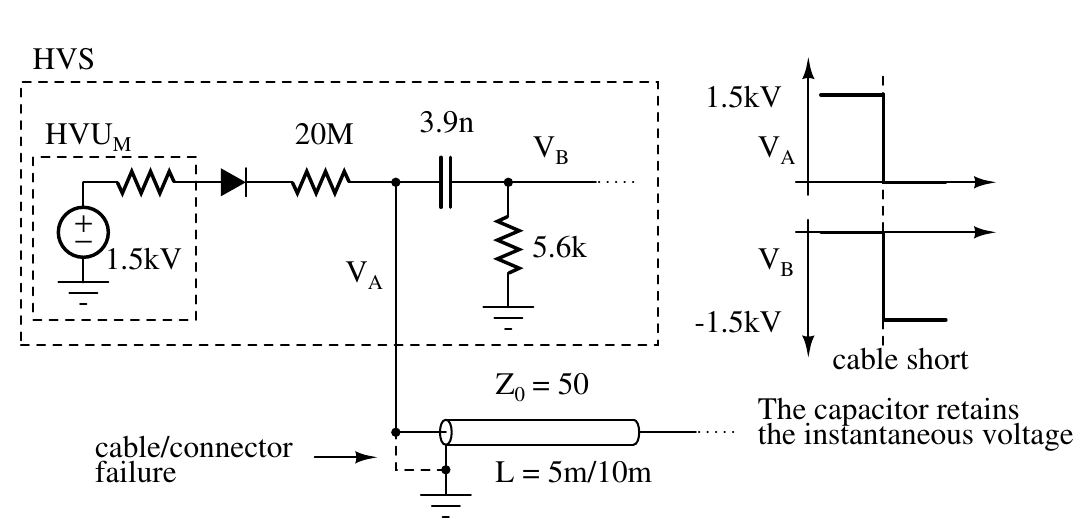}}}\\
	\subfloat[][Single-channel schematic with protection diodes and positive rail bias.]
	{\resizebox{0.95\linewidth}{!}{\includegraphics{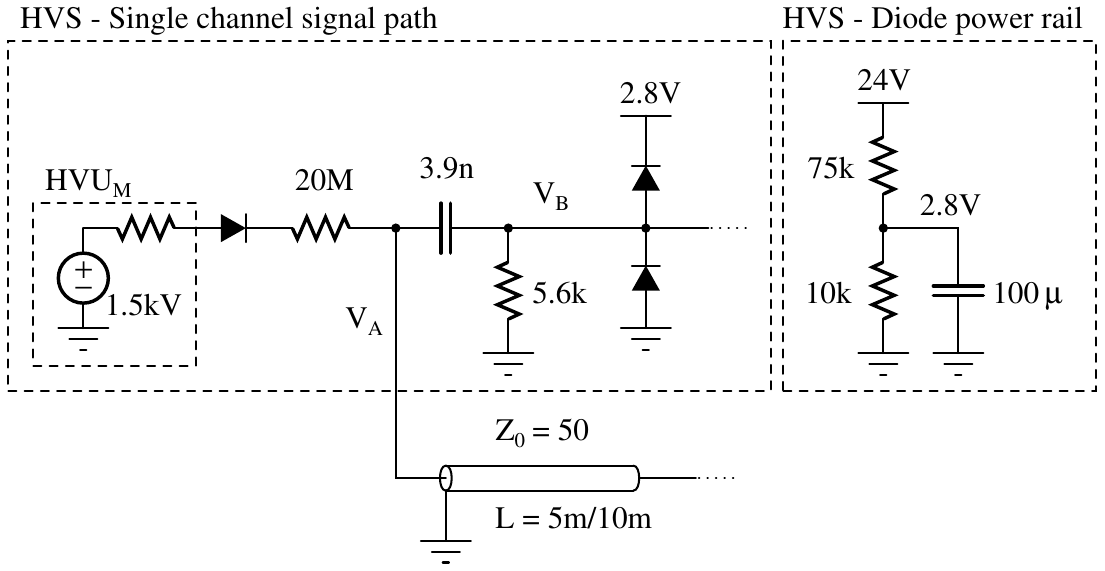}}}
	\caption{Single-channel schematic with and without protection diodes.}
	\label{failure_mode}
\end{figure}

For illustration purposes, let us consider the simplified schematic of Figure \ref{failure_mode}(a), which corresponds to a partial schematic of a single channel of the HVS without protections. When a short occurs in node $V_A$ and the voltage suddenly drops to zero, the instantaneous voltage of the capacitor is retained, causing a sudden drop in voltage in node $V_B$ equal in magnitude to the voltage stored in the capacitor and of negative amplitude. This is effectively electrostatic discharge (ESD), and it is common practice to use diodes that limit this effect, providing a low impedance path for the charge to dissipate without damaging components.

After this incident, the design of the HVS was revised to include protection diodes, physically located near the interface with the ABC. Figure \ref{failure_mode}(b) shows the schematic for the protection diodes (PN: BAV99S) used on the HVS board, with two diodes connected in reverse-bias on each channel. During normal operation, the relatively small amplitude of the negative signal pulses generated by the SPMT is not sufficient to forward-bias the diode connected to ground. However, when the voltage at node $V_B$ becomes too large in either positive or negative amplitude, the diodes become forward-biased, providing a low-impedance path for the charge to dissipate. 

The diode connected to the positive power rail is biased at $2.8\mathrm{\:V}$, as shown in Figure \ref{failure_mode}(b). Given that the HVS is powered by a single $24\mathrm{\:V}$ power supply, the $2.8\mathrm{\:V}$ were generated using a resistive divider and a relatively large $100\mathrm{\:\upmu F}$ bypass capacitor, to provide a low-impedance path to signal ground for fast transients.

\subsubsection{Discharge transient waveform}

\begin{figure}[t]
	\centering
	\subfloat[][HVS output voltage during an ESD event for different SPMT bias voltages.]
	{\resizebox{1\linewidth}{!}{\includegraphics{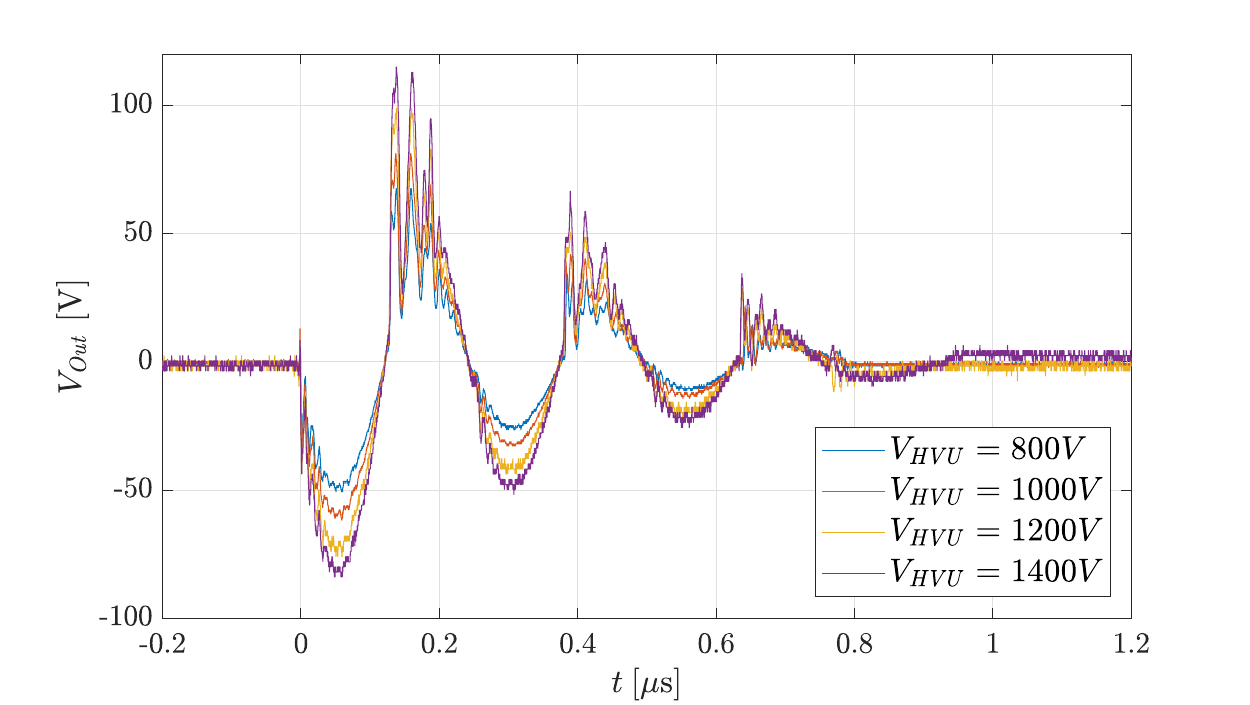}}}\\
	\subfloat[][Single-channel circuit schematic with path inductance.]
	{\resizebox{0.95\linewidth}{!}{\includegraphics[trim={3cm 0.5cm 0.3cm 0.5cm},clip]{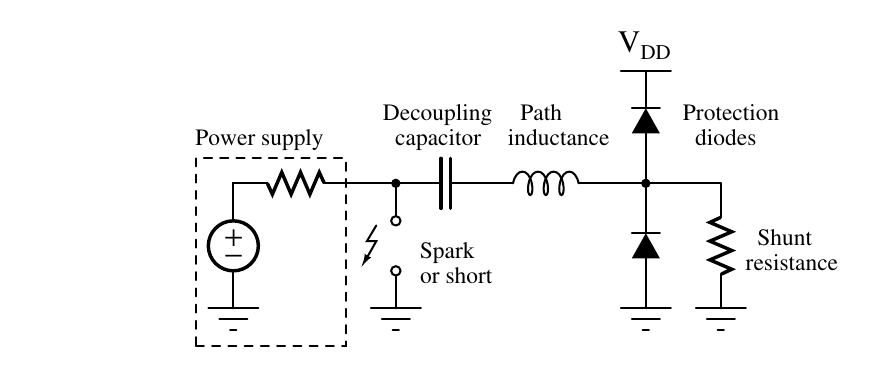}}}
	\caption{Induced shorts and ESD measurements.}
	\label{protection_meas}
\end{figure}

Measurements were done to better understand the behavior and effectiveness of the protection diodes during an ESD event. Shorts were induced on the SPMT connectors located at node $V_A$ in Figure \ref{failure_mode}(b) while the HVUs were outputting high voltage. This replicated the event described earlier, and the output signal of the HVS was measured using a high-voltage oscilloscope probe. The measurement results are shown in Figure \ref{protection_meas}(a) for different bias voltages.

The waveform has two distinguishable components: a lower frequency damped oscillation and a higher frequency component. The systematic high-frequency artifacts present in the plots are not fully understood. Although they appear only on the positive half-cycle of the waveform---seemingly linked to the forward-biased protection diode connected to the positive rail---the effect could not be replicated in simulation. The low-frequency damped oscillation, however, can be explained by considering the path inductance, as shown in Figure \ref{protection_meas}(b). As the accumulated charge dissipates through the diodes, it energizes the path inductance, which then charges the capacitor, forming a tank circuit that oscillates while dissipating energy.

Although the protection diodes significantly reduce the voltage measured on the output of the HVS, the measured voltage remains high enough to raise concerns about potential damage to the CATIROC chips, requiring further testing.

\subsubsection{Testing and validation}

\begin{figure}[t]
	\centering
	\subfloat[][Emulation circuit]
	{\resizebox{0.417\linewidth}{!}{\includegraphics{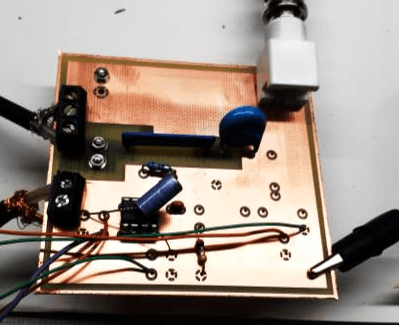}}} \hspace{1mm}
	\subfloat[][Spark gap]
	{\resizebox{0.501\linewidth}{!}{\includegraphics{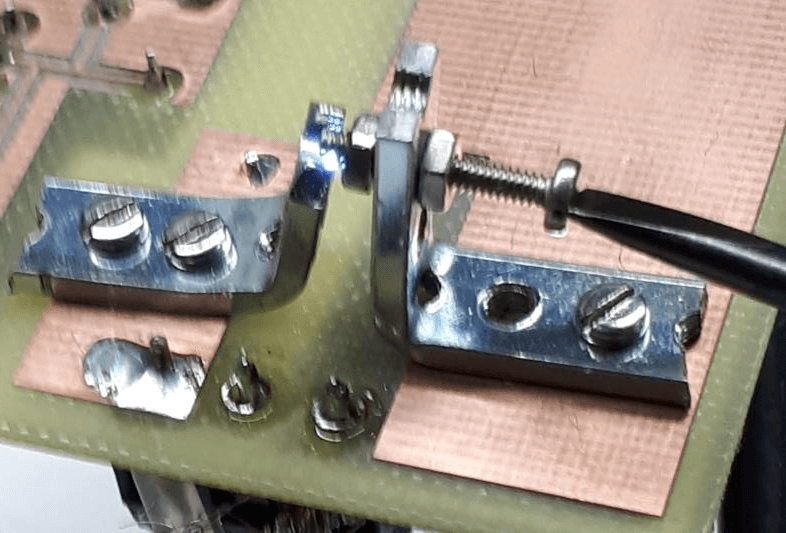}}}
	\caption{Emulation PCB used to test the protection diodes.}
	\label{emulation_pcb}
\end{figure}

Given the limited availability of the custom CATIROC ASICs, the protection diodes were first evaluated through simulation and an emulation circuit to avoid subjecting the chips to a potentially destructive test. To this end, a simple PCB was manufactured to emulate the circuit schematic shown in Figure \ref{protection_meas}(b) using an explicit inductor, plus an operational amplifier (PN: AD818) to emulate the input circuit of the ABC board and CATIROC chip, in the same configuration as the circuit shown in Figure \ref{simplified_sch}. The testing PCB is shown in Figure \ref{emulation_pcb}. After confirming that the protection diodes successfully protected the operational amplifier, a combined spark test between the ABC board and the HVS was conducted.

The combined spark test involved the complete front-end electronics assembly, including the ABC, GCU, and HVS boards. For the test, one of the 16-channel subgroup underwater connectors was assembled with exposed conductors to induce arcing. The voltage was raised to $1.2\mathrm{\:kV}$ and maintained for one minute, during which numerous audible sparks were observed. After the test, the ABC board was initially unresponsive, which was eventually diagnosed as corrupted or unconfigured FPGA firmware. After reprogramming the FPGA, the ABC board was confirmed to be fully functional, with no damage to the CATIROC chips.

\section{Physical-level design}
\label{section:layout}

\subsection{Board specifications}

\begin{figure*}[t]
	\centering
	\includegraphics[width=0.8\linewidth]{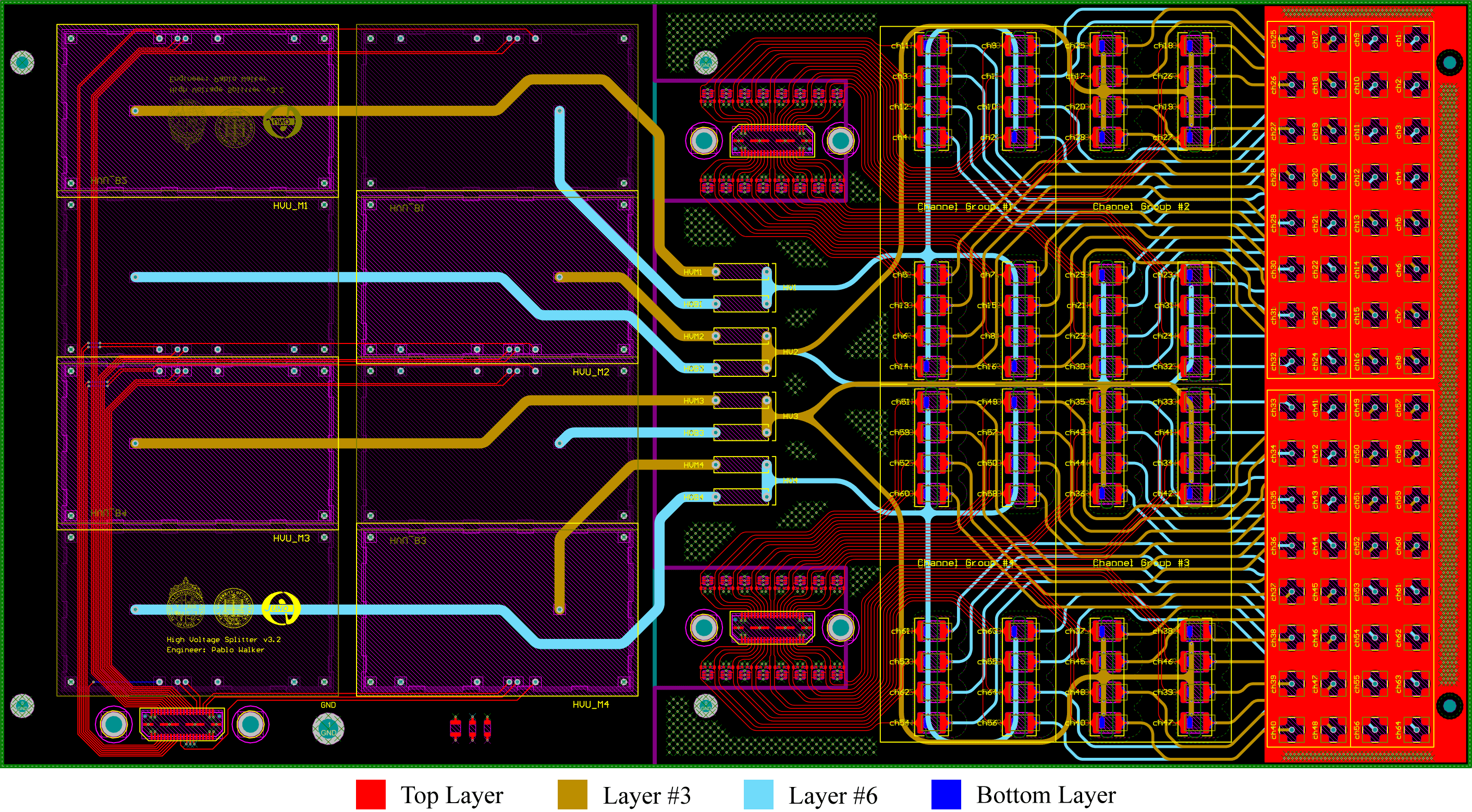}
	\caption{Layout of the HV-Splitter version 3.2. The color coding at the bottom references the signal layers.}
	\label{HVS_layout_full}
\end{figure*}

Table \ref{gen_specs} provides a summary of the PCB specifications for the HV-Splitter, while Figure \ref{HVS_layout_full} shows an image of the layout design. The board services 64 SPMTs, and has a maximum operational voltage of $1.42\mathrm{\:kV}$ when connected to an SPMT channel subgroup requiring a $1.3\mathrm{\:kV}$ bias voltage (the relationship between these values is described in Section \ref{subsection:resistors}). The dimensions of the HVS are $365\textrm{\:mm} \times 190\textrm{\:mm}$, it is $4.1\textrm{\:mm}$ thick and has eight layers.

\begin{table}[t]
	\centering
	\caption{General PCB specifications for HV-Splitter.}
	\resizebox{1\linewidth}{!}{
		\begin{tabular}{c|c}
			\textbf{Specifications} & \textbf{Details} \\
			\hline\hline
            Operational voltage & 1.42 kV \\ \hline
            Number of channels & 64 \\ \hline
			Dimensions & 365~mm $\times$ 190~mm \\ \hline
			Material & FR-4 High-TG \\ \hline
			Surface finish & HASL \\ \hline
			Number of layers & 8 \\ \hline
			Dielectric thickness & 0.87~mm between HV and LV layers\\ \hline
			Finished thickness & 4.1~mm \\ \hline
			Blind vias & Between L3 and L8 \\ \hline
			Impedance control & 50~$\Omega$ with $\pm$10\% accuracy \\ \hline
			Clearance rules & 4.77~mm (HV to LV) and 2.4~mm (HV to HV)\\ \hline
			Controlled-depth milling & 0.5~mm depth underneath capacitors \\ \hline
			Compound filling & Around capacitors and under MCX connectors \\ \hline
			Conformal coating & On exposed high-voltage pads\\
			\hline\hline
		\end{tabular}
	}
	\label{gen_specs}
\end{table} 

The PCB is constructed from High-TG FR-4, which has a higher glass transition temperature than regular FR-4, resulting in a board that is less prone to warping during soldering and thermal cycling. The use of alternative materials with higher insulation properties was deemed unnecessary (e.g., ceramic-filled epoxy), as FR-4 is more cost-effective and has a dielectric strength that is more than sufficient for this application.

Component and interface placement played a major role in determining the size of the HVS, as the HVUs, capacitors, and connectors are of considerable size. This is particularly true for the HVUs, which occupy more than a third of the board's area, as visible on the left-hand side of Figure \ref{HVS_layout_full}. The HVUs are mounted on the HVS by through-hole pins that require hand soldering—to avoid exposing the entire body of the component to very high temperatures during the soldering process—and are therefore distributed in a way that leaves their pins exposed and easily accessible.

Trace routing also played a significant role in determining the size and layer count of the board, primarily due to the use of high voltage, which requires large planar clearances between high- and low-voltage traces. To maintain proper clearances, it was necessary to use an eight-layer design and distribute the trace routing across multiple layers. The use of multiple signal layers requires proper high-voltage insulation between layers, which led to the use of thick dielectric layers and an overall thick board.

The following sections describe and expand on these---and more---considerations and design decisions related to the physical design of the HVS, the trade-offs between them, and how they interact with the system-level constraints described in Section \ref{section:system}.

\subsection{Layer stackup and voltage insulation}
\label{subsection:layer_stackup}

\begin{figure}[t]
	\centering
	\includegraphics[width=1\linewidth]{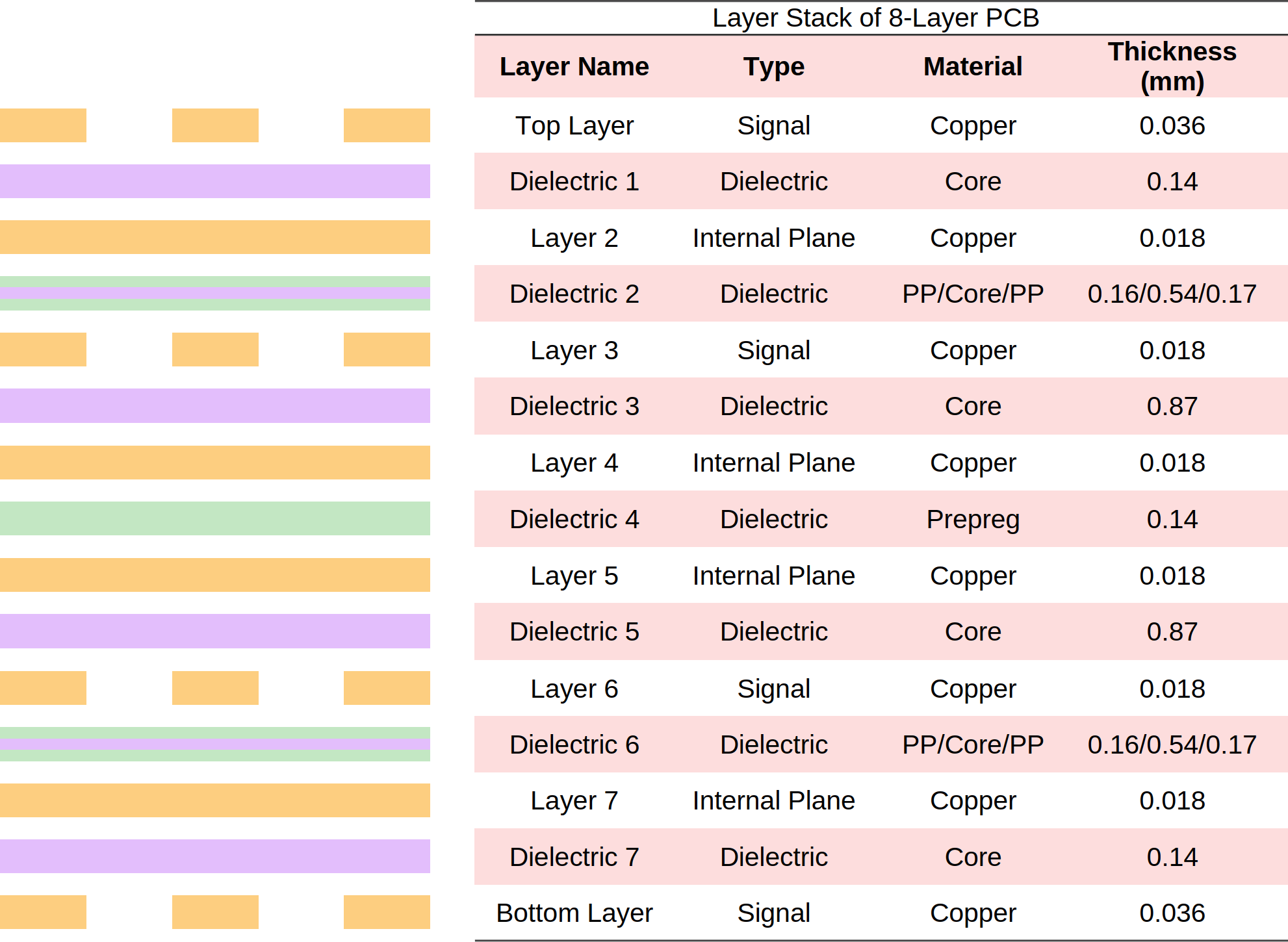}
	\caption{Layer stackup}
	\label{layer_stack}
\end{figure}

The HVS consists of eight layers: four are dedicated to high-voltage and signal routing, while the remaining four primarily serve as grounding layers. The layer stackup of the board is shown in Figure \ref{layer_stack}. Notably, dielectric layers 2, 3, 5, and 6 have a thickness of $0.87\mathrm{\:mm}$, providing high-voltage insulation between the signal and ground layers. A standard value for the dielectric strength of glass-filled epoxy, such as FR-4, is $360\mathrm{\:V/mil}$ \cite{Harper2004}, or roughly $14\mathrm{\:kV/mm}$, although actual values may vary between manufacturers. Wide safety margins are often used to account for potential material imperfections and aging. For an operational voltage of $1.42\mathrm{\:kV}$, a 0.87-mm-thick dielectric requires a dielectric strength greater than $1.64\mathrm{\:kV/mm}$, which is several times smaller than the dielectric strength of FR-4.

Multilayer PCBs are traditionally built from a base of copper-plated FR-4 cores of fixed thickness, sandwiched between layers of FR-4 prepreg with variable thickness. In the case of dielectric layers 2 and 6, their thickness was too large to be constructed entirely with prepreg, and, as per the recommendation of the manufacturer, were built using a combination of unplated FR-4 cores and prepreg, as shown in Figure \ref{layer_stack}.

\begin{figure}[t]
	\centering
	\subfloat[][Ground cutouts underneath SMD decoupling capacitors]
	{\resizebox{0.235\textwidth}{!}{\includegraphics{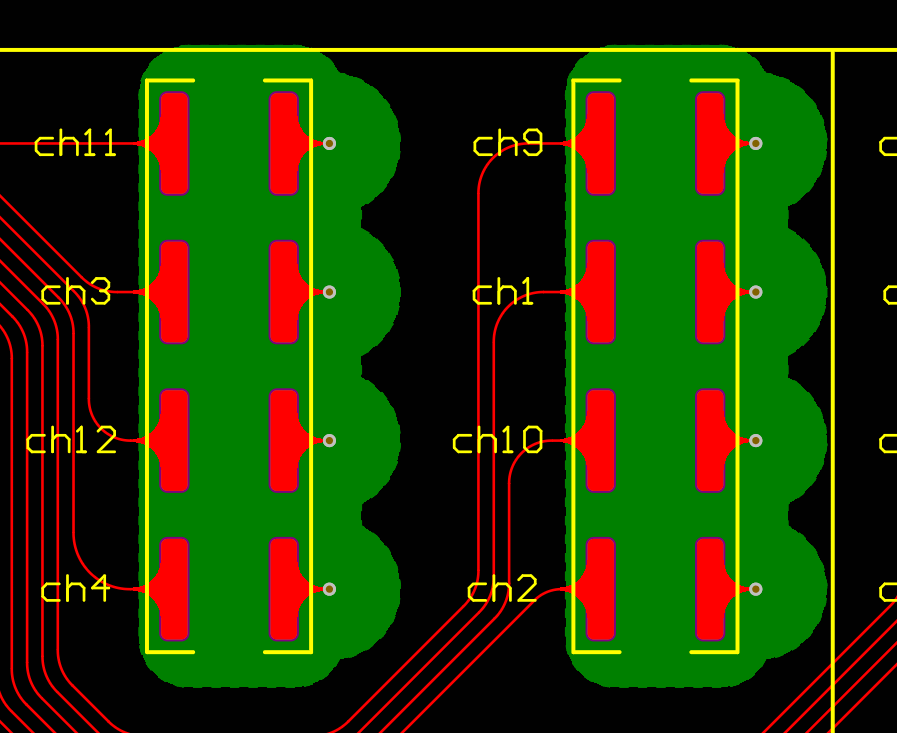}}} \hspace{1mm}
	\subfloat[][Internal plane clearance for THD diode pads]
	{\resizebox{0.212\textwidth}{!}{\includegraphics{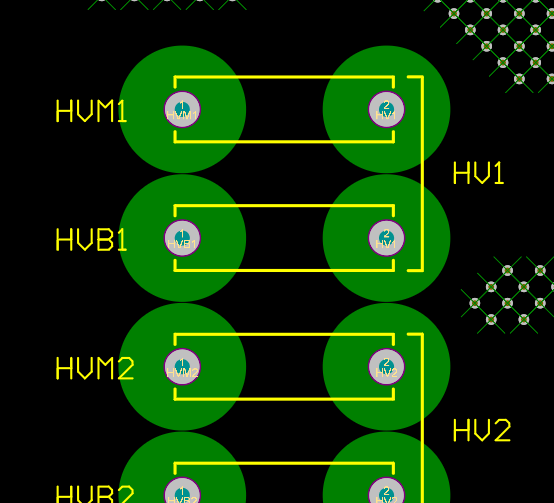}}}
	\caption{Cutouts and clearances on the ground planes for high-voltage components. The green areas represent the absence of copper on the internal layers. Similarly, clearances are also used for high-voltage vias.}
	\label{gnd_cutouts}
\end{figure}

The two outer layers are used for low-voltage signal routing and component placement, including high-voltage components. The low-voltage analog signal traces are routed on the top layer, separated by a thin, 0.14-mm-thick dielectric from the nearest ground plane in order to reduce trace width---which is linked to the thickness of the dielectric to achieve the desired characteristic impedance---and to minimize crosstalk---which is directly proportional to the distance from the nearest ground plane. Since there are high-voltage surface-mounted components on the outer layers, ground cutouts are used underneath them to insure high-voltage insulation, an example of which is shown in Figure \ref{gnd_cutouts}(a). Therefore, for the high-voltage components on the outer layers, the effective dielectric thickness separating them from the nearest ground layers is approximately $1\mathrm{\:mm}$, the sum of the two outermost dielectrics. For through-hole components and vias, appropriate clearances are used between the multilayer conductor and the ground layers, an example of which is shown in Figure \ref{gnd_cutouts}(b). An analysis of conductor clearances is presented in Section \ref{subsection:hv_clearances}.

The two inner signal layers are used exclusively for high-voltage routing. In fact, except for the placement of high-voltage component on the outer layers, as well as high-voltage multilayer components and vias, all high-voltage traces are routed on layers 3 and 6. Each one of these two layers is sandwiched between two 0.87-mm-thick dielectrics. Blind vias\footnote{Blind vias are a type of multilayer via that connect outer layers to inner layers without penetrating the entire board}---between layer 3 and the bottom layer---are used to connect the high-voltage traces on layers 3 and 6 to the high-voltage series resistors on the bottom layer, which allows the design to be more compact.

\subsection{High-voltage clearances}
\label{subsection:hv_clearances}

\subsubsection{Conductor clearances and design standards}

In PCB design, clearance rules are a set of constraints defined by the designer and enforced by CAD software that limit the minimum separation between conductive components on the same layer. These rules may be determined by various factors, such as achieving a specific crosstalk target or ensuring voltage insulation. When voltage insulation is the primary factor, clearances are typically more stringent on outer layers due to traces being separated by only a thin layer of soldermask (if applicable) and air, which provides lower insulation compared to the FR-4 material on inner layers. 

Under ideal conditions, the dielectric strength of air is $3\mathrm{\:kV/mm}$ \cite{Rigden1996}. In practice, this value may be significantly lower due to ambient humidity and surface contaminants. Experiments on PCB surface finishes conducted by the IPC 3-11g committee indicate that the breakdown voltage between uncoated parallel conductors subjected to high voltage is primarily a function of conductor spacing \cite{Cullen2004}. Based on the results of those experiments, the UL-796 Printed Wiring Boards standard specifies
a withstand voltage\footnote{The withstand voltage is the maximum voltage that an insulating material can safely endure during a specified test period, typically set below the breakdown voltage, without experiencing dielectric failure} between uncoated parallel conductors of $1.6\mathrm{\:kV/mm}$, under conditions of $35^\circ \mathrm{C} \pm 2^\circ \mathrm{C}$ and $87.5\% \pm 2\%$ relative humidity \cite{UL796}.

When voltage insulation is a concern in the definition of clearance rules, PCB designers often refer to the available safety and design standards, which have a proven record of proper functionality. One such standard is the IPC-2221B Generic Standard for PCB design \cite{IPC2221B}, which specifies minimum required conductor clearances based on the operational voltage of the circuit and the type of coating used. As an example, the standard specifies an $8\mathrm{\:mm}$ clearance between external uncoated conductors with an operational voltage of $1.6\mathrm{\:kV}$, without specification of temperature or relative humidity. This is eight times larger than the $1.6\mathrm{\:kV/mm}$ withstand voltage specified in UL-796, which, while not intended to represent operational conditions, is evidence of the built-in safety margins of the clearances specified by the IPC-2221B standard.

\subsubsection{High-voltage to low-voltage trace clearances}
\label{subsubsection:HV2LV}

The minimum trace clearance used in the design of the HVS between high-voltage and low-voltage conductors in both outer and inner layers is $4.77\mathrm{\:mm}$. This value was determined based on IPC-2221B guidelines for external coated conductors\footnote{On Table 6.1 of the IPC-2221B standard, categories B4, A5 and A7 refer to external conductors with polymer coating (soldermask), external conductors with conformal coating, and external component leads with conformal coating, respectively, and all share the same clearance guidelines for voltages higher than $500\mathrm{\:V}$}, using an over-specified voltage rating of $1.8\mathrm{\:kV}$. This rating was selected not only to provide an additional safety margin, but to allow the board to operate at higher voltages should it be necessary in the future due to PMT aging.

While it may seem wasteful to enforce this over-specified rule in the inner layers, in practice there were only a handful of instances where the design was constrained by the rule, having only a small impact on the overall size of the board.

\subsubsection{High-voltage to high-voltage trace clearances}
\label{subsubsection:HV2HV}

Under normal conditions, clearances between high-voltage traces should consider only the voltage difference between them, which is typically not large. However, the need for system reliability requires that, in the event of a channel failure that produces a significant voltage drop, that failure should not propagate to healthy channels on the HVS through arcing or tracking. Therefore, clearances must be sufficient even between high-voltage channels to mitigate the risk of failure propagation. Nevertheless, it is not practical to adhere to the same clearance rule described in Section \ref{subsubsection:HV2LV}, as this would lead to an unnecessarily large board. 

The HVS is subjected to its highest operational voltage when it is connected to an SPMT channel subgroup requiring a $1.3\mathrm{\:kV}$ voltage bias, under which conditions there would be $1.42\mathrm{\:kV}$ on the output node of the HVUs and $1.3\mathrm{\:kV}$ on the input node of the SPMTs. The traces for these nodes are primarily routed on the internal layers of the HVS, with two notable exceptions mentioned in Section \ref{subsubsection:hv_exceptions}, and the clearance rules were based on the IPC-2221B guidelines for internal conductors. The clearance between traces with a maximum voltage of $1.42\mathrm{\:kV}$ and any other trace was set at $2.8\mathrm{\:mm}$, which corresponds to a $1.52\mathrm{\:kV}$ rating. Similarly, the clearance between traces with a $1.3\mathrm{\:kV}$ maximum voltage and other traces was set at $2.4\mathrm{\:mm}$, equivalent to a $1.36\mathrm{\:kV}$ rating. These clearances were chosen based on IPC-2221B guidelines for voltages above operational levels while being limited by the available board space. This combination of clearance rules accounts for potential failures while enabling a relatively compact design given the channel multiplicity.

\subsubsection{High-voltage components}
\label{subsubsection:hv_exceptions}

The clearance rules described in Sections \ref{subsubsection:HV2LV} and \ref{subsubsection:HV2HV} were not applied to the high-voltage decoupling capacitors and the MCX connectors, although the reasons differ for each component. For the capacitors, the distance between the pads of neighboring-channel capacitors was set to $2.4\mathrm{\:mm}$, which does not fully account for potential channel failures, as it does not meet the IPC-2221B guidelines for external conductors at $1.3\mathrm{\:kV}$. This clearance was selected mainly due to space availability.

Similarly, the MCX connectors (which are discussed in Section \ref{subsection:interfaces}), by design, have a small separation between the central conductor---connected to the SPMT bias voltage---and the outer body of the connector---connected to ground. In the HVS, there is a $1.2\mathrm{\:mm}$ distance between the high- and low-voltage pads of the footprint of the component, being the singular point of minimal clearance on the HVS. Even though their separation is small, these connectors have been qualified by the manufacturer to have a $1.5\mathrm{\:kV}$ rating.

To compensate for the lower-than-ideal clearances, and to provide additional insulation to these critical components, both the capacitors and MCX connectors were encased in an insulating compound with a much higher dielectric strength than air, the details of which are described in Section \ref{subsection:compound_conformal}.

\subsection{Impedance control}

When high-frequency signals and long signal paths are involved, it is critical that the source, signal path, and load share the same characteristic impedance to minimize unwanted reflections that can cause ringing, overshoot, and other types of signal degradation. In the case of the SPMT system, all components are matched to a characteristic impedance of $50\:\Omega$, including cables, connectors, terminations, and PCB traces on the HVS and ABC board.

The impedance of a PCB trace depends on several factors, including transmission line type (i.e., microstrip or stripline), copper trace width, copper weight\footnote{Copper weight is a term used to refer to the thickness of the copper lamination on a PCB}, dielectric type, and dielectric thickness. In the HVS design, FR-4 was selected as the dielectric based on system-level considerations; copper weight---which has a limited impact on trace impedance---was chosen based on material availability from the manufacturer; and the thickness of some dielectric layers was determined by other constraints, such as high-voltage insulation, discussed in Section \ref{subsection:layer_stackup}.

Microstrips are a type of PCB transmission line routed on the outer layers over a large reference ground plane, whereas striplines are routed on the inner layers, sandwiched between two reference ground planes and typically equidistant from both (i.e., symmetric stripline). In the HVS, the analog signal path is separated into high- and low-voltage sections, both of which require a $50\:\Omega$ characteristic impedance. A decision was made to route all low-voltage analog traces on the top layer, while limiting the high-voltage analog trace routing exclusively to the inner layers. This is mainly due to two factors: microstrips generally require thicker traces to reach the same characteristic impedance as striplines; and high-voltage clearance rules are less stringent on inner layers, as discussed in Section \ref{subsection:hv_clearances}. This means that if the analog high-voltage traces were routed on the outer layers, they would not only be thicker but would also require larger clearances, becoming very ineffective in terms of space.

Consequently, the high-voltage sections of the analog signal path were routed as striplines and the low-voltage sections as microstrips. For the microstrips, the minimum dielectric thickness available from the manufacturer (i.e., $0.14\mathrm{\:mm}$) was selected between the outer layers and the nearest ground planes. This choice not only reduces the necessary trace width to achieve the desired characteristic impedance---allowing for a more efficient use of the available space---but also minimizes potential crosstalk. The striplines, on the other hand, were sandwiched between two 0.87-mm-thick dielectric layers, the thickness of which was selected to favor voltage insulation, as discussed in Section \ref{subsection:layer_stackup}. The result of these design decisions is that the inner-layer analog traces have a thickness larger than $0.73\mathrm{\:mm}$ to achieve a $50\:\Omega$ characteristic impedance.

As discussed in Section \ref{subsection:layer_stackup}, given that the outer dielectric layers are thin, ground cutouts were placed underneath high-voltage components, as shown in Figure \ref{gnd_cutouts}(a). These ground cutouts introduce discontinuities and asymmetry on the reference ground planes, which may influence the characteristic impedance of the traces. Similarly, harsh transitions between the impedance-controlled traces and a component footprint may introduce discontinuities in the signal path, which is why pad tapering is used in the HVS design—a gradual transition from the width of the trace to the size of the pad. Furthermore, the components themselves (e.g., capacitors), interfaces, and transmission line stubs (e.g., series and shunt resistors) also introduce discontinuities which may affect the impedance of the line. Ultimately, the effects of the design of the HVS on signal integrity were determined experimentally, and are presented in Section \ref{subsection:reflections}.

\subsection{Interfaces}
\label{subsection:interfaces}

\begin{figure}[t]
	\centering
	\includegraphics[width=1\linewidth]{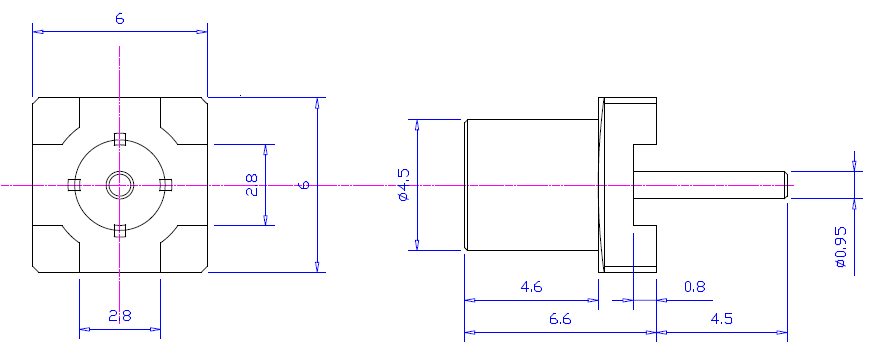}
	\caption{Technical drawing of custom-made hybrid THD/SMD MCX connector with a long central pin (reproduced with permission from the Axon Cable Company).}
	\label{mcx_drawing}
\end{figure}

The HVS interfaces with the SPMTs, the ABC, and the GCU. The connection to the SPMTs is made through MCX connectors, visible on the right-hand side of Figure \ref{HVS_top_bottom}(a), where 64 of them are arranged in a $4\times16$ grid. Figure \ref{mcx_drawing} shows a technical drawing of the MCX connectors used. This component was custom-made by Axon for the HVS specifications and features a hybrid mounting style, with a surface-mounted body and a $4.5\mathrm{\:mm}$ through-hole central pin. The long through-hole central pin ensures the connectors are securely attached to the PCB, avoiding the risk of tear-off that is common with surface-mount soldering. The hybrid mounting style provides a robust connection while keeping the inner layers around the mounting area free of low-voltage pads and traces, reducing clearance overheads and maintaining a compact design.

\begin{figure}[t]
	\centering
	\subfloat[][Connector and standoffs]
	{\resizebox{0.235\textwidth}{!}{\includegraphics{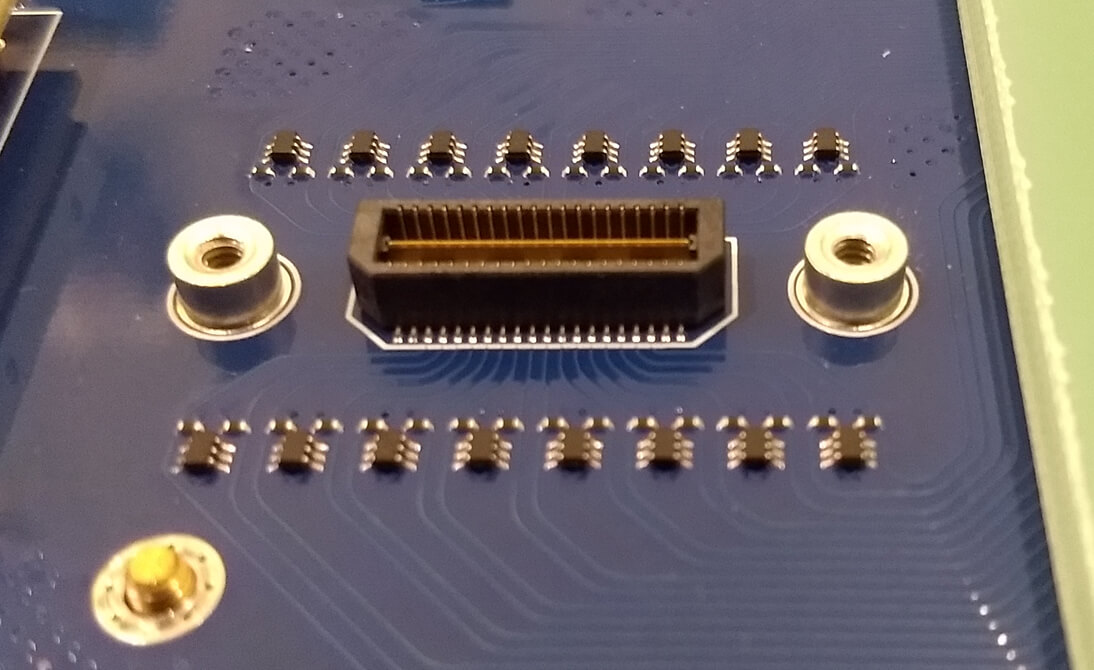}}} \hspace{0.5mm}
	\subfloat[][Cable screwed to HVS]
	{\resizebox{0.226\textwidth}{!}{\includegraphics{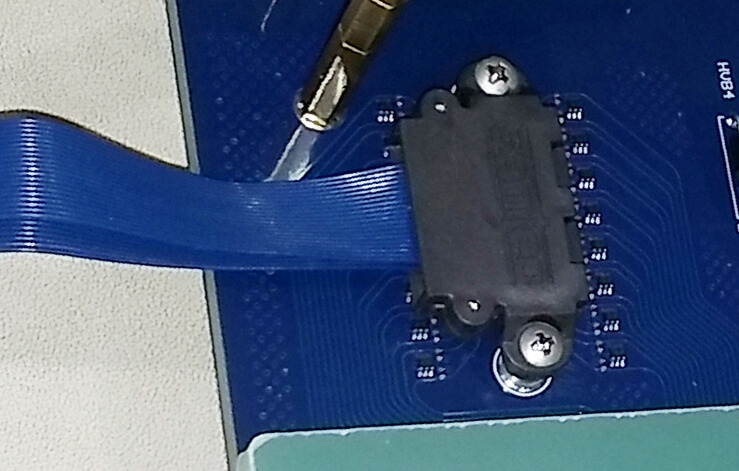}}}
	\caption{QTE connector and mating cable, used for the interfaces between the HVS and the ABC (shown in the image) and the GCU.}
	\label{qte-conn}
\end{figure}

The interface with the ABC and the GCU is made using the same 40-pin Samtec QTE connectors (PN: QTE-020-01-F-D-A) shown soldered to the HVS in Figure \ref{qte-conn}. This connector was selected for its 40-pin density, $9\mathrm{\:GHz}$ bandwidth, characteristic impedance of $50\:\Omega$, and robust connection method, which uses standoffs and screws (as shown in Figure \ref{qte-conn}), all of which match with the specifications of the SPMT readout system. There are three QTE connectors visible in Figure \ref{HVS_top_bottom}(a): two located along the middle of the board that interface with the ABC and carry the decoupled analog signals from the SPMTs, and one at the bottom left, which interfaces with the GCU and carries power and RS-485 communication lines for the HVUs.

\subsection{Compound and conformal coating}
\label{subsection:compound_conformal}

To enhance the electrical insulation of the decoupling capacitors and the MCX connectors---both of which are subjected to high voltage across their closely spaced terminals during normal operation---the colorless organosilicon compound Pentelast-712 was used to encase these components. This compound has excellent properties as an electrical insulator at the operational temperature of the HVS ($\sim$$21^\circ\mathrm{C}$), with a very high volume and surface resistivity ($10\mathrm{\:T\Omega}$ and $10\mathrm{\:T\Omega\cdot cm}$, respectively) and dielectric strength (>$15\mathrm{\:kV/mm}$) \cite{Pentelast712}.

\begin{figure}[t]
	\centering
	\includegraphics[width=0.7\linewidth]{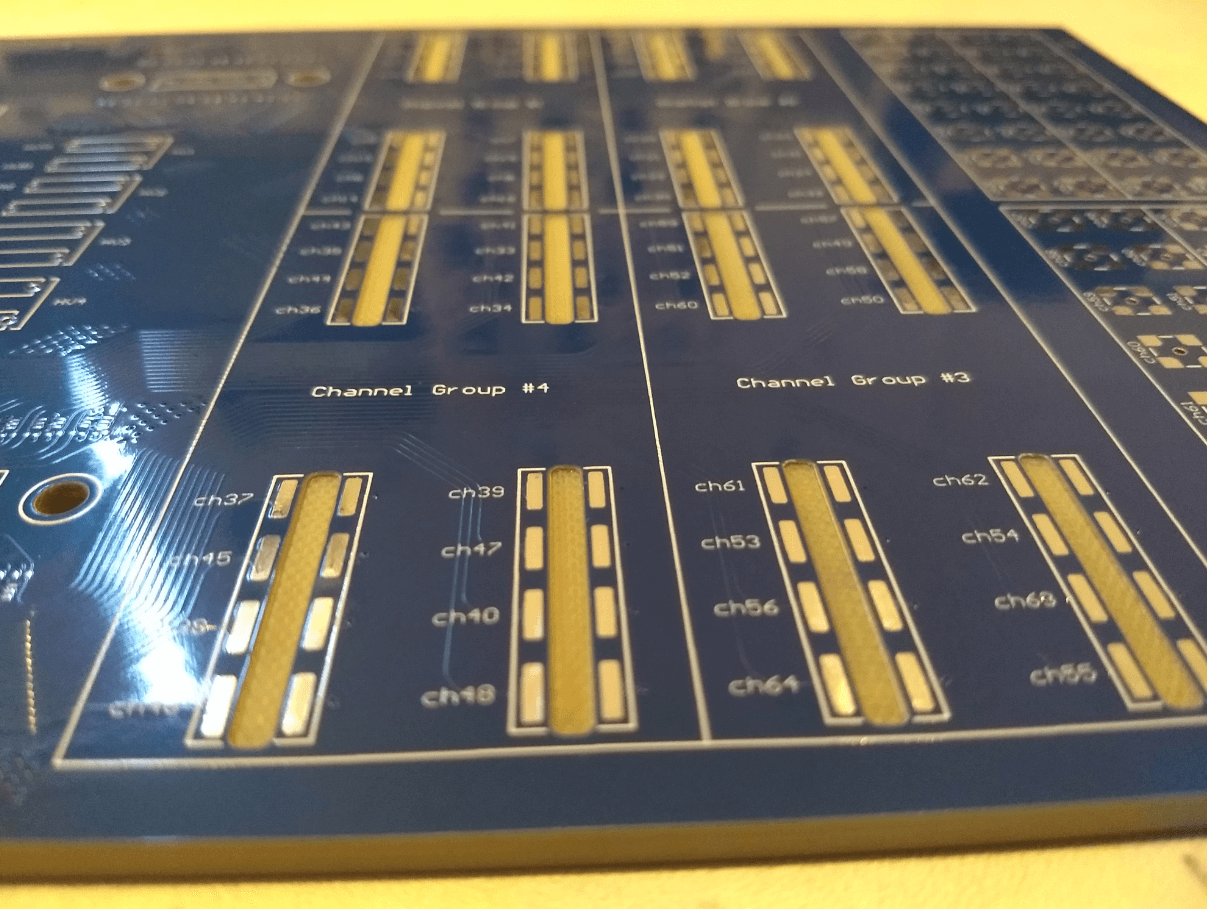}
	\caption{Controlled-depth milling slots.}
	\label{milling_slots}
\end{figure}

To ensure the liquid compound could properly settle underneath the capacitors during the curing process, a $0.5\mathrm{\:mm}$ controlled-depth milling slot was included in the design of the HVS, visible in Figure \ref{milling_slots}. Without this slot, the compound would not flow adequately underneath the capacitors, potentially leaving air pockets once it had cured. Similar milling slots are not necessary for the MCX connectors, which feature a $0.8\mathrm{\:mm}$ cavity in their base, visible in Figure \ref{mcx_drawing}, allowing the compound to settle between the central conductor and the connector body.

\begin{figure}[t]
	\centering
	\subfloat[][Capacitors and milling slot]
	{\resizebox{0.268\textwidth}{!}{\includegraphics{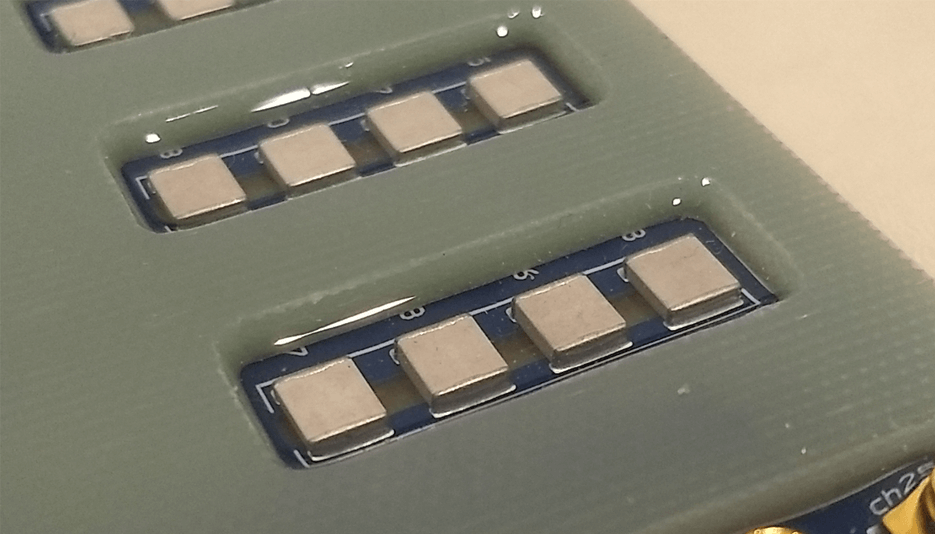}}} \hspace{0.5mm}
	\subfloat[][MCX connectors]
	{\resizebox{0.205\textwidth}{!}{\includegraphics{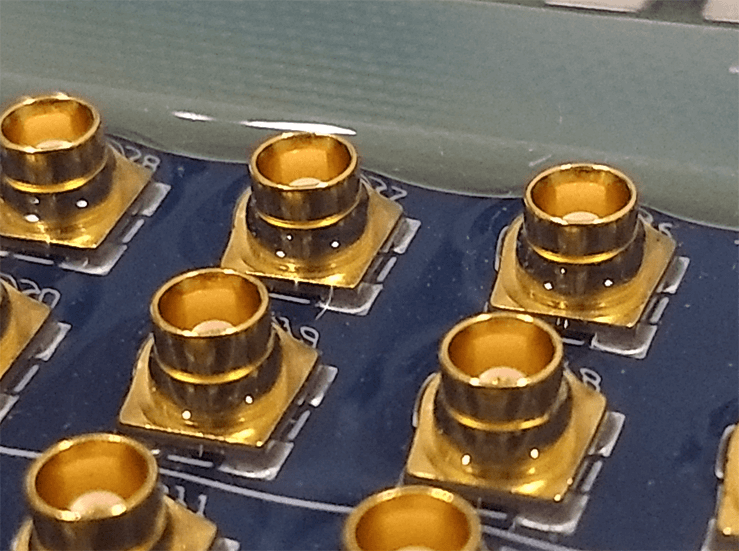}}}
	\caption{Colorless organosilicon compound Pentelast-712 applied to the decoupling capacitors and MCX connectors. A 6-mm-thick FR-4 mold is used for pouring the compound. Controlled-depth milling slots are used underneath the capacitors for the compound to flow, while the MCX connectors feature an open slot underneath by design.}
	\label{compound_example}
\end{figure}

To pour the compound, a 6-mm-thick FR-4 piece with cutouts to match the positioning of the decoupling capacitors and MCX connectors was glued to each HVS during production, which is visible on the right-hand side of Figure \ref{HVS_top_bottom}(a). Figure \ref{compound_example} shows an example of the mold and the applied compound on both the capacitors and the MCX connectors.

The other components soldered to the HVS that are subjected to high voltage are the HVUs, high-voltage diodes, and series resistors. By design, the high-voltage pads for these components have significant clearance from any low-voltage pads or traces, particularly on the outer layers. The singular exception would be a scenario in which one end of a series resistor is shorted to ground due to a failure elsewhere in the circuit, as described in Section \ref{subsection:resistors}. However, the resistor would prevent arcing by providing a lower-resistance path to ground. It is difficult to predict all failure scenarios given the long operational lifetime of the HVS, so to further insulate against unforeseen events (e.g., humidity, debris), conformal coating was applied to the exposed high-voltage pads of the HVUs, the bottom-side of the MCX connectors, the high-voltage diodes, and the series resistors.

\section{Performances}
\label{section:performances}

One of the two main functions of the HVS is to decouple the signal pulse generated by the SPMT from the high-voltage bias. A marker of good performance is achieving this without introducing any significant signal distortion, which could affect energy readings and potentially trigger false positives. The requirement for the analog path of the SPMT system, previously described in Section \ref{subsubsection:integrity}, is that all forms of unwanted effects---noise and distortion included---must be kept below 1/10 if the amplitude of an SPE, or roughly $200\mathrm{\:\upmu V}$, to avoid inducing false positives.

During the development of the HVS, three sources of signal distortion directly influenced by the board's design were identified: overshoot due to the capacitance value of the decoupling capacitor (as described in Section \ref{sec:cap_overshoot}), signal reflections due to impedance mismatch, and layout-induced crosstalk. The following sections show the performance of the HVS in relation to these types of distortions for both low- and high-energy events.

\subsection{Measurement setup}

\begin{figure}[t]
	\centering
	\includegraphics[width=0.95\linewidth]{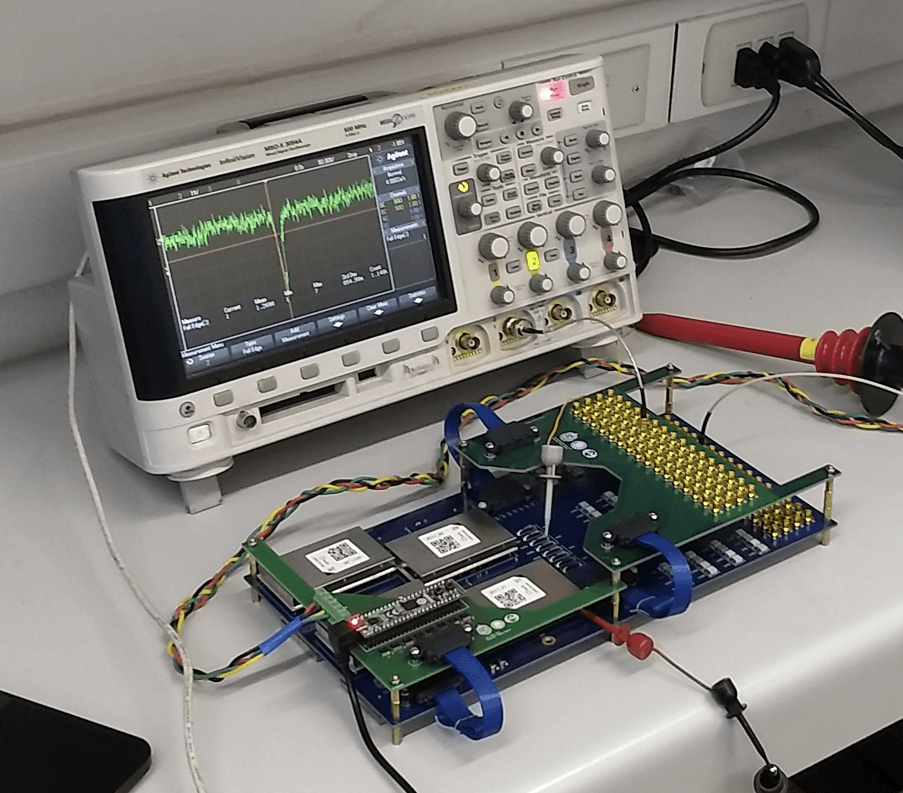}
	\caption{Signal measurement setup. Although the version of the HVS shown in the image is older, the same setup was used throughout the development process.}
	\label{meas_setup}
\end{figure}

Figure \ref{meas_setup} shows the setup used to make signal measurements on the HVS. As shown in the image, two complementary boards are used: mounted on the left side of the HVS there is a PCB containing an FPGA development board (CMOD S6), used to used to communicate and control the HVUs through a PC; and mounted on the right side of the HVS is an adapter board, a PCB with $50\mathrm{\:\Omega}$ impedance-controlled traces used to adapt the connectors which interface with the ABC to MCX connectors. The outputs of the adapter board are connected directly to an oscilloscope through MCX-BNC coaxial cables, and the oscilloscope channels are configured to have a $50\mathrm{\:\Omega}$ termination resistance to emulate the shunt resistor on the ABC. A single SPMT is connected at a time through the custom $50\:\Omega$ coaxial cable developed by Axon, and $5\mathrm{\:m}$ and $10\mathrm{\:m}$ variants were used for the different measurements.

The use of the adapter board---and indeed, any adapting element---can introduce slight mismatches in the impedance matching of the signal line and facilitate additional coupling between adjacent channels, which can influence reflection and crosstalk measurements. Consequently, the performance of the HVS in relation to these types of distortions should be better than the results presented in the following sections.

\subsection{Single photoelectrons}

\begin{figure*}[t]
	\centering
	\subfloat[][]
	{\resizebox{0.325\linewidth}{!}{\includegraphics{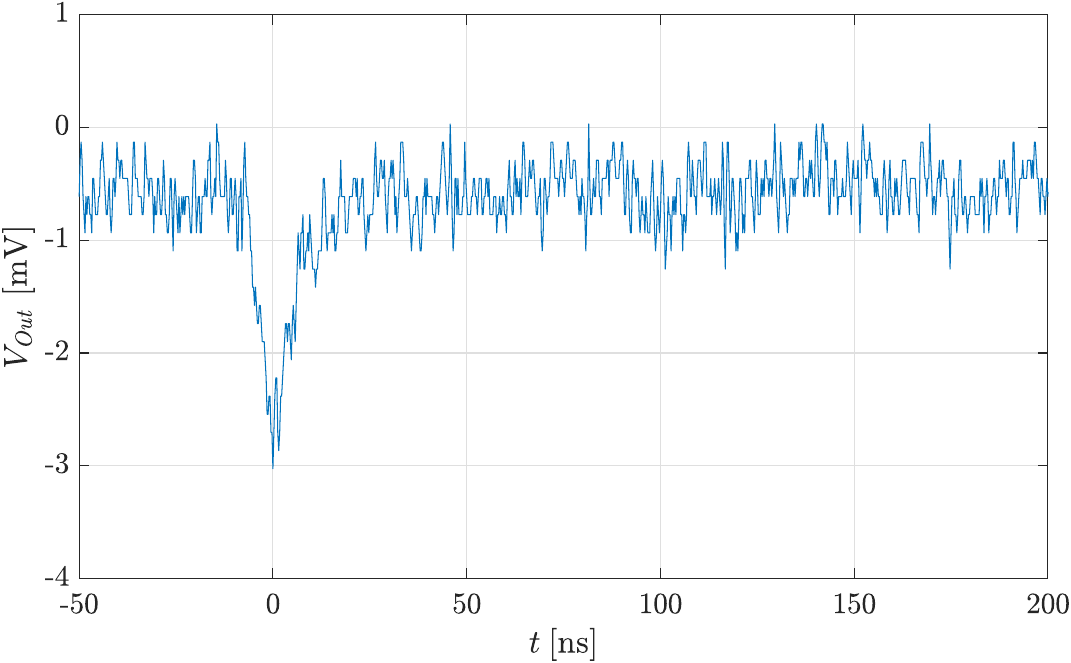}}}\hspace{1mm}
	\subfloat[][]
	{\resizebox{0.325\linewidth}{!}{\includegraphics{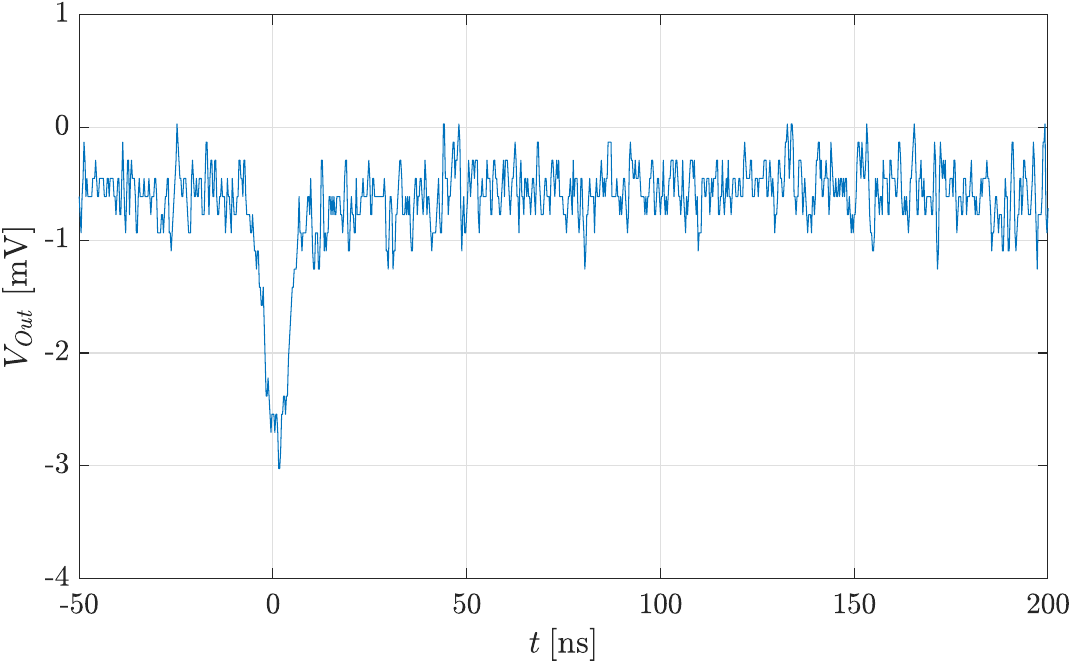}}}\hspace{1mm}
	\subfloat[][]
	{\resizebox{0.325\linewidth}{!}{\includegraphics{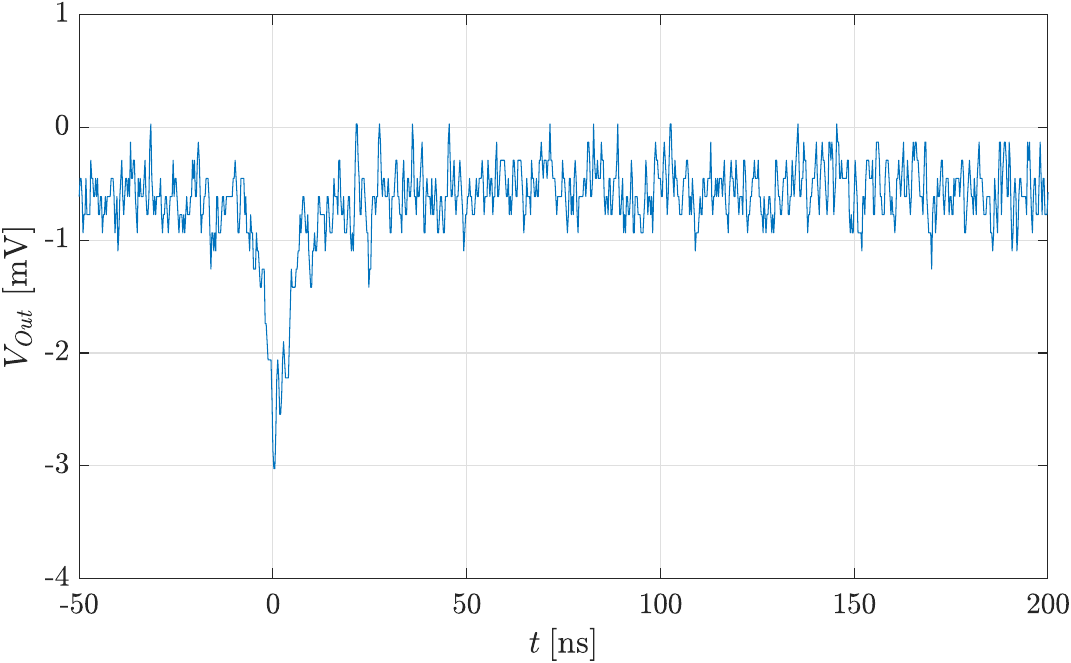}}}
	\caption{Individual SPE events measured using a 10m coaxial cable. The noise present in these measurements does not represent the noise of the JUNO SPMT system, as it is influenced by power supply noise (i.e., the 24~V power supply used to power HVUs) and electromagnetic interference in the experimental setup.}
	\label{spe_events}
\end{figure*}

\begin{figure}[t]
	\centering
	\subfloat[][Single photoelectron waveform.]
	{\resizebox{0.95\linewidth}{!}{\includegraphics{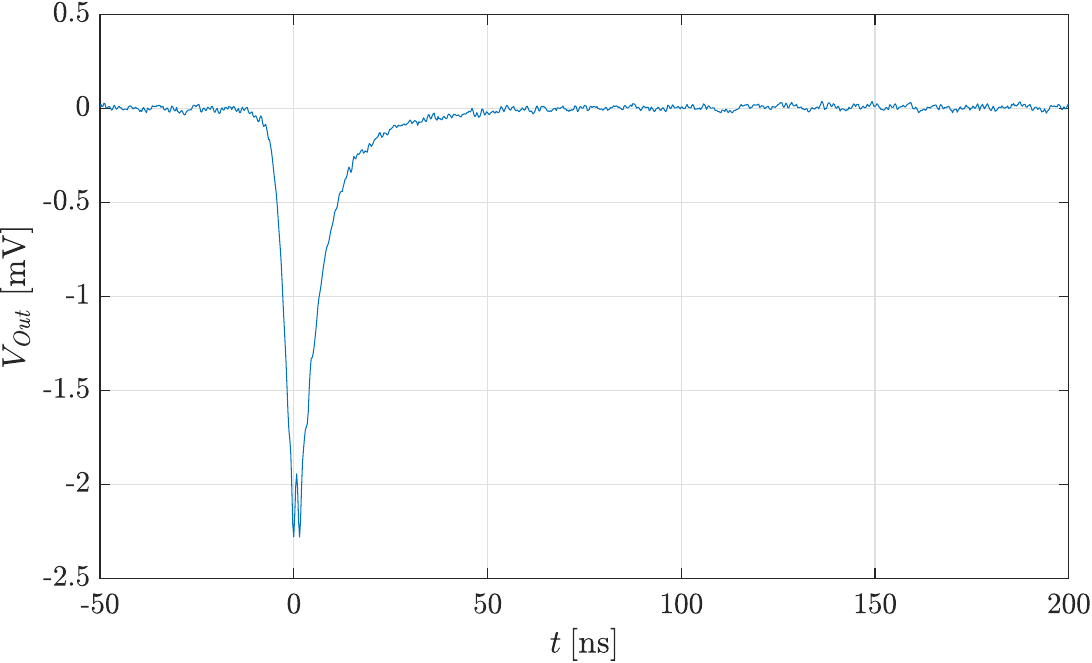}}}\\
	\subfloat[][The 64 channels on the HVS superimposed.]
	{\resizebox{0.95\linewidth}{!}{\includegraphics{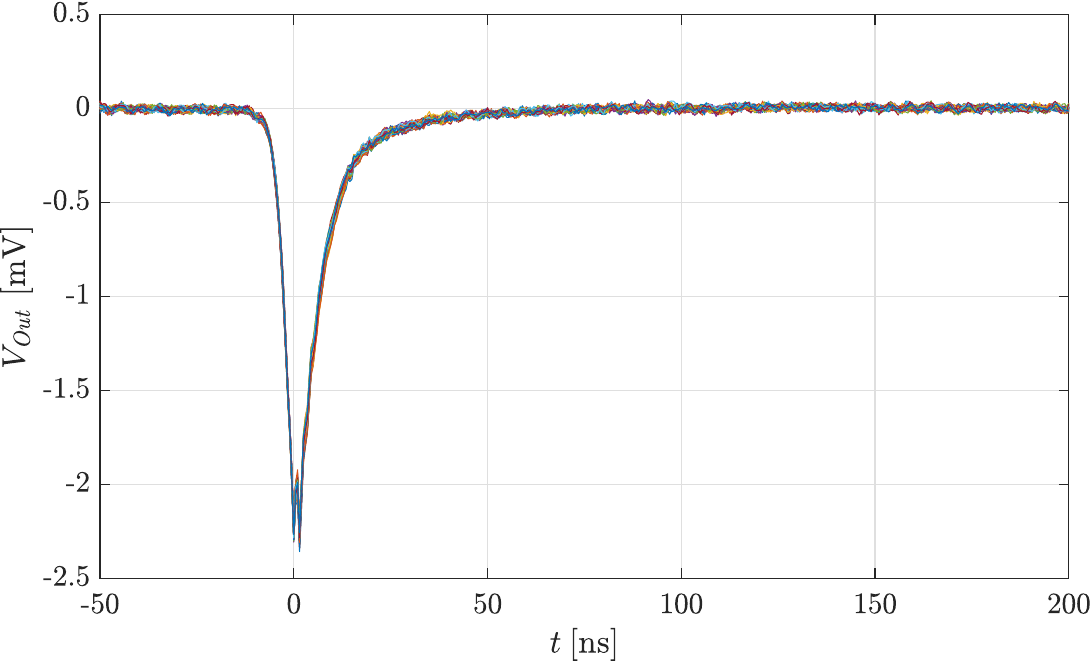}}}
	\caption{Average SPE waveform for a 10m coaxial cable. Every curve was computed by averaging 1000 individual SPE signals and removing the systematic offset.}
	\label{hvs_SPE}
\end{figure}

The lowest energy and most common events that are detected by the SPMTs are SPEs. At nominal gain, $3\times 10^6$ electrons of charge are generated by the SPMT, and the current is effectively entirely captured by the circuit branch containing the decoupling capacitor and the $50\mathrm{\:\Omega}$ shunt resistor, generating a relatively short voltage pulse of roughly $2\mathrm{\:mV}$ in amplitude. Figure \ref{spe_events} shows three examples of typical SPE events. The non-zero baseline visible in the plots is attributed to the systematic offset of the oscilloscope.

Figure \ref{hvs_SPE}(a) shows the average SPE waveform at nominal gain measured on an arbitrarily selected channel on the HVS, while using a $10\mathrm{\:m}$ coaxial cable to connect to the SPMT. This was computed by capturing and averaging 1000 individual SPEs, similar to those shown in Figure \ref{spe_events}, and removing the systematic offset. The same measurements were also taken for all channels on the HVS, and for both $5\mathrm{\:m}$ and $10\mathrm{\:m}$ coaxial cables. The average SPE waveform using a $5\mathrm{\:m}$ cable is practically identical to the one shown in Figure \ref{hvs_SPE}(a), and is therefore not shown. Figure \ref{hvs_SPE}(b) shows the average SPE waveform measured on all 64 channels of the HVS superimposed in the same plot, demonstrating that there are no obvious outliers among the channels.

The plot shown in Figure \ref{hvs_SPE}(a) shows no noticeable distortion, however it does display the remnants of what appears to be white noise, which can obfuscate low-amplitude artifacts.

\subsection{Reflections and overshoot}
\label{subsection:reflections}

The custom coaxial cables used to connect the SPMTs to the HVS are long enough to effectively behave as transmission lines, meaning that they can introduce reflections into the circuit if the signal path is not matched to the characteristic impedance of the cable, which is $50\mathrm{\:\Omega}$. As a consequence of this, signal traces on the PCBs, connectors and termination resistances must have the same impedance as the cable to minimize reflections.

\begin{figure}[t]
	\centering
	\includegraphics[width=0.9\linewidth]{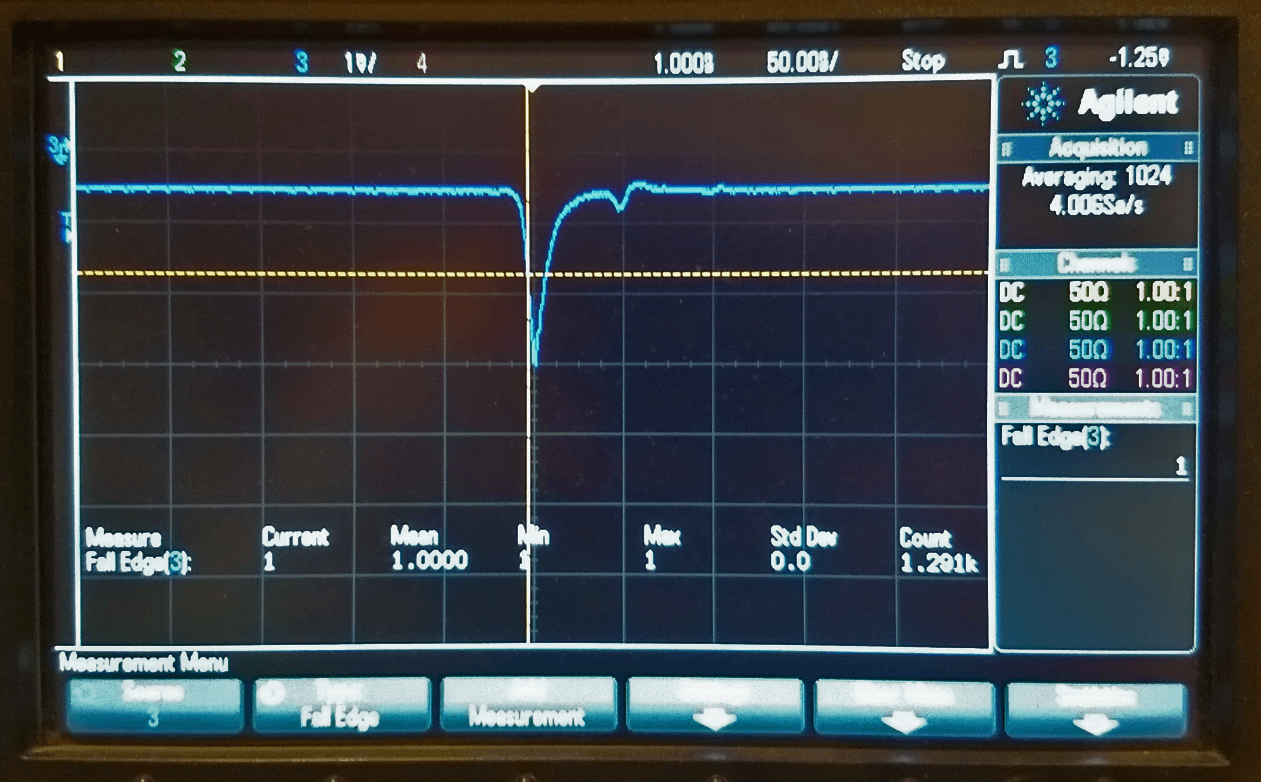}
	\caption{Average SPE waveform for a 5m coaxial cable on a board without impedance matching. The SPMT used is biased at slightly higher than nominal gain.}
	\label{no_matching}
\end{figure}

Figure \ref{hvs_SPE}(a) shows no obvious signs of signal reflection due to the fact that the analog signal traces on the HVS are impedance matched to $50\mathrm{\:\Omega}$. During the development of the HVS, to better appreciate the effects of the impedance matching of the HVS on the overall setup, one of the prototypes was fabricated with thicker dielectrics---for larger insulation margins---and without impedance control. Figure \ref{no_matching} shows an SPE measurement on that older version of the HVS without impedance matching, while using a $5\mathrm{\:m}$ cable to connect to the SPMT. The image shows clear signs of reflection at roughly the $50\mathrm{\:ns}$ mark, which corresponds to twice the signal propagation delay through a $5\mathrm{\:m}$ copper coaxial cable.

\begin{figure}[t]
	\centering
	\includegraphics[width=0.95\linewidth]{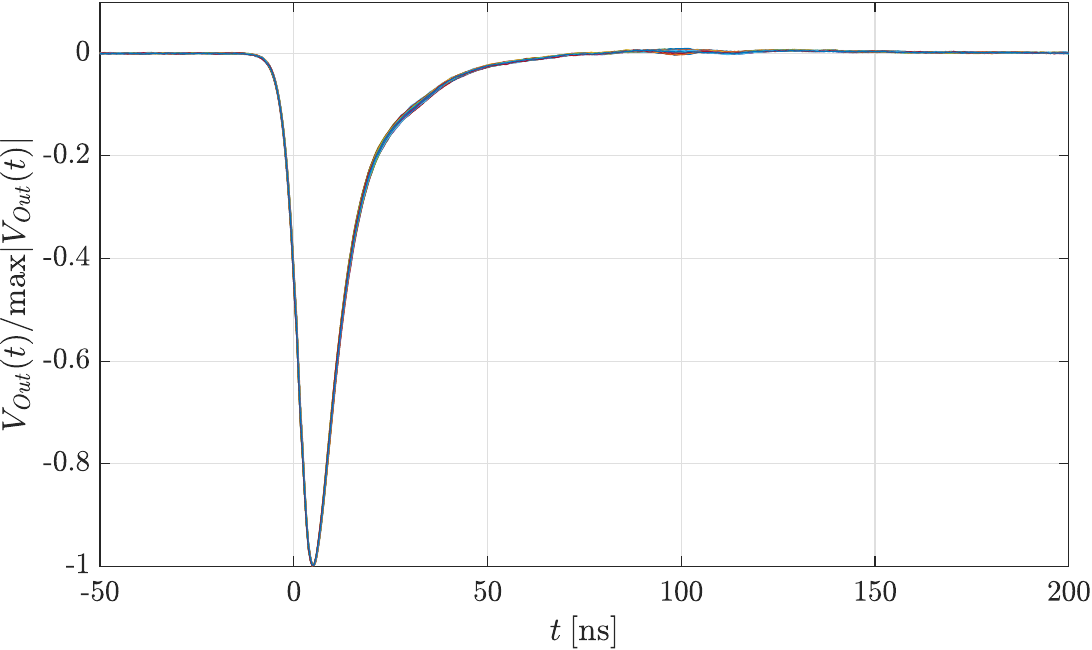}
    \caption{Average normalized waveform for a 10m coaxial cable for all 64 channels on the HVS superimposed. Each curve was computed by averaging 1000 events---and each event was captured with the scope trigger set at -8~mV, adjusted by offset. The resulting waveforms were normalized to unit amplitude.}
	\label{normalized_64ch}
\end{figure}

Although reflections might not be apparent in Figure \ref{hvs_SPE}(a), that does not mean that they are not present in the circuit even with controlled-impedance traces on the PCB. By capturing higher energy events and averaging the results, it is possible to to improve the SNR to better appreciate low-amplitude effects. Figure \ref{normalized_64ch} shows the average waveform of 1000 events while using a $10\mathrm{\:m}$ coaxial cable and setting the oscilloscope trigger at $-8\mathrm{\:mV}$, after accounting for the systematic offset. The resulting waveforms do not have a uniform amplitude, unlike SPE signals, and therefore were normalized to unit amplitude for an easier comparison between channels.

Figure \ref{normalized_64ch} also shows the waveforms for all 64 channels superimposed, showing remarkable similarity on the waveform between all channels. However, there are some signal artifacts at roughly the $100\mathrm{\:ns}$ mark, which corresponds to twice the propagation delay through a $10\mathrm{\:m}$ copper coaxial cable, and are attributable to reflections due to slight impedance mismatches on the signal line. 

\begin{figure}[t]
	\centering
	\subfloat[][]
	{\resizebox{0.95\linewidth}{!}{\includegraphics{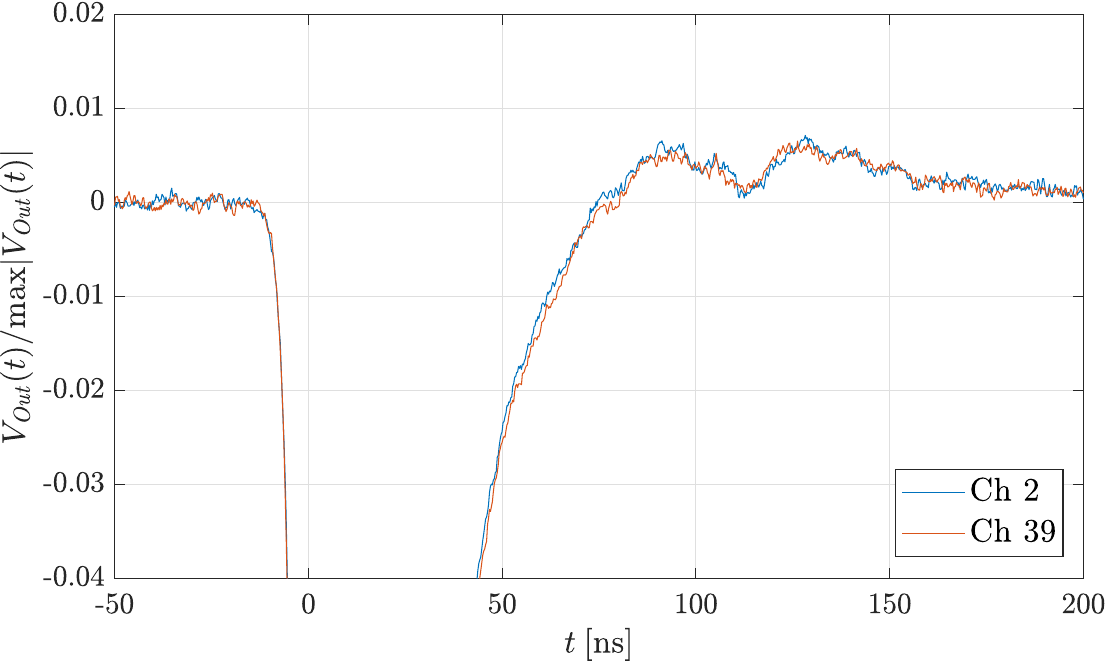}}}\\
	\subfloat[][]
	{\resizebox{0.95\linewidth}{!}{\includegraphics{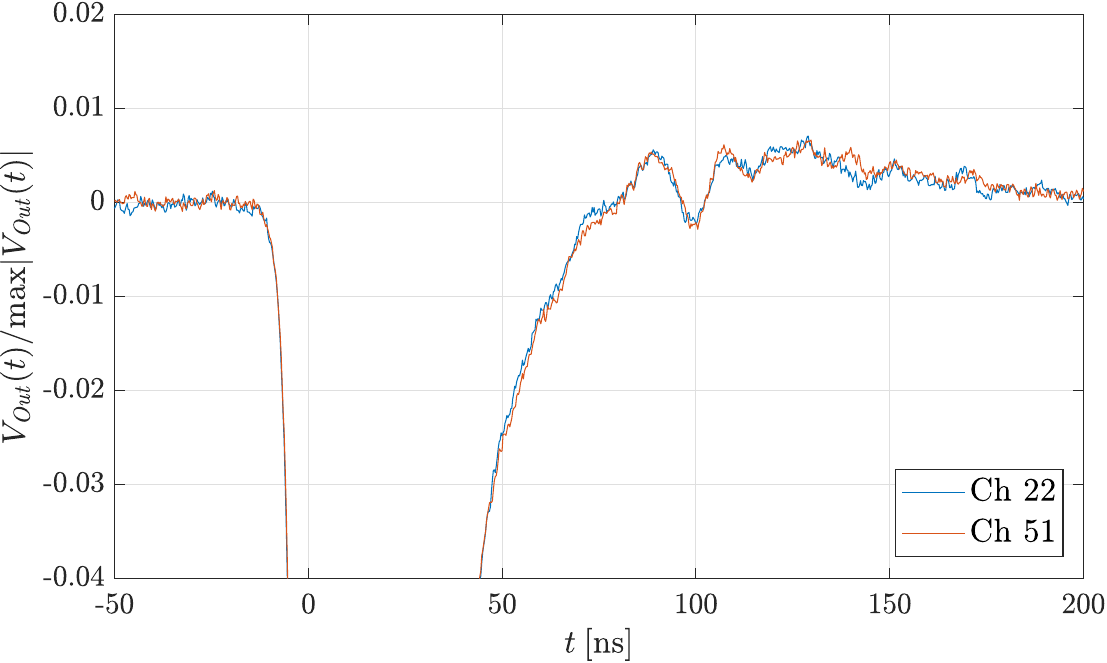}}}
	\caption{Close-up of the normalized waveform for two mirrored-channel pairs.}
	\label{sister_channels}
\end{figure}

Figure \ref{sister_channels} show a close-up of the normalized waveform for two pairs of channels. The layout of the analog traces on the HVS is symmetrical, and these pairs of channels were selected due to them having an identical layout, although mirrored. From the plots, it appears that there are still some residual overshoot and reflections, but the amplitude of these artifacts is small enough to be practically inconsequential. Furthermore, it can be seen that the behavior of the channels on the pair is identical between them, suggesting that the shape of the artifacts is a function of the particular layout for that pair of channels. The largest positive and negative artifact peaks measured on the board are roughly $0.9\%$ and $0.3\%$ of the peak amplitude of the signal, respectively.

\subsection{Crosstalk}

On a PCB, crosstalk is caused by the capacitive, inductive and resistive coupling between traces of different channels. Common practices to reduce this effect include the use of inner-layer routing (stripline), the increase of trace spacing and the reduction of the distance of signal traces to reference planes.

On the HVS, the routing of the 64 analog channels is distributed through three layers: the top layer and two inner layers. Throughout the testing of the multiple HVS prototypes, it quickly became apparent that interlayer crosstalk was minimal, and that the worst cases for intralayer crosstalk was between neighboring channels on the top layer. As it was not possible to omit entirely the routing through the top layer, some measures were taken in an attempt to minimize crosstalk: by minimizing the distance of the top layer to the nearest ground plane; and by increasing trace spacing given the available space, the high-voltage clearance and the fixed trace width due to impedance matching.

\begin{figure*}[t]
	\centering
	\subfloat[][Aggressor and victim line]
	{\resizebox{0.46\linewidth}{!}{\includegraphics{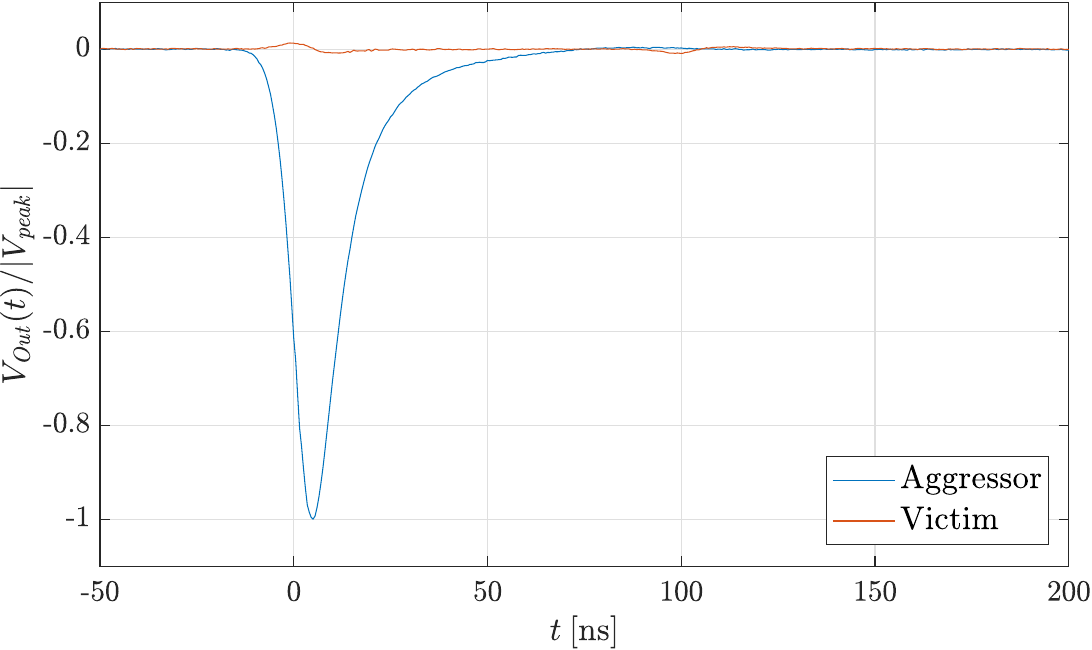}}}\hspace{1mm}
	\subfloat[][Zoom-in on the waveform of the victim line]
	{\resizebox{0.46\linewidth}{!}{\includegraphics{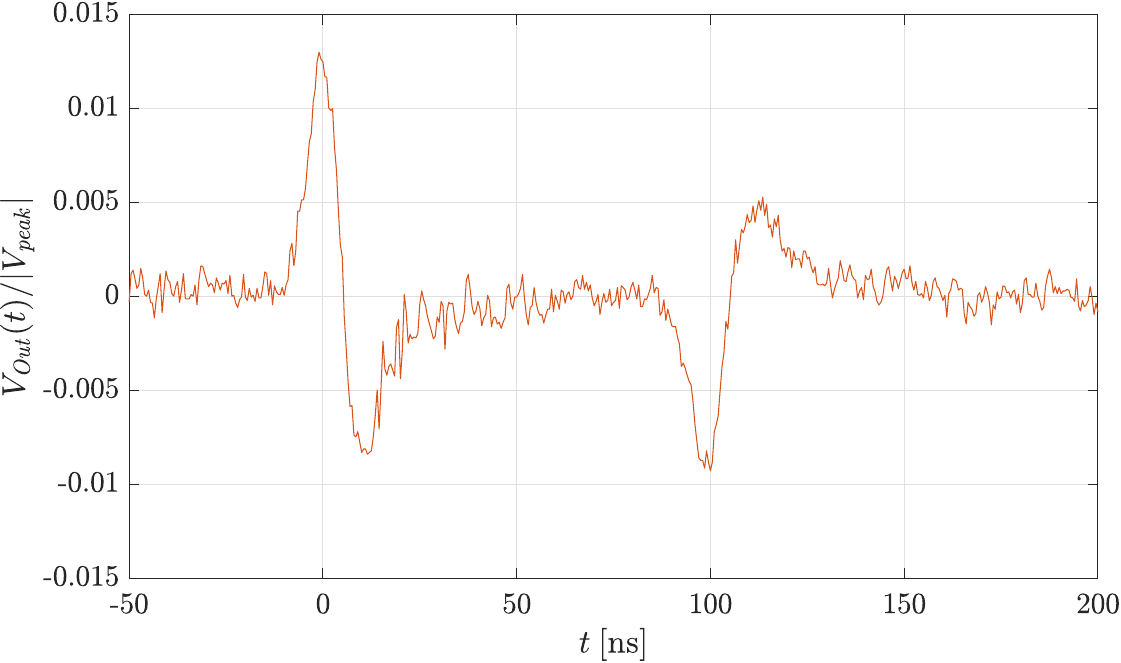}}}
	\caption{Crosstalk on a victim-aggressor pair.}
	\label{xtalk_agg_vic}
\end{figure*}

To qualify the crosstalk behavior of the HVS, two channels were measured simultaneously with an oscilloscope: an aggressor channel in which signal was injected with an SPMT, and an idle victim channel without an SPMT source. Figure \ref{xtalk_agg_vic} shows an example of crosstalk between a pair of channels which are adjacent on the top layer. The waveforms in Figure \ref{xtalk_agg_vic} are normalized by the peak amplitude of the aggressor signal ($|V_\mathit{peak}|$). From \ref{xtalk_agg_vic}(b) it can be seen that there are two crosstalk artifacts: the first coincides with the timing of the aggressor signal and has a positive peak followed by a negative one, and the second occurs at roughly $100\mathrm{\:ns}$, which has a negative peak followed by a positive one. The same crosstalk behavior was observed for all the channel pairs that were measured.

\begin{figure}[t]
	\centering
	\includegraphics[width=0.95\linewidth]{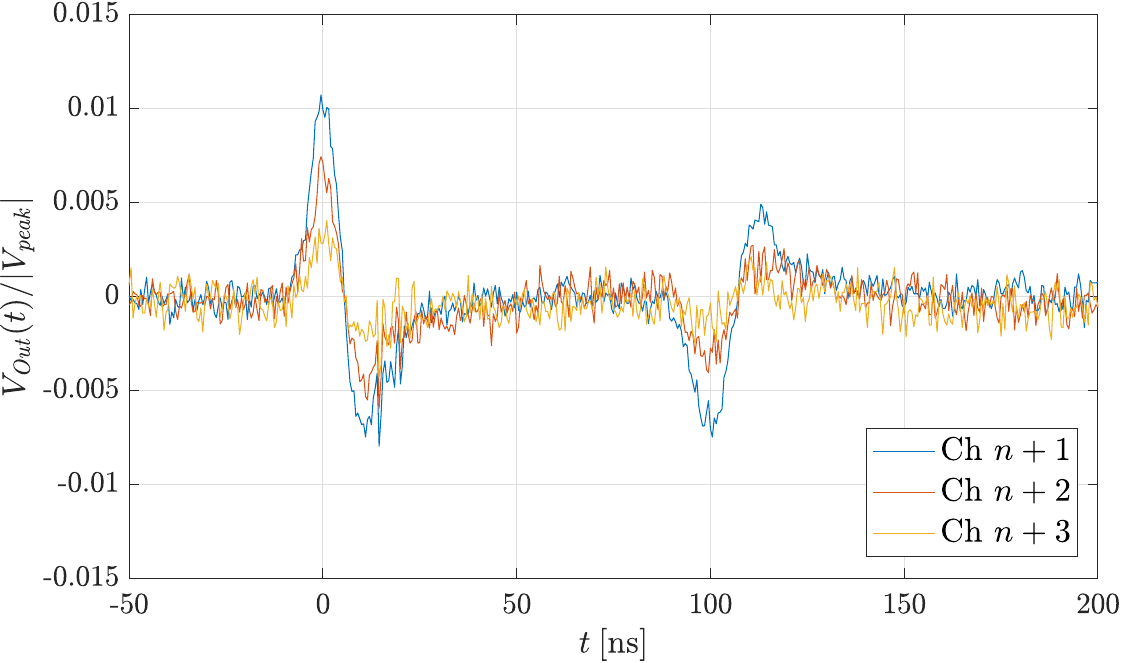}
	\caption{Crosstalk between an aggressor channel, namely channel $n$, and adjacent channels with increased channel separation.}
	\label{xtalk_separation}
\end{figure}

The reasons as to why the crosstalk waveform has two distinct artifacts is not immediately apparent. The second artifact appears to be a reflection of the first one with inverted polarity, which is a sign of a transmission line terminated with a short circuit or low impedance. Furthermore, the delay between the first and second artifacts is roughly twice the propagation delay through a $10\mathrm{\:m}$ cable. However, it is not clear from the circuit and the measurement setup the mechanism by which the reflection is being generated.

To further understand the behavior of the observed crosstalk, measurements were taken for limited selection of channels while adjusting the oscilloscope trigger, and it was observed that the resulting crosstalk amplitude linearly increased with the amplitude of the aggressor signal. Measurements were also taken for a limited selection of second-order neighbors, that is, channels that are parallel an adjacent to the aggressor line, but with increased channel separation. Figure \ref{xtalk_separation} shows the crosstalk for second-order neighbors with up to three channels of separation. The amplitude of the crosstalk artifacts naturally decreases with increasing separation, but the effect is still noticeably present in near channels.

\begin{figure}[t]
	\centering
	\includegraphics[width=0.95\linewidth]{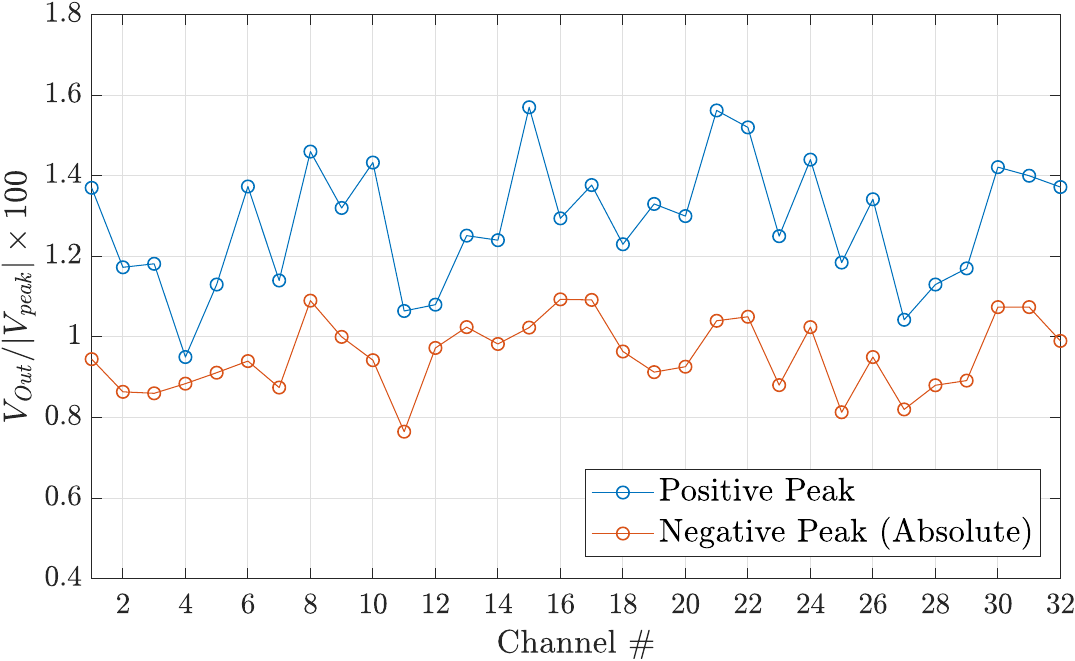}
	\caption{Peak value of the positive and negative crosstalk peaks measured on each channel displayed as a percentage of the aggressor signal peak. Since the routing geometry of the HVS is symmetrical, only the first 32 channels are displayed.}
	\label{xtalk_peak}
\end{figure}

Since measuring the crosstalk between any arbitrary pair of channels would have been too time consuming, the systematic study of this effect was limited to channels that were direct neighbors in any of the three signal layers, as these correspond to the worst-case scenarios. The crosstalk waveform in Figure \ref{xtalk_agg_vic}(b) has four clearly discernible peaks, and the maximum and the minimum values measured on each channel were recorded and plotted, as shown in Figure \ref{xtalk_peak}. In some cases, channel neighbors on the top layer and inner layers were different, and therefore only the worst case was considered in the plot. Since the routing geometry of the HVS is symmetrical, only the crosstalk of the first 32 channels are displayed on the plot.

The values on Figure \ref{xtalk_peak} are expressed as a percentage of the peak signal of the aggressor channel. Positive crosstalk peaks can have an effect on energy reading if they coincide with real events, but of particular relevance are the negative peaks, as they can induce false positives. For example, as the CATIROC discriminator is set at 1/3 of the peak amplitude of an SPE, then an event with an amplitude of 33.3 SPEs could trigger a false positive on a victim channel with a $1\%$ peak negative crosstalk.

As visible in Figure \ref{xtalk_peak}, the worst cases measured peaked at 1.6\% in positive polarity and 1.1\% in negative polarity. Although the HVS is the limiting factor in the crosstalk budget of the SPMT front-end electronics, the performance falls well within system specifications.

\section{Reliability and aging}
\label{section:reliability}

\subsection{Definitions}

\subsubsection{Acceleration factor}

In order to test the reliability of a component over a long period of time, it is common practice to use methods for accelerated aging, as it can be impractical to do so through real-time testing. One such method, and the one used to assess the reliability of the HVS and its electronic components, is the Arrhenius High Temperature Operating Life (HTOL) method, which is used to evaluate the long-term reliability of electronic components under normal operating conditions but at elevated temperatures \cite{Nelson2004}. This method is based on the Arrhenius model for time acceleration ($A_f$), given by the following formula:
\begin{equation}
A_f = \exp\left[ \frac{E_a}{k} \left( \frac{1}{T_a} - \frac{1}{T_e}\right)\right]
\label{eq:acc_factor}
\end{equation}
where $T_a$ is the operational temperature, $T_e$ is the experimental temperature, $E_a$ is the activation energy, and $k$ is the Boltzmann constant.

The activation energy ($E_a$) is an empirical value defined as the minimum energy required to initiate a failure mode process in a given material. It is commonly derived from Accelerated Life Testing (ALT) of a component, an aging method that introduces additional stress variables to induce component failure, and by recording the time to failure of the component at multiple elevated temperatures, which makes it possible to derive $E_a$ from linear regression. In absence of empirical data, generic values dependent on the type of material are sometimes used, such as $0.7\mathrm{\:eV}$ for diode-type semiconductors.

\subsubsection{Failure rate and FIT}

The main figure of merit used to quantify the reliability of components is the failure rate ($\lambda$), defined as the probability that a component will fail within a unit of time. It is calculated by dividing the total number of component rejected by the total time of operation. For the HTOL method, total time of operation is referred to as Equivalent Device Hours ($\textit{EDH}$), and is given by the following formula:
\begin{equation}
\textit{EDH}=D\cdot H \cdot A_f
\end{equation}
where $D$ is the number of devices tested, $H$ is the number of hours the test was run, and $A_f$ is the acceleration factor defined in (\ref{eq:acc_factor}). The failure rate is then defined as:
\begin{equation}
\lambda_h = \frac{r}{\textit{EDH}}
\label{eq:failure_rate}
\end{equation}
where $\lambda_h$ is the failure rate \textit{per hour}, and $r$ is the total number of components rejected.

Frequently, after performing accelerated aging tests on a group of components, either none or a very small number of them will have exhibited any failures over the time of operation, which is often the case when the sample size is relatively small. However, it is still possible to make statistical observations about the failure rate of a component with a certain confidence level. To do so, the number of rejects $r$ in (\ref{eq:failure_rate}) is replaced by a probability function using the chi-squared distribution ($\chi^2$):
\begin{equation}
r \sim \frac{\chi^2(\alpha, \nu)}{2}
\label{eq:n_rejects}
\end{equation}
where $\alpha$ is the confidence level or probability that the \textit{real} number of failures over an extended period of time is below the calculated number, and $\nu$ is the degrees of freedom of the $\chi^2$ distribution. The reliability calculation employed here uses $\nu = 2(r+1)$, where $r$ is the number of rejects.

From (\ref{eq:failure_rate}) and (\ref{eq:n_rejects}), a more statistically significant value of $\lambda_h$ is given by:
\begin{equation}
	\lambda_h = \frac{\chi^2(\alpha, \nu)}{2 \cdot \textit{EDH}}
	\label{eq:failure_rate_chi}
\end{equation}

The Failure in Time (FIT) is an standard industry value which is often used instead, defined as the failure rate per billion hours of operation:
\begin{equation}
	\mathrm{FIT} = \lambda_h \cdot 10^9
\end{equation}

\subsection{JUNO requirements}

As described in Section \ref{subsubsection:reliability}, the system-level requirement for the acceptable number of failed UWBs was set at a maximum of $10\%$ loss after 6 years of operation, which is equivalent to a failure rate $\lambda_h$ of:
\begin{equation}
	\lambda_h < \frac{0.1}{6\times 365\times 24} \left[\frac{\mathrm{fail}}{\mathrm{hour}}\right]
\end{equation}
Therefore, the FIT requirement is:
\begin{equation}
	\mathrm{FIT} < 1900
\end{equation}
Since component failure rate estimations can be imprecise due to the exponential acceleration factor used in the calculations, this global FIT value serves as an order of magnitude upper bound for the different electronic components tested.

\subsection{Diodes}

\begin{figure}[t]
	\centering
	\includegraphics[width=0.85\linewidth]{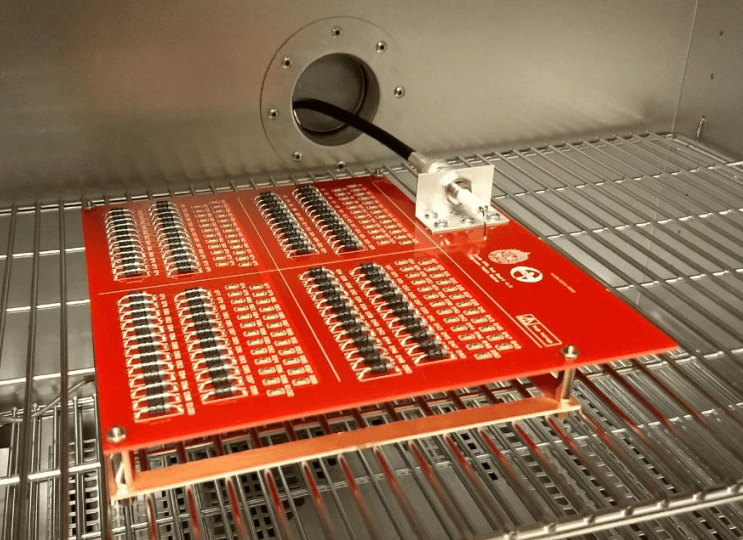}
	\caption{PCB containing a 100 high-voltage diodes placed inside an burn-in oven.}
	\label{diodes_oven}
\end{figure}

\begin{figure}[t]
	\centering
	\includegraphics[width=0.5\linewidth]{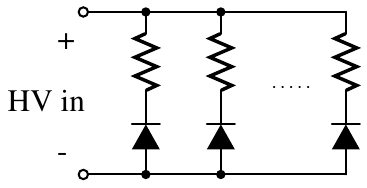}
	\caption{Schematic of the diode board, where 100 resistors and diodes are connected in parallel.}
	\label{diodes_schematic}
\end{figure}

In order to assess the reliability of the high-voltage diodes (PN: R5000F), a PCB containing 100 of these components was manufactured, as shown in Figure \ref{diodes_oven}. The diodes on this board are placed in reverse bias, and are connected through resistors to an external high-voltage power supply through a single SHV coaxial connector. A simplified schematic of the circuit is shown in Figure \ref{diodes_schematic}.

\begin{table*}[t]
	\centering	
	\caption{Variables for the calculation of the failure in time (FIT) of HV-Splitter components, using experimental results.}
	\begin{tabular}{c|c|c|c|c}
		\multirow{2}{*}{\textbf{Variable}} & \multirow{2}{*}{\textbf{Nomenclature}} & \multicolumn{3}{c}{\textbf{Value}} \\ \cline{3-5}
		&  & \textbf{Diodes} & \textbf{Capacitors} & \textbf{HVUs} \\ \hline\hline
		Operational temperature & $T_a$ & \multicolumn{3}{c}{295.15} \\ \hline
		Experimental temperature & $T_e$ & 373.15~K & 373.15~K & 343.15~K \\ \hline
		Applied voltage & $V_\mathit{DC}$ & 1.75~kV & 1.2~kV & 2.7~kV \\ \hline
		Number of devices & $D$ & 100 & 100 & 186 \\ \hline
		Number of hours & $H$ & 598.5 & 122.8 & 7,300 \\ \hline
		Observed number of failures & $N$ & 0 & 0 & 1 \\
		\hline\hline
	\end{tabular}
	\label{component_summary}
\end{table*}

The test consisted of placing the board in a burn-in oven at $100^\circ \mathrm{C}$ while powered with a high-voltage power supply set at $1.75\mathrm{\:kV}$. This would emulate the normal operation of the diode---when reversed biased---under accelerated aging conditions. The test ran for a total of $598.5$ hours. During that time, the voltage of the power supply was routinely monitored to look for changes that could indicate a short, which did not occur. Table \ref{component_summary} shows a summary of the conditions of the test and the obtained results.

Two simple measurements were done on the diodes to qualify if their proper operation: (i) check if the diode is either shorted or in open circuit and (ii) measure the diode with a multimeter in diode mode to check for anomalies. The measurements were done before and after the board was subjected to accelerated aging, and all components were deemed to be working properly after the test.

To compute the FIT value, a confidence level of 90\% was used, and an activation energy of $1\mathrm{\:eV}$ was considered, following the recommendation of the manufacturer. With these considerations, the FIT value can be computed to be $10.4$. Furthermore, the FIT was also computed using the manufacturer-provide data shown in Table \ref{diode_summary_rectron}, yielding a value of $0.1$ \cite{Rectron2019}.

\begin{table}[t]
	\centering	
	\caption{Variables for the FIT calculation of the high-voltage diodes, provided by the manufacturer (Rectron) \cite{Rectron2019}.}
	\resizebox{0.9\linewidth}{!}{
		\begin{tabular}{c|c|c}
			\textbf{Variable} & \textbf{Nomenclature} & \textbf{Value} \\ \hline\hline
			Experimental temperature & $T_e$ & 398.15~K \\ \hline
			Applied voltage & $V_\mathit{DC}$ & 5~kV \\ \hline
			Device hours & $D\cdot H$ & 1{\small,}080{\small,}000 \\ \hline
			Observed number of failures & $N$ & 0 \\
			\hline\hline
		\end{tabular}
	}
	\label{diode_summary_rectron}
\end{table}

\subsection{Capacitors}

\begin{figure}[t]
	\centering
	\subfloat[][Printed circuit board.]
	{\resizebox{0.85\linewidth}{!}{\includegraphics{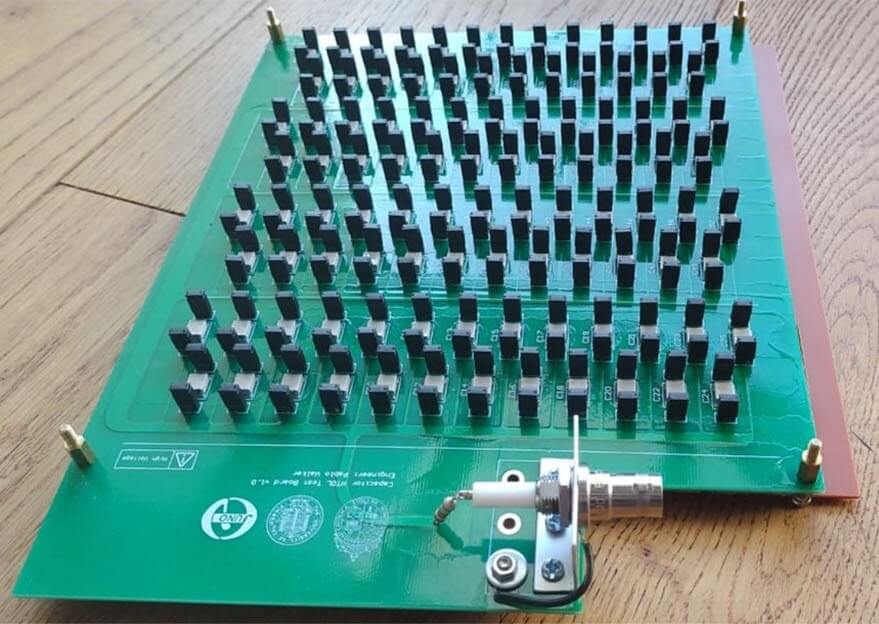}}}\\
	\subfloat[][Zoom of a capacitor with jumpers on both ends.]
	{\resizebox{0.6\linewidth}{!}{\includegraphics{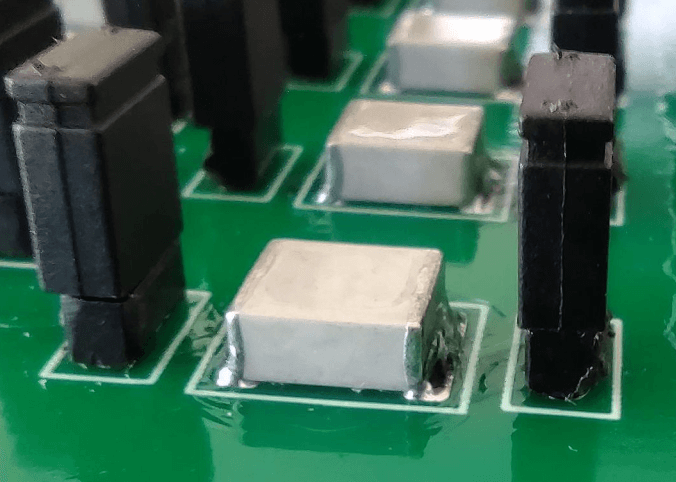}}}
	\caption{PCB containing 100 high-voltage capacitors connected in parallel. Jumpers are placed on both ends of each capacitor in order to disconnect them from the circuit if necessary.}
	\label{capacitors_oven_board}
\end{figure}

A procedure similar to the one described in the previous section was used to assess the reliability of the high-voltage capacitors (PN: C2220C392MGGACTU). A PCB containing 100 capacitors connected in parallel was used for the accelerated aging test, as shown in Figure \ref{capacitors_oven_board}(a). Although all capacitors are connected in parallel, jumpers were placed at both ends of each component, as shown in Figure \ref{capacitors_oven_board}(b), to allow them to be disconnected from the circuit when needed. This feature can be used in the event of a failure to isolate the component, but more importantly, it was used for in-circuit measurements of the capacitors without needing to desolder them. The board has a single SHV coaxial connector, which is used to connect it to an external high-voltage power supply.

The test consisted of placing the board in a burn-in oven at $100^\circ \mathrm{C}$ while powered with a high-voltage power supply set at $1.2\mathrm{\:kV}$. The test ran for a total of $122.8$ hours. Table \ref{component_summary} shows a summary of the conditions of the test and the obtained results.

For the duration of the experiment, the voltage and current of the power supply were routinely monitored and recorded. At the start of the experiment, the current showed a measurement in the tens of $\upmu\mathrm{A}$, but it slowly decreased, reaching submicron levels at around the 35 hour mark. The cause of the leakage is unknown, but the most likely cause was either ambient humidity or the capacitors themselves. Regardless, as the current leakage eventually subsided, it was not deemed problematic.

\begin{figure}[t]
	\centering
	\includegraphics[width=0.95\linewidth]{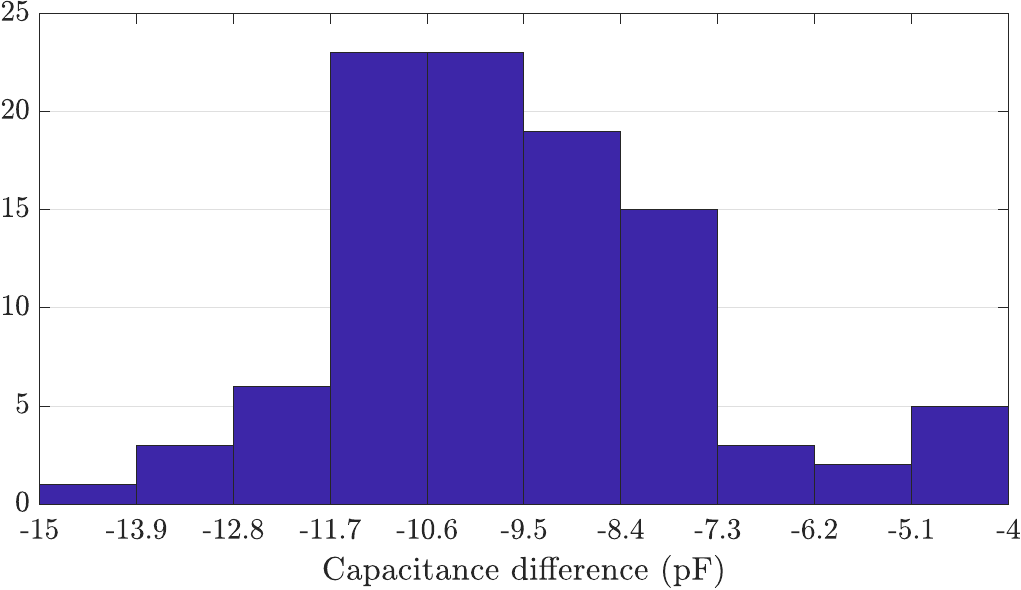}
	\caption{Capacitance changes after aging.}
	\label{cap_changes}
\end{figure}

The capacitance of the components was measured both before and after the test, and the changes in capacitance due to aging were tabulated and plotted, as shown in Figure \ref{cap_changes}. The capacitance changes were small enough as to be effectively inconsequential to the performance of the HVS, being much smaller than the capacitance differences that can be expected due to the $20\%$ tolerance of the component. It was deemed that no component had failed through the accelerated aging test.

The vendor provided reliability data for the family of components that the capacitor forms part of, and the data for one of the months is displayed in Table \ref{capacitor_summary_kemet} \cite{Kemet2020}. However, the information provided did not include the activation energy of this particular component. To compute the FIT value, a confidence level of 90\% was used, and in absence of empirical data, an activation energy of $1\mathrm{\:eV}$ was considered. With these considerations, the FIT value can be computed to be $10.4$ and $0.4$ based on the experimental and vendor-provided data, respectively. Naturally, both of these values are well below the FIT requirement of 1900.

\begin{table}[t]
	\centering	
	\caption{Variables for the FIT calculation of the high-voltage capacitors, provided by a KEMET representative \cite{Kemet2020}. A voltage that is twice the voltage rating of the capacitors was applied during the test.}
	\resizebox{0.9\linewidth}{!}{
		\begin{tabular}{c|c|c}
			\textbf{Variable} & \textbf{Nomenclature} & \textbf{Value} \\ \hline\hline
			Experimental temperature & $T_e$ & 398.15~K \\ \hline
			Applied voltage & $V_\mathit{DC}$ & 4~kV \\ \hline
			Number of devices & $D$ & 3{\small,}200 \\ \hline
			Number of hours & $H$ & 2{\small,}000 \\ \hline
			Observed number of failures & $N$ & 2 \\
			\hline\hline
		\end{tabular}
	}
	\label{capacitor_summary_kemet}
\end{table}

\subsection{High Voltage Units}
\label{subsection:hvu_reliability}

The reliability of the HVUs was extensively tested during their development for their deployment in the LPMT system, the details of which are out of the scope of this document. The activation energy of the HVUs was determined experimentally to be $0.63\mathrm{\:eV}$. The results of the high-temperature reliability test conducted on the HVUs are summarized in Table \ref{component_summary}. From these results, and considering a confidence level of 90\%, the FIT of the HVUs can be computed to be 53.

During testing, it was observed that the communication with the HVUs becomes erratic at temperatures higher than $70^\circ\mathrm{C}$, and they stopped working entirely at around $85^\circ\mathrm{C}$, which is why the experimental temperature of the HVUs was limited to $70^\circ\mathrm{C}$. This can be explained, in part, due to the melting of the Pentelast-712 compound used in their construction.

\subsection{Printed circuit board}

\begin{figure*}[t]
	\centering
	\subfloat[][HVS without mold and compound.]
	{\resizebox{0.5\linewidth}{!}{\includegraphics{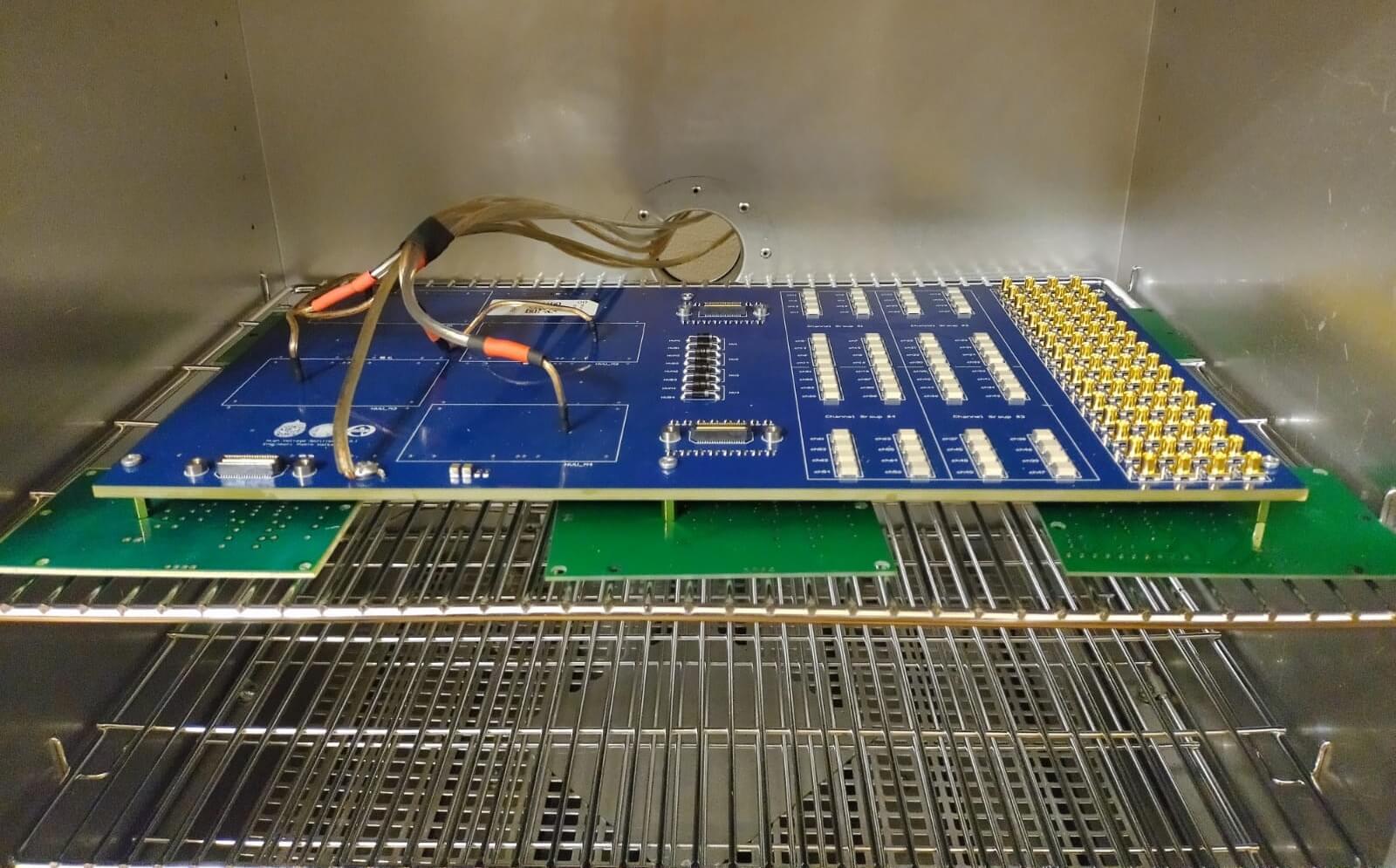}}} \hspace{2mm}
	\subfloat[][HV-Units placed outside the oven.]
	{\resizebox{0.3035\linewidth}{!}{\includegraphics{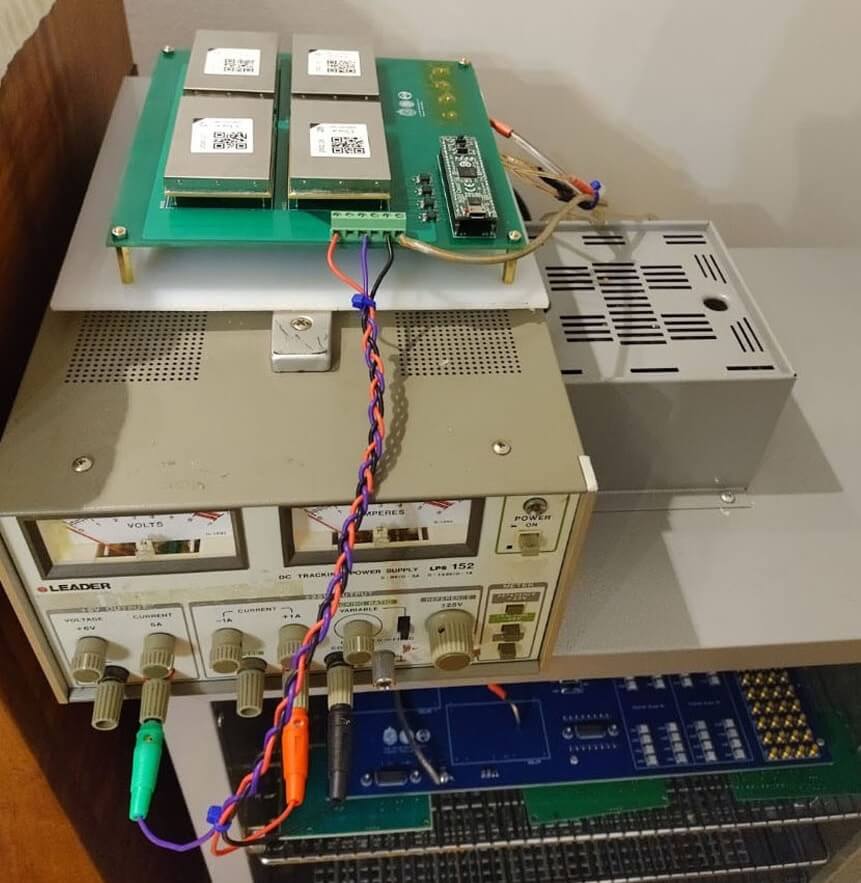}}}
	\caption{Single PCB used for the accelerated aging test of the HV-Splitter at 100$^\circ$C. The use of the Pentelast-712 compound was omitted and the HVUs placed outside the oven, as temperatures above 70$^\circ$C can melt the compound.}
	\label{hvs_single_aging}
\end{figure*}

Due to limitations in budget and schedule, only a single HVS PCB was used to test the reliability of the board itself. The primary objective of this test was to assess the effects of dielectric aging and determine whether aging could degrade the board's electrical insulation to the point of dielectric failure.

The experimental setup used in the accelerated aging test of the HVS is shown in Figure \ref{hvs_single_aging}. As described in Section \ref{subsection:hvu_reliability}, due to issues encountered during the high-temperature testing of the HVUs, the maximum experimental temperature to which they were subjected was limited to $70^\circ\mathrm{C}$. To increase the temperature that can be applied to the HVS and therefore increase the time acceleration factor, a custom PCB was designed to house the HVUs outside of the oven for the duration of the test, while the high-voltage output and ground of the HVUs was connected to the HVS through cables. 

The test consisted of placing the board in a burn-in oven at $100^\circ \mathrm{C}$ while applying high voltage with the HVUs. The test ran for a total of $689.5$ hours, which would be equivalent to aging the board for 291 years. The applied voltage was $1\mathrm{\:kV}$ at the start of the experiment, and it was gradually increased to $1.4\mathrm{\:kV}$, maintaining that value until the end.

\begin{figure}[t]
	\centering
	\subfloat[][Normalized waveforms.]
	{\resizebox{0.95\linewidth}{!}{\includegraphics{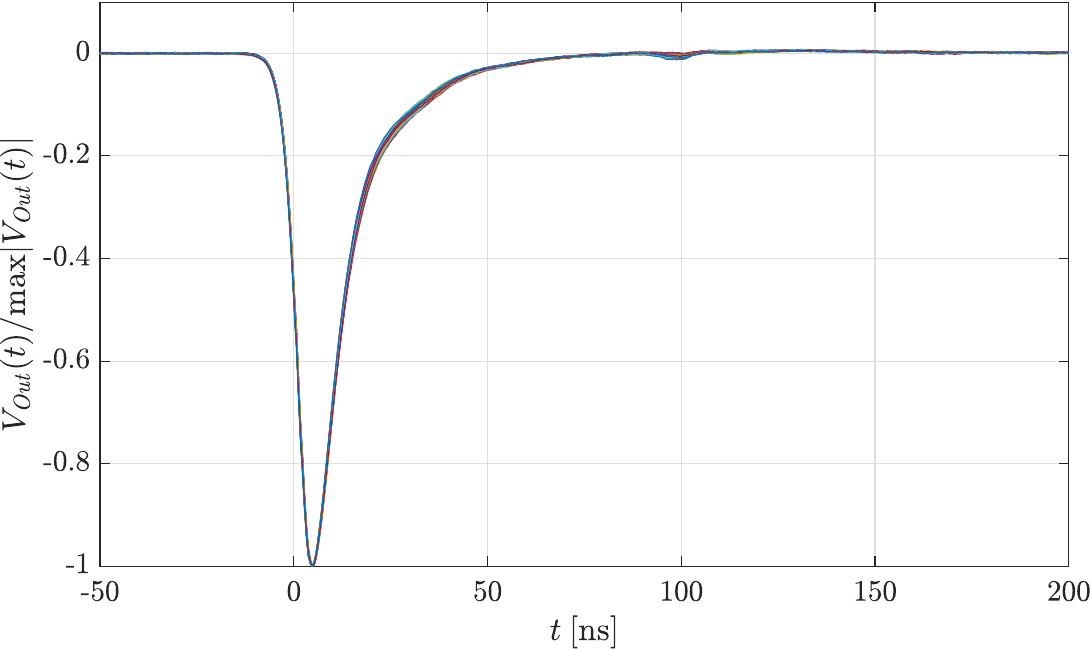}}}\\
	\subfloat[][Zoom-in on signal reflection.]
	{\resizebox{0.95\linewidth}{!}{\includegraphics{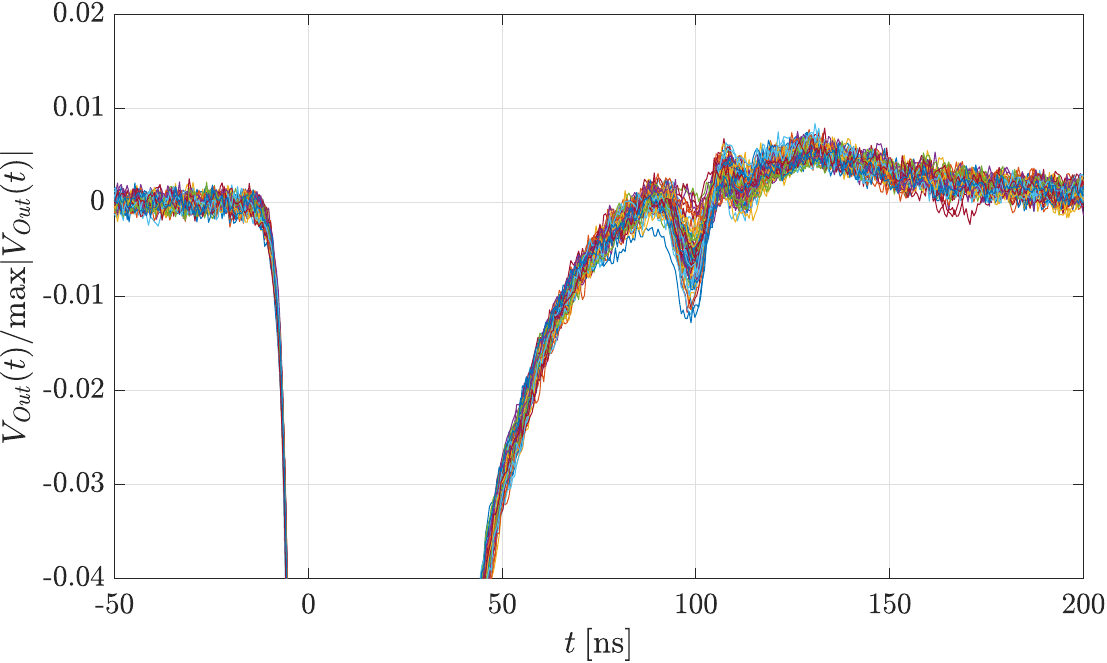}}}
	\caption{Average normalized signal measurements for a 10m coaxial cable for all 64 channels of the HVS superimposed, performed after the accelerated aging test.}
	\label{aging_signal}
\end{figure}

The voltage was monitored and recorded with the HVUs for the duration of the test, and no unusual behavior was recorded---particularly voltage drops that could indicate leakage. Before the experiment, the SPMT signal waveforms were captured and recorded for all channels, resulting in a plot effectively identical to the one shown in Figure \ref{normalized_64ch}. The measurements were repeated after the experiments, the results of which are shown in Figure \ref{aging_signal}. From the plot, it can be seen that the effects of signal reflection are considerably more pronounced than the ones shown in Figure \ref{normalized_64ch}---going from a maximum $0.3\%$ negative peak to several channels over $1\%$--- which suggest a deterioration on the impedance matching of the board. This, in turn, could be explained due to the aging of the dielectric materials. However, it is hard to say if the aging caused by thermal cycling---likely caused by the differential material expansions---are perfectly analogous to what can be expected due to time alone. Nonetheless, the observed changes in the accelerated aging test are small enough to not be a cause for concern. Additionally, no issues were observed in relation to the high-voltage insulation of the board.

Although the sample size is a single board, it is still possible to compute a FIT value for the HVS. In absence of empirical data, an activation energy of $1\mathrm{\:eV}$ was considered for the board as a whole, given that the same value was used for the diode and capacitor calculations---the most critical components on the board. Considering a confidence level of 90\%, the FIT value can be computed to be 900.1.

\section{Production and Testing}
\label{section:production}

\subsection{Manufacturing yield}

The HVS printed circuit board (PCB) was produced by the China-based company Shenzhen Zsipak Technologies Co., Ltd. in 2022. A total of 435 HVS boards were manufactured---35 more than the minimum requirement of 400. Of these additional boards, eight will be installed in the JUNO-TAO experiment \cite{Lombardo2023}, while the remaining boards will serve as spares.

To ensure high reliability, as part of the quality control procedure, all HVS boards underwent burn-in and functional testing, during which a small number of PCB defects were detected and repaired. Ultimately, 430 boards were deemed fully functional, while five failed to pass the functional test---described in Section \ref{subsection:functional_testing}---resulting in a yield of approximately 99\%. Of the five problematic boards, two had HVUs that were not outputting high voltage, one had an unresponsive HVU, one did not show signals in a few channels, and one failed the short-circuit test.

\subsection{Burn-in Testing}

A common practice in the production of PCBs for applications requiring high reliability is burn-in testing, where boards are subjected to temperature stress to remove early failures, i.e., units that, due to component defects, are prone to failure early in their operating life. All the HVS boards underwent burn-in testing during production.

The burn-in testing for the HVS was conducted following the guidelines of the GJB899-90 Chinese military standard \cite{GJB899-90}. Each HVS board was subjected to 13 thermal cycles, each ranging from 0--$70^\circ$C. A cycle starts at $0^\circ$C, where it dwells for 15 minutes, then heats up over 7 minutes at a rate of $10^\circ$C/min, dwells at $70^\circ$C for 15 minutes, and finally cools down over 7 minutes at a rate of $10^\circ$C/min. The maximum temperature was limited to $70^\circ$C, as temperatures above this would cause the Pentelast-712 compound in both the HVS and HVUs to start melting, which could result in permanent damage to the boards.

While being subjected to temperature cycling, the main HVUs were set to output $1.2\mathrm{\:kV}$ and the backup units were set to output $1.15\mathrm{\:kV}$ while their output was loaded. The internal voltage reading of the HVUs was monitored during the process, and the test concluded without any major issues.

\begin{figure}[t]
	\centering
	\subfloat[][HVS boards connected and ready to undergo burn-in.]
	{\resizebox{1\linewidth}{!}{\includegraphics{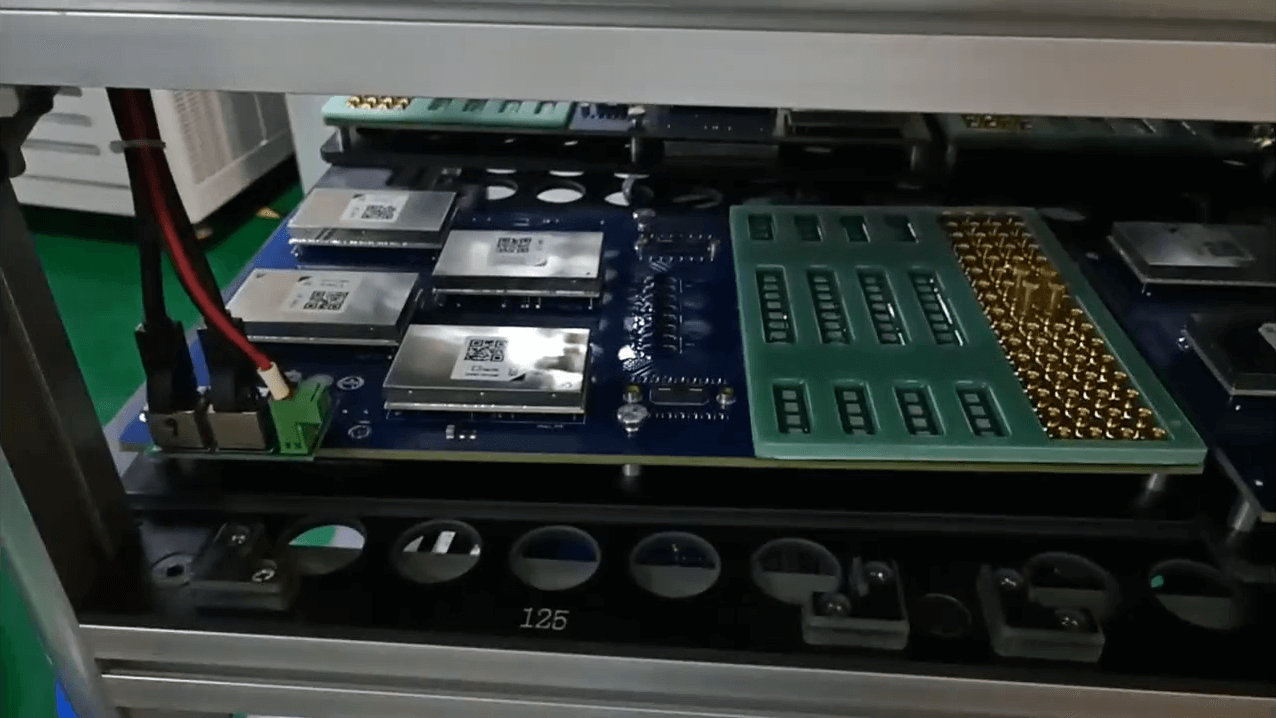}}}\\
	\subfloat[][The Burn-in Control Board. LEDs show that the HVUs are outputting high voltage correctly]
	{\resizebox{1\linewidth}{!}{\includegraphics{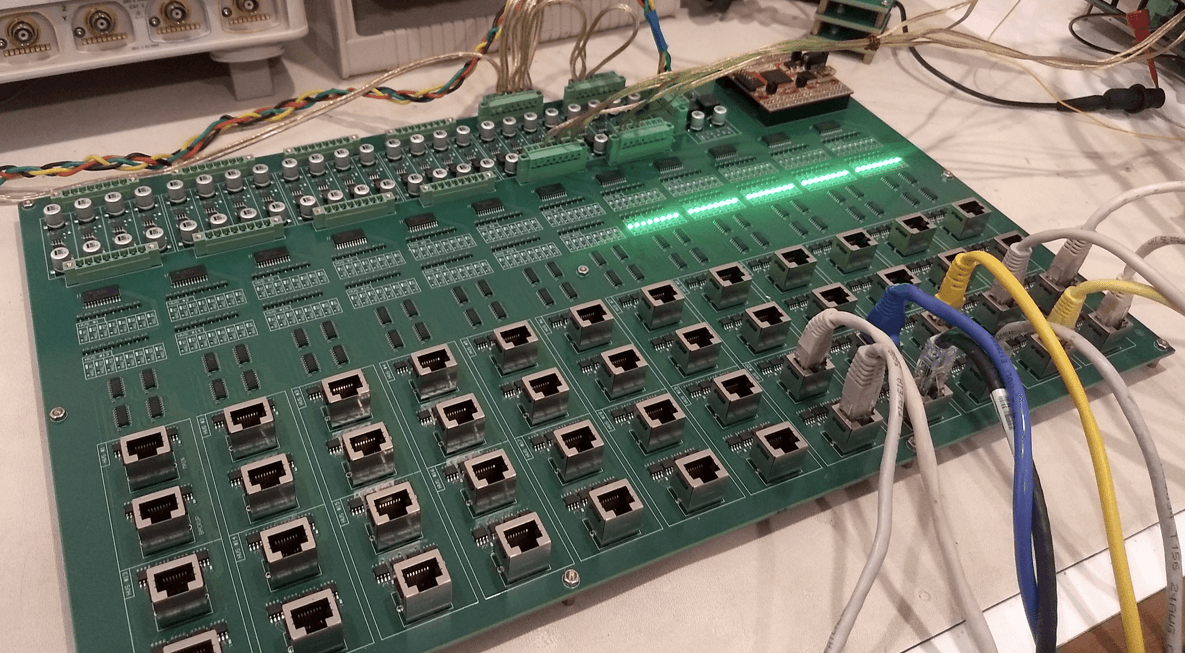}}}
	\caption{Hardware used for burn-in testing.}
	\label{burn-in_hw}
\end{figure}

Figure \ref{burn-in_hw} shows one HVS connected to the burn-in setup. A single $50\:\Omega$ termination is used per group of 16 channels to load each HVU, as opposed to loading each individual channel, to simplify the connection and testing process. The communication and power delivery are handled by a custom-designed PCB named Burn-in Control Board (BCB), shown in Figure \ref{burn-in_hw}(b), which can be connected with up to 24 HVS boards simultaneously. LEDs on the BCB are used to quickly check whether the voltage readout of the HVUs is outputting the correct value.

\subsection{Functional testing}
\label{subsection:functional_testing}

The final stage of production involved verifying the proper operation of the HVS boards through functional testing after they had undergone burn-in. For this purpose, a custom PCB was designed, referred to as the HVS Test Board (HVSTB), along with the development of custom firmware and software to run a fully automated test to qualify the most important aspects of the HVS's operation.

\begin{figure}[t]
	\centering
	\includegraphics[width=1\linewidth]{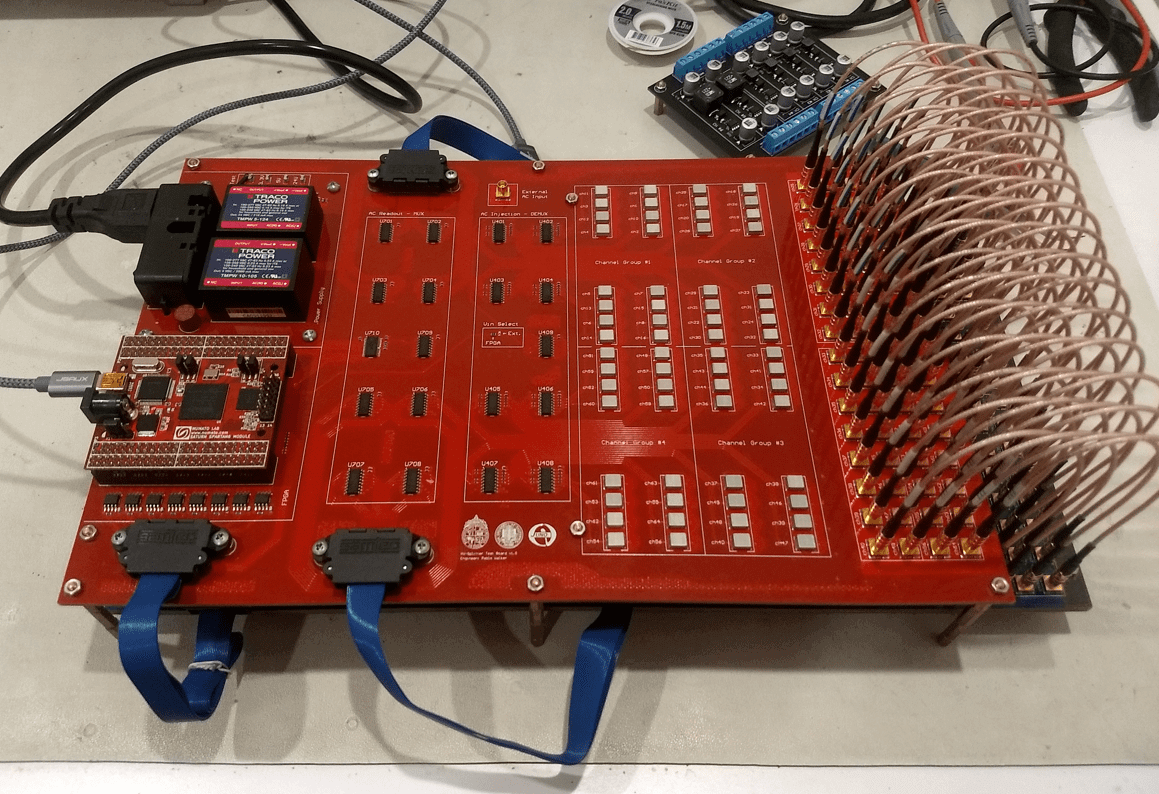}
	\caption{The HVS Test Board connected to and mounted on top of a HVS board.}
	\label{HVSTB_setup}
\end{figure}

The functional test routine consisted of three individual tests: (i) the HVU communication test, which evaluates the proper functionality of the HVUs; (ii) the AC injection test, which evaluates the continuity and integrity of the physics signal path; and (iii) the DC readout test, which checks for voltage drops and leakage.

\subsubsection{Physical design and interfaces}

Figure \ref{HVSTB_setup} shows the HVSTB mounted on top of and connected to the HVS board. The HVSTB shares some design elements with the HVS, such as part of the layout and components of the high-voltage section, which are prominently visible on the right side of Figure \ref{HVSTB_setup}. The HVSTB has a Spartan 6 FPGA development board mounted in the bottom-left corner, as visible in Figure \ref{HVSTB_setup}, which handles all the digital processing and connects to a PC through USB.

The HVSTB connects to the HVS via its intended high-voltage I/Os using short RG-178 cables with MCX connectors, providing a robust and safe connection between the boards. Alternative methods of connection which could have helped expedite the testing process were analyzed and ultimately rejected, such as test points and spring-loaded pins, as the high voltage and insulation on the HVS made them difficult to integrate into the testing process.

\subsubsection{Circuit schematic}

\begin{figure}[t]
	\centering
	\includegraphics[width=1\linewidth,trim={0 0.7cm 0 2cm},clip]{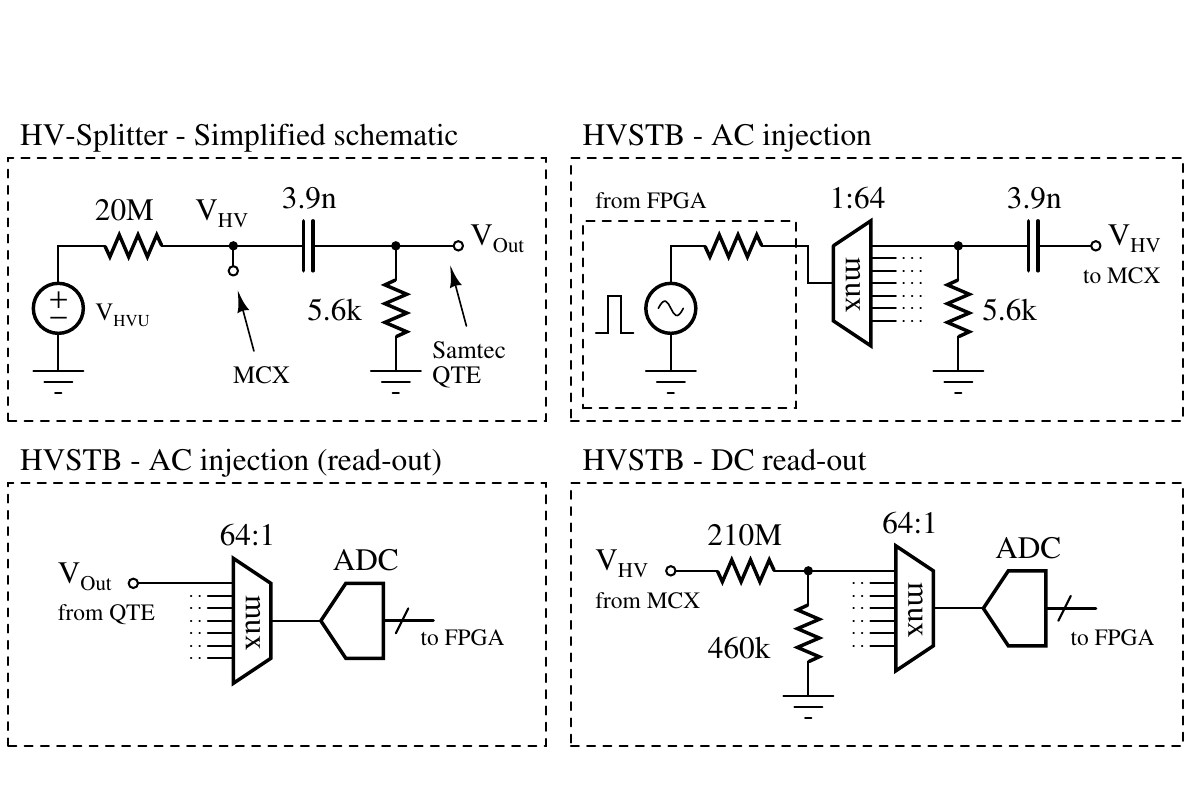}
	\caption{Simplified schematic of the HVS and the testing subcircuits of the HVSTB. The multiplexers shown in the schematic are in fact multiplexer banks, and the combined operation of each bank is that of a single analog multiplexer with a 64:1 ratio.}
	\label{HVSTB_sch}
\end{figure}

Figure \ref{HVSTB_sch} shows a simplified schematic of the HVS and HVSTB, detailing the subcircuits of the HVSTB related to the AC injection and DC readout tests.

The signal path for the AC injection test starts at the circuit block shown on the top-right side of Figure \ref{HVSTB_sch}, followed by a connection to the HVS, a simplified schematic of which is shown on the top-left, and ends at the circuit block shown in the bottom-left corner. A square pulse is generated by the FPGA, which connects to the high-voltage node of one of the channels on the HVS through a multiplexer and a coupling capacitor, is decoupled on the HVS from the high-voltage DC component, and is directed to an ADC for sampling by another multiplexer on the HVSTB. Given that the use of multiplexers allows for the selection of any combination of channels between the injection and readout paths, this circuit was also used to check for short circuits between adjacent channels by selecting pairs of channels known to be neighbors at the layout level on the HVS.

The circuit block corresponding to the DC readout test is shown in the bottom-right of Figure \ref{HVSTB_sch}. For this test, each channel has a resistive divider to lower the voltage within the dynamic range of the ADC. The component tolerances of the $20\mathrm{\:M}\Omega$ resistor on the HVS and the $210\mathrm{\:M}\Omega$ and $460\mathrm{\:k}\Omega$ resistors on the HVSTB add random variation to the voltage measured by the ADC, which were taken into account when analyzing the results.

\subsubsection{Test routine and acceptance criteria}

Before the test routine begins, the FPGA configures and turns on the high-voltage output of the HVUs. A minute is given for the voltage to stabilize, after which the test sequence is initiated. The process is automated, with the FPGA firmware and software working together as follows:

\begin{figure}[t]
	\centering
	\includegraphics[width=0.85\linewidth]{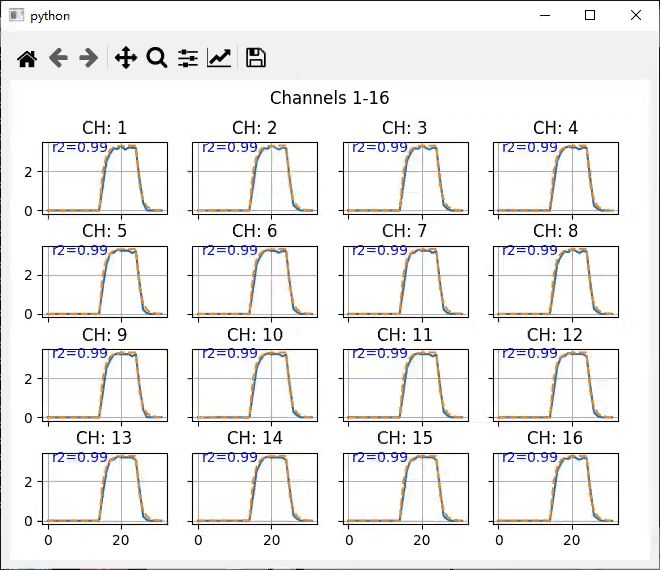}
	\caption{Plots generated by the functional testing software for the AC injection test under normal conditions. The x-axis represents the vector index number, and the y-axis represents voltage. The sampling period is 40~ns, and the injected voltage pulse has a width of 400~ns.}
	\label{ACinj_plots}
\end{figure}

\begin{enumerate}[(i)]
	\item \textbf{HVU Communication:} The software retrieves monitoring voltage values from the HVUs to verify that the communication is working properly and the high-voltage output is within expected ranges.
	
	\item \textbf{AC Injection:} A low-voltage square pulse is injected into all HVS channels---one channel at a time---travels through the physics signal path, and is sampled by the HVSTB ADC. This test checks AC continuity and signal integrity by comparing the waveform to a standard example. If significant deviations are found, the board is marked for further inspection. An example of the type of results obtained through this test is shown in Figure \ref{ACinj_plots}.
	
	\begin{itemize}
	\item \textbf{Short-circuit detection:} A subroutine is performed in which signal is injected and measured between neighboring channels on the top layer to check for short circuits due to soldering defects.
	\end{itemize}
	
	\item \textbf{DC Readout:} The high-voltage output on each channel on the HVS is read through a resistive divider followed by an ADC on the HVSTB. If the measured value deviates from the expected thresholds calculated from component tolerances, it could indicate current leakage and the board is marked for further inspection.
\end{enumerate}

\subsubsection{Issues encountered}

During the functional testing process at the factory, several multiplexer chips used for the AC injection test were damaged, resulting in test signals that either had reduced amplitude or were entirely absent. Due to scheduling constraints, an in-depth investigation of the issue was not possible, so the damaged components were replaced, and functional testing continued until all HVS boards were qualified. Although the source of the issue remains unclear, it is likely that the components were damaged by ESD or another form of overcurrent during an unidentified transient. Attempts were made to replicate the issue on a different setup, but it could not be reproduced.

\section{Conclusions}

The High Voltage Splitter (HVS) board has been designed, qualified, and produced for installation in the JUNO detector as part of the SPMT system. Its primary functions are to provide high voltage for SPMT biasing and to decouple the physics signal from said voltage. Every aspect of the design, from electrical performance to physical layout, was analyzed and refined through prototyping and iteration, ensuring a balance between channel density, signal integrity, insulation, and long-term reliability in the underwater detector environment.

This paper provides a thorough description and analysis of the design, performance, and manufacturing process. Through careful component selection, optimized clearances, and protective coatings, the HVS is expected to operate reliably throughout JUNO’s 20-year operational lifetime. Its design and validation offer valuable insights for future high-voltage electronics in large-scale neutrino experiments and other particle physics applications.

\section*{Acknowledgments}

We thank the Axon Cable Company for the development and production of custom cables and connectors for the SPMT system. This work was supported by the Agencia Nacional de Investigación y Desarrollo (ANID)---under the Fondecyt and Quimal programs---and the Instituto Milenio SAPHIR (ICN2019\_044) in Chile; by the Chinese Academy of Sciences (CAS) in China---under the Strategic Priority Research Program; and by the National Science Foundation (NSF) and the University of California, Irvine (UCI) in the USA.

\bibliographystyle{elsarticle-num}


\end{document}